# SOME ASPECTS OF LOW ENERGY PROPERTIES OF NUCLEONS

**A**
**THESIS**
**SUBMITTED FOR THE AWARD OF THE DEGREE OF**

## DOCTOR OF PHILOSOPHY

**IN**
**PHYSICS**

**BY**

**ALKA UPADHYAY**

**DEPARTMENT OF PHYSICS**
**FACULTY OF SCIENCE**
**THE M.S. UNIVERSITY OF BARODA**
**VADODARA-390002, GUJARAT, INDIA**

**DECEMBER 2005**

## *ACNOWLEDGEMENT*

*I take this opportunity to place on record my heartfelt thanks to all those who helped me materialize this arduous task. I wish to express my deep sense of gratitude and indebtness to my research guide Dr. J.P. Singh, Department of Physics whose suggestions and criticism have been indispensable in the successful completion of this work. His valuable guidance, admirable patience and constant inspiration to me during my research work has motivated me during my research work.*

*I would like to thank Prof. C.F. Desai, Head, and Prof. D.R.S Somayajulu, former Head, Department of Physics, for the Departmental facilities.*

*I greatly acknowledge the support and encouragement of my colleagues and friends at M.S University of Baroda, and how can I forget the best interaction which I had with teachers and friends at various institutes like TIFR, HRI, II.Sc, SINP, who have been a constant motivators for my research work.*

*I owe my special thanks to Profs. J. Pasupathy, N.L. Singh, A.C Sharma, and Dr. J. S. Bandukwala for their encouragement, support and persuasion to complete this work.*

*I also wish to thank all the members of the teaching staff and non teaching staff of the Physics Department for the help rendered by them from time to time.*

*Special thanks to my best friends Sarita, Pomita, Vikram, Kumar Rao, Mukesh, Rehman, Rohit, Avdhesh, Nilam, Ritesh, Mounik who have helped me whenever their need was felt. I also acknowledge the cooperation and company of my all seniors/co-researchers, since their presence along all this way would be an unforgettable experience.*

*I find myself unable to express in words the extreme love, affection, blessings, I have received from my parents. I consider it too formal to thank my brothers, sister and other close relatives and friends. But in a way I have registered thanks to them also. I also express my hearty gratitude to all my beloved family members for their constant encouragement and support.*

*Alka Upadhyay*

# ***PREFACE***

This thesis is devoted to the study of the properties of nucleons at low energies. The vast available data on nucleus indicate that a nucleon is a complicated system of strongly interacting quarks and gluons. Naturally, we require either some nonperturbative approach or some model which can capture relevant physics. Chapter I is introductory, in which we review the relevant developments in the subject which will be useful in the course of investigating the problem discussed in the following chapters.

In chapter II, we investigate some properties of nucleons related to their spin using a statistical model. The various quark–gluon Fock states of a nucleon have been decomposed in a set of states in which each of the three-quark core and the rest of the stuff, termed as a sea, appears with definite spin and color quantum number, their weight being determined, statistically, from their multiplicities. We have also considered two modifications of this model with a view to reduce the contributions of the sea components with higher multiplicities. With certain approximations, we have calculated the quark contributions to the spin of the nucleon, the ratio of the magnetic moments of nucleons, their weak decay constant, and the ratio of SU(3) reduced matrix elements for the axial current.

In chapter III, we have investigated isospin breaking in the diagonal pion-nucleon coupling constant ($g_{\pi NN}$) using conventional QCD sum rule. The effect of quark mass dependent terms, $\pi^0$-$\eta$ mixing and electromagnetic corrections to meson-quark vertices have been included as well. Some of the implications of the isospin splitting have also been discussed.

In chapter IV, we investigate gluonic contributions to the nucleon self-energy in an effective theory. The couplings of the topological charge density to nucleons give rise to OZI violating η-nucleon and η'-nucleon interactions. The one-loop self-energy of a nucleon arising due to these interactions have been calculated regularizing the divergences using form factors.

In chapter V, we have studied anomaly-anomaly correlator using QCD sum rule. This has been used to evaluate the first derivative of the topological susceptibility at zero momentum, a quantity which is useful in the discussion of the proton-spin problem.

Finally, we conclude with some future outlooks.

# *LIST OF PUBLICATIONS*

# ***SYNOPSIS***

To understand the observed properties of nucleons from the underlying theory of strong interaction, quantum chromodynamics (QCD), is a challenging task since the QCD remains intractable at low-energy. Among the attempts made in dealing with the strong interaction at low energy scales is the QCD sum rule approach. Here the hadronic parameters are determined with a controllable accuracy by circumventing the difficulty of performing direct nonperturbative QCD calculations through a combination of operator product expansion and phenomenological identification of nonperturbative quantities as quark condensate, gluon condensate, etc. It is also possible to construct QCD based effective theory which describes the processes of strong interaction at low energies. This theory is self consistent in terms of expansion in powers of particle momenta, and masses of Goldstone bosons and the η' mass in the chiral limit. In another approach, there are models which are believed to represent various aspects of nonperturbative structure of QCD. In one such model, nucleon is considered as a statistical system, and a nucleon state is expanded in terms of quark and gluon Fock states. In the present thesis, we have used the above three approaches to study various low-energy properties of a nucleon appropriate for the method concerned.

The composition of nucleons in terms of fundamental quark and gluon degrees of freedom has been modeled variously to account for their observed properties. In the first part of this thesis, we work in a statistical model in which a nucleon is taken as an ensemble of quark-gluon Fock states. A spin up nucleon state has been expanded in Fock states consisting of three valence quarks and a sea consisting of quarks, antiquarks and gluons, and containing up to five constituents which have definite spin and color quantum numbers. The expansion of a Fock state into spin and color states has been done using the assumption of equal probability for each sub state of such a state. We also use the approximation in which a quark in the core is not antisymmetrized with an identical quark in the sea, and have treated quarks and gluons as non-relativistic particles moving in S-wave motion. We have not taken into account any contribution of s-quark and other heavy quarks, and have covered only ~ 86% of the total Fock states. The remaining Fock states have been assumed to be decomposed in approximately same proportion as the earlier discussed case. With these approximations, we have calculated the quark contribution to the spin of the nucleons, the ratio of the magnetic moments of the nucleons, their weak decay constant and the ratio of SU(3) reduced matrix elements for the axial current. All of these quantities give the integrated results of the Bjorken variable. We have also considered two modifications of the above statistical approach with a view to reduce the contributions of the sea components with higher multiplicities, and done the above calculations for these two cases as well. Our results of calculation hold good for a typical hadronic energy scale~$1GeV^2$. The use of the Melosh rotation, which takes care of relativistic effect of the quark intrinsic transversal motion inside the nucleon, makes agreement with the data better.

Determination of meson-nucleon coupling is of particular interest in the study of nucleonic properties. It serves as an useful test of low-energy behaviour of QCD, and is an important parameter in the construction of effective field theories with nucleons and mesons as explicit degrees of freedom. In the second part of this thesis, we investigate the isospin splitting in the diagonal pion-nucleon coupling constant $\delta g$, by studying the vacuum-to-pion matrix element of the correlation function of interpolating fields of a nucleon in the frame work of conventional QCD sum rule. QCD sum rule has also been used earlier in the literature to investigate the pion-nucleon coupling constant ($g_{\pi NN}$). In the existing calculation, we have included quark mass dependent terms, $\pi^0$-$\eta$ mixing term and electromagnetic correction to the quark-meson vertices. In order to reduce the direct dependence of $\delta g$ on the isospin splitting of quark condensates, the sum rules for the proton and the neutron have been divided with their respective chiral-odd mass sum rules. Taking into account the different ranges of values used in the literature for quark condensate, gluon condensate, twist-4-parameter, and continuum threshold, we obtain a range of $\delta g$ and $g_{\pi NN}$: $\delta g = -(4.99 \pm 1.97) \times 10^{-2}$ and $g_{\pi NN} = 11.44 \pm 2.76$. These results have been compared with the corresponding results found in the literature. Contributions to $\delta g$ for its central value coming from various symmetry breaking parameters (mixing angle, fine structure constant, quark mass difference, nucleon mass difference and the isospin splitting of the quark condensate) have also been calculated and it has been found that they individually add up almost linearly to give final value of $\delta g$ when these are all taken to be non-zero.

We have also calculated the difference of pp- and nn- scattering lengths by using the above found result of $\delta g$ and $g_{\pi NN}$ in the phenomenological Argonne $v_{18}$ potential disregarding the electromagnetic potential part. This gives a range of the values of the difference of scattering lengths, which covers the experimental value.

In the third part of our thesis, we investigate nucleon self-energy due to gluonic interaction in an effective theory. The one-particle irreducible coupling of the topological charge density (Q) to the nucleon ($g_{QNN}$) is, in part, related to the amount of the spin carried by polarized gluons in a polarized proton, and is expected to be large. The mixing of the gluonic term Q to the flavor singlet would-be-Goldstone boson $\eta_0$ generates masses largely to the η’ and to some extent to the η. OZI violation in the η’-nucleon system is a probe of the role of the gluons in a dynamical chiral symmetry breaking in low-energy QCD. We have calculated the nucleonic mass arising due to the above gluonic interaction within the framework of heavy baryon chiral perturbation theory which includes the η’ as well. This calculation has been done by restricting to one-loop diagrams of the η and η’ with the vertices arising due to gluonic interaction with the nucleon. Divergences arising from loop diagrams have been regularized using various types of phenomenological form factors. The non-trivial structure of the QCD vacuum has also been taken into account. The gluonic contribution to the nucleon self-energy, obtained this way, is over and above the contributions associated with meson exchange models. This gives a contribution to the nucleon mass which is (2.5-7.5)% of the nucleon mass and negative as compared to the one-loop pion contribution which is typically (10-20)% of the nucleon mass and negative.

In the fourth part of the thesis, we have studied the anomaly-anomaly correlator, using QCD sum rules. Using the matrix element of anomaly between vacuum and pseudoscalars π, η, and η’, the derivative of the correlator at zero momentum $\chi'(0)$ has been evaluated and found to be $\approx 1.82\times10^{-3}\text{GeV}^2$. The singlet axial charge of a nucleon is related to $\chi'(0)$ and hence its evaluation is useful for the discussion of the proton-spin problem. Assuming that $\chi'(0)$ has no significant dependence on quark masses, the mass of η’ in the chiral limit is found to be ≈723MeV. The same calculation also yields for the singlet pseudoscalar decay constant in the chiral limit a value of ≈178MeV.

We conclude that nucleon is a many body complex system whose low-energy behaviour is determined mainly by strong interaction. Non-perturbative approach to QCD, such as QCD sum rule and the QCD based effective theory, and the models such as a statistical model, have a complementary role in exposing different aspects of nucleonic properties.

# *CONTENTS*

# CHAPTER-I

## NUCLEONIC PROPERTIES: AN INTRODUCTION AND OVERVIEW

### 1.1 Introduction

Dramatic progress has been made in particle physics during the past four decades [1]. A series of important experimental discoveries have firmly established the existence of sub-nuclear worlds of quarks and leptons. The nucleons, i.e, proton and neutron which form nuclei are no longer regarded as elementary particles but are found to be made up of quarks. Later on, the quark structure of the nucleon was directly observed in deep inelastic electron scattering experiments.

The dynamics of quarks and leptons can be described by an extension of the sort of quantum field theory (QFT) that proved successful in describing electromagnetic interaction of charged particles, QED. To be more precise, the fundamental interactions are widely believed to be described by QFT possessing local gauge symmetry [2], whereby the interaction between quarks and leptons are being discussed through the exchange of gauge field quanta, mainly photons, gluons and weak bosons. The short range attractive force is responsible for binding the nucleons inside the nucleus. The fact that the large variety of nuclei are constructed out of nucleons makes their study interesting. Hence the internal structure of the nucleon is of fundamental importance in nuclear and particle physics, to both experimentalists and theorists.

In 1933, Frisch and Stern performed the first experiment for measuring the magnetic moment of the proton. These measurements are the experimental evidence for the internal structure of the nucleon which says that nucleon is not a point-like particle. The anomalous magnetic moment of the proton was determined to be 2.5 times as large as one would expect for a spin 1/2 Dirac particle (the actual value is 2.793 nucleon magneton).

In 1935 Yukawa proposed, in analogy with quantum electrodynamics (QED), that the nuclear forces were due to the exchange of quanta of finite mass, a meson. The interaction between two nucleons would proceed via the exchange of a virtual intermediate meson. A simple calculation based on the uncertainty principle shows that for a range of $1.4\times10^{-15}$ m for the strong force the exchanged meson must have a mass of about 140 $MeV/c^2$ in contrast to the infinite range of the electromagnetic force which is due to the fact that photon is massless.

As far as the fundamental constituents of matter were concerned, it appeared, by 1939, that the proton, neutron, electron and neutrino are the constituents of matter, supplemented by photon and the hypothesized Yukawa meson as the mediating particles of the electromagnetic and strong interaction respectively.

Low energy properties of nucleons can be studied in various approaches. There are various models of nucleons in terms of their elementary constituents which are suitable to study some aspects of their properties. The QCD sum rules have been extensively used to investigate nonperturbative regions of hadronic physics. Effective theories based on symmetries of QCD are fit to study the low energy interactions among hadrons. In the present thesis, we shall use some of these approaches to study some aspects of low energy properties of nucleons. In the following we shall give a brief introduction of relevant developments in the subject which will be useful in the course of investigating the problems discussed in the following chapters.

## 1.2 The Quark Model

In 1963 Gell-Mann and Zweig [3] proposed a model that explained the spectrum of strongly interacting particles (i.e hadrons) in terms of elementary constituents called quarks. The quark model was developed to account for the regularities observed in the hadron spectrum, with hadrons interpreted as bound states of localized but essentially non-interacting quarks. It provides us a simple picture of internal structure of hadrons and an effective way to describe their dynamics at high energy. Much of the success of the model lies in the circumstance that to a reasonably good approximation we can regard quarks as free or weakly interacting particles (except for the confining mechanism). Mesons were expected to be quark-antiquark bound states. Baryons were interpreted as bound states of three quarks. The quark constituents of the baryons are assigned to have spin ½ from the observed spins of low-lying baryons.

The low-lying baryons were interpreted in the quark model as symmetric states of space, spin and $SU(3)_f$ flavor degrees of freedom. However, Fermi-Dirac statistics requires a total antisymmetry of the wave function. The resolution of this dilemma come through the introduction of color degree of freedom. The baryon wave functions are totally anti-symmetric in the color degree of freedom. Of course, the introduction of another degree of freedom would lead to a proliferation of states, so the color degree of freedom had to be supplemented by a requirement that only color singlet states exist in nature. Hence proton would be a bound state of (uud) and neutron would be a bound state of (udd) quarks which makes them color singlet. This model had great success in predicting new hadronic states, and in explaining the strength of electromagnetic and weak interaction transitions among different hadrons. In particular, it naturally incorporates the most important symmetry relations among hadrons.

Once quark structure of hadrons got some acceptance, it became natural to look for the dynamics obeyed by the quark system responsible for the composition of hadrons as well as for hadronic reactions. In order to get experimental information on quark dynamics, the most sensible way, is to probe the inside of hadrons, (e.g., proton) by applying a beam of structureless particles such as leptons. We need much higher energies and larger momentum transfers for the study of hadronic structure to have higher resolutions. The electromagnetic form factors are key ingredients to the understanding of the internal structure of composite particles like the nucleon, since they contain the information about the distributions of charges and currents. The knowledge of hadron form factors, especially for the nucleons and the pions, represent an important source of information about their electromagnetic structure. By varying the momentum transfer, large as well as small distances can be explored, allowing one to learn about hadronic physics. De-Broglie wavelength of an electron becomes much shorter than the size of a typical nucleus at sufficiently high energies in GeV range. In such cases, the scattering result is dominated by the charge distributions within individual nucleons. The primary interest of scattering at these energies shifts to the structure of nucleon rather than that of nucleus.

The quarks are classified as "light" or "heavy" depending on their entries in the mass matrix m of QCD Lagrangian equation. These masses are "running" as well: they depend on the scale μ at which they are determined. The masses of the lightest (u and d) quarks, $m_{u,d} <$ 10MeV (estimated at a renormalization scale μ~1 GeV) are very small compared to typical hadron masses of order 1 GeV, such as those of the ρ meson or the nucleon. The strange quark mass, $m_s$； (100 -150) MeV is an order of magnitude larger than $m_{u,d}$ but still counted as "small" on hadronic scales. The charm quark mass $m_c$； (1.1-1.4) GeV takes an intermediate position while the b and t quarks $m_b$ ； (4.1-4.4)GeV, $m_t$=(174±5)GeV) fall into the "heavy" category. These different quark masses set a hierarchy of scales, each of which is governed by distinct physics phenomena.

## 1.3 The Parton Model

The first series of experiments to study the structure of proton was initiated in 1960’s at SLAC and the process was called electron-proton deep inelastic scattering (DIS). For DIS, the momentum transfer squared $q^2$ is so large so that the spatial resolution for observing the target nucleon (proton) by projectile electron is high. DIS experiments are of utmost importance since it helps in revealing the internal structure of the proton. The finite size of the proton was measured to be about 0.8 fm.

In 1969 Bjorken [4] reported the scaling property of structure function in electron-nucleon scattering which was expected in the deep inelastic region where momentum transfer squared $q^2$ and energy transfer ν of electron are very large with the ratio $q^2/\nu$ kept fixed. It is claimed that structure function in the deep inelastic region depend only on the ratio $q^2/\nu$ rather than on two independent variable $q^2$ and ν. Bjorken scaling is obtained by assumption of the existence of free independent point-like particles (partons) inside proton. Conversely, it suggests that the quark dynamics must have the property of asymptotic freedom, i.e, the coupling constant decreases at short distances, hence quark interaction gets weaker at short distances.

The correlation pattern of energy and angular distribution of the scattered leptons in the DIS can be described simply by Feynman's parton model [5]. The essence of the parton model is the assumption that, when a sufficiently high momentum transfer reaction takes place, the projectile, be it a lepton or a parton inside a hadron, sees the target as made up of almost free constituents, and is scattered by a single, free, effectively massless constituent. Moreover the scattering from individual constituents is incoherent. The picture thus looks much like the subnuclear version of the impulse approximation of high energy scattering of composite particles with weakly bound constituents. The inclusive scattering is viewed as due to incoherent elastic scattering from point-like constituents of the nucleons: partons. The final state partons then recombine somehow into hadronic states. These partons were later identified as quarks, since experimentally it was suggested that their quantum numbers such as charges and spins were practically the same as those of quarks.

## 1.4 Quantum Chromodynamics (QCD)

Quantum chromodynamics (QCD) is the theory of strong interaction with interacting quarks and gluons. It is well tested in the high energy regime where perturbative QCD is applicable. Understanding confinement and hadronic structure in the non-perturbative region of QCD remains a challenge. It describes the interactions of quarks, via their color quantum numbers. It is an unbroken gauge theory and the gauge bosons are gluons.

It is a consistent quantum field theory with a simple and elegant underlying Lagrangian, based entirely on the invariance under a local gauge group, $SU(3)_{color}$. Out of this Lagrangian emerges an enormously rich variety of physical phenomena, structures and phases. Exploring and understanding these phenomena is undoubtedly one of the most exciting challenges in modern science.

In QCD, which is to some extent similar to QED, the fundamental interactions are between spin ½ quarks and massless spin 1 gluons. The quarks and gluons carry a new quantum number called color. Each quark can exist in three different color states and each gluon in eight color states. Under an SU(3) group of transformations which mixes up colors, the quarks and gluons are said to transform as a triplet and an octet respectively. No physical particle with the attribute of color has ever been found, so it is believed that all particles are 'color neutral'. By this we mean that all physical states must be invariant, or singlets under color transformations.

The elementary spin- ½ particles of QCD, the quarks, come in six species, or flavors, grouped in a field $\psi(x) = (u(x), d(x), s(x), c(x), b(x), t(x))T$. Each of the u(x), d(x), etc., is a four-component Dirac spinor field. Quarks experience all three fundamental interactions of the Standard Model [2]: weak, electromagnetic and strong. Their strong interactions involve $N_c$ = 3 "color" charges for each quark. These interactions are mediated by the gluons, the gauge bosons of the underlying gauge group of QCD, $SU(3)_{color}$.

The Lagrangian density of QCD [6] in terms of quark and gluon degrees of freedom for interacting quarks with masses $m_i$ is given by the equation

$$L_{QCD} = -\frac{1}{4} F^{A}_{\mu\nu} F^{\mu\nu}_{A} + \sum_{i}^{n} \overline{q^{a}_{i}} (i\not{D} - m_i)_{ab} q^{b}_{i} - \frac{1}{2\lambda} ( {}^{\mu} A^{A}_{\mu})^2 + L_{Ghost} \tag{1.1}$$

Here $q^a_i$ are quark fields with mass $m_i$, $A^A_\mu$ is the gluon field and the covariant derivative is given by

$$(D_\mu)_{ab} = \delta_{ab} \; {}_{\mu} + i g_s (t^c A^c_\mu)_{ab} \tag{1.2}$$

Under local gauge transformations they transform as ($t^a = \frac{\lambda^a}{2}$ are Gell-Mann matrices of SU(3) group).

$$q_a(x) \quad q'_a(x) = \exp(it.\theta(x))_{ab} q_b(x) = \Omega(x)_{ab} q_b(x), \quad t.\theta = t^c \theta^c \tag{1.3}$$

$$t.A_\mu \quad t.A'_\mu = \Omega(x) t.A_\mu \Omega^{-1}(x) - \frac{1}{ig} ( \; {}_{\mu} \Omega(x)) \Omega^{-1}(x) \tag{1.4}$$

$$D_\mu q(x) \quad D'_\mu q'(x) = \Omega(x) D_\mu q(x) \tag{1.5}$$

The non-Abelian field strength tensor is given by

$$F^A_{\mu\nu} = \; {}_{\mu} A^A_\nu - \; {}_{\nu} A^A_\mu - g_s f^{ABC} A^B_\mu A^C_\nu \tag{1.6}$$

which transforms as

$$t.F_{\mu\nu} \quad t.F'_{\mu\nu} = \Omega(x) t.F_{\mu\nu} \Omega^{-1}(x) \tag{1.7}$$

With the transformations (1.3), (1.5) and (1.7), it is easy to see that $L_{QCD}$ remains invariant under local gauge transformations.

The extra term in $F^{A}_{\mu\nu}$ makes it invariant under non-Abelian gauge transformation. This extra term has profound consequences for the theory: it means that gluons are self-interacting through three- and four-point vertices. This will turn out to give rise to asymptotic freedom at high energies and strong interactions at low energies, among the most fundamental properties of QCD.

Finally, it turns out that in a non-Abelian gauge theory, it is necessary to add one extra term to the Lagrangian density, related to the need for ghost particles. Basically they arise because when a non-Abelian gauge theory is renormalized it is possible for unphysical degrees of freedom to propagate freely. These are cancelled off by introducing into the theory an unphysical set of fields, the ghosts, which are scalars but have Fermi statistics. For practical purposes it is enough to know that there exist Feynman rules for ghosts and that in every diagram with a closed loop of internal gluons, we must add a diagram with them replaced by ghosts. It is worth noting that in physical gauges, as the name suggests, ghost contributions always vanish and they can be ignored.

QCD has similar structure as QED, but with one important difference; the gauge group is non-Abelian SU(3), and gluons are self interacting. The non-linear three- and four-point couplings of the gluon fields $A^{A}_{\mu}$ with each other are at the origin of the very special phenomena encountered in QCD and strong interaction physics. Hence the theory is asymptotically free (i.e coupling constant decreases at short distances) at high-energy and grows strong at low energies. These interactions are confining and dictates that quarks must be confined within a region of about ~1 Fermi in radius to give a hadron, so one would expect that as two or more nucleons approach each other within a nucleus, quarks and gluons should take over the dynamics and show up in observables. The only stable color singlets are quark-antiquark pairs, mesons, and three quark states, baryons.

There exist two limiting situations in which QCD is accessible with "controlled" approximations. At momentum scales exceeding several GeV (corresponding to short distances, r<0.1 fm), QCD is a theory of weakly interacting quarks and gluons (perturbative QCD). At low momentum scales considerably smaller than 1GeV (corresponding to long distances, r>1 fm), QCD is characterized by confinement and a non-trivial vacuum (ground state) with strong condensates of quarks and gluons. Confinement is believed to be behind the spontaneous breaking of a symmetry which is exact in the limit of massless quarks: chiral symmetry. Spontaneous chiral symmetry breaking in turn implies the existence of pseudoscalar Goldstone bosons. For two flavors ($N_f = 2$) they are identified with the isotriplet of pions ($\pi^+$, $\pi^0$, $\pi^-$). For $N_f = 3$, with inclusion of the strange quark, this is generalized to the pseudoscalar meson octet. Low-energy QCD is thus realized as an Effective Field Theory (EFT) in which these Goldstone bosons are the active, light degrees of freedom.

## 1.5 Asymptotic Freedom and Confinement

The property of QCD that led directly to its discovery as a candidate theory of the strong interaction is asymptotic freedom, i.e., coupling strength decreases at short distance [7]. This property is due to the presence of gluons which carry color charge and have spin one. It can either be explained as a dielectric or a paramagnetic effect. In first case, one calculates the dielectric properties of the vacuum and ascribes the asymptotic freedom of the theory to the self interaction of the gluon field. In the later one, asymptotic freedom is explained as a paramagnetic effect due to the spin of the gluons.

The success of QCD in describing the strong interactions is summarized by two terms i.e asymptotic freedom and confinement and their importance can be better understood by recalling certain facts about strong interaction. Asymptotic freedom refers to the weakness of short distance interaction, while the confinement of quarks follows from its strength at long distances.

Confinement has a relatively simple interpretation for heavy quarks and the “string” of (static) gluonic field strength that holds them together, expressed in terms of a static potential. When light quarks are involved, the situation is different. Color singlet quark antiquark pairs prop out of the vacuum as the gluon fields propagate over larger distances. Light quarks are fast movers: they do not act as static sources. In this case the potential picture is not applicable. The common features of the confinement phenomenon can nevertheless be phrased as follows: non-linear gluon dynamics in QCD does not permit the propagation of colored objects over distances of more than a fraction of a Fermi. Beyond the one-Fermi scale, the only remaining relevant degrees of freedom are color-singlet composites (quasiparticles) of quarks, antiquarks and gluons.

Hadron spectra are very well described by the quark model, but quarks have never been seen in isolation. Any effort in scattering experiment leads only to the production of the familiar mesons and baryons. Evidently, the forces between quarks are strong. In QFT, when higher order effects in perturbation theory are taken into account, then couplings acquire momentum dependence. An isolated charge in vacuum polarizes the surrounding medium in virtual electron-positron pairs, which, in turn, screen its charge. Hence, when the charge of such a particle is measured by scattering another charged particle on it, the charge depends on the distance between these particles: the smaller the distance, larger is the charge since then the test charge can penetrate inside the charge cloud. In quantum theory, separation is inversely proportional to the momentum transferred. Thus, the result of scattering experiment can be summarized as:

$$\frac{d}{dt_0}\alpha(t_0) > 0 \,, \qquad (1.8)$$

where $\alpha$ is the fine structure constant and $t_0 = -\vec{k}^2$ is the momentum transferred. For QED, however, the charge is so small that $\alpha(t_0)$ does not become large until $t_0$ is of astronomical scale. In QCD, in addition to the processes which are already there in QED, we also have to include the processes arising out of three-gluon couplings. This makes a very important difference. The emission of a gluon "leaks away" the color charge of the heavy particle into the cloud of virtual particles. Thus, for small $t_0$, when the two heavy particles stay far apart, they are actually more likely to see each other's true charge. As $t_0$ increases, they penetrate further and further into each other's charge cloud and are less and less likely to measure the true charge. For this reason, we expect "antiscreening" for QCD:

$$\frac{d}{dt_0}\alpha_s(t_o) < 0 \,. \qquad (1.9)$$

To be more quantitative, let us define ($\mu^2 = -t_0$)

$$\mu \frac{dg_s(\mu)}{d\mu} = \beta(g_s(\mu)) , \tag{1.10}$$

where $g_s$ is the strong coupling constant and $\mu$ is the renormalization scale. It has been found that

$$\beta(g_s) = -g_s[\frac{\alpha_s}{4\pi}\beta_1 + (\frac{\alpha_s}{4\pi})^2\beta_2 + ........] , \tag{1.11}$$

$\beta_1 = 11 - 2n_f/3$.

Here $n_f$ is the number of quark flavors. For $n_f = 6$, $\beta_1$ is positive and $\beta$ negative. Differential equation (1.10) can be solved and, to the lowest-order, $\alpha_s(\mu^2)$can be written in terms of a single variable as

$$\alpha_s(\mu^2) = \frac{4\pi}{\beta_1 \ln(\frac{\mu^2}{\Lambda^2})} \tag{1.12}$$

where $\Lambda = \Lambda_{QCD}$, is a free parameter which sets the scale for the running coupling. The QCD scale parameter Λ is determined empirically (Λ； 0.2 GeV for $N_f = 4$). The fact that $\alpha_s$ decreases with increasing μ leads to the property known as "asymptotic freedom" in the domain μ？ 1 GeV in which QCD can indeed be treated as a perturbative theory of quarks and gluons. The theoretical discovery of asymptotic freedom was honored with the 2004 Nobel Prize in Physics. At the scale of the Z-boson mass, $\alpha_s(M_Z)$ ； 0.12 [8] . So while $\alpha_s$ is small at large μ, it is of order one at μ < 1 GeV. At low energies and momenta, an expansion in powers of $\alpha_s$ is therefore no longer justified: we are entering the region commonly referred to as non-perturbative QCD.

## 1.6 Operator Product Expansion

The operator-product expansion is a technique in which the singularities of the operator products are expressed as a sum of nonsingular operators with the coefficients being singular c-number functions [9]. The physical basis for this expansion is that a product of local operators at distances small compared to the characteristic length of the system should look like a local operator. In theories like QCD, the functions describing the singularities in this expansion have a momentum dependence governed by renormalization group equations; hence due to the asymptotic freedom, they can be calculated at large momenta using perturbation theory. Secondly, these functions exhibit the full symmetry of the underlying theory by possible spontaneous symmetry breaking.

It also enables us to extract a short distance piece in the scattering cross sections, which is calculable through the QCD Lagrangian by using renormalization group method. OPE can be defined using proper renormalization scale μ which is used to separate hard and soft momenta.

Wilson [9] hypothesized that the singular part as $x \to y$ of the product A(x)B(y) of two operators is given by a sum over other local operators $O_i$:

$$A(x)B(y) \qquad C_i(x-y)O_i(\frac{1}{2}(x+y)) \tag{1.13}$$

where $C_i(x-y)$ are singular c-number functions called Wilson coefficients. It has been proven for renormalizable theories that such expansions are valid as $x \quad y$ to any finite order of perturbation theory. The short distance behaviour of the Wilson coefficients is expected to be that obtained, up to a logarithmic multiplicative factor, by dimensional counting (x<<1/m)

$$C_i(x) \quad x^{d_i - d_A - d_B} (\ln xm)^p [1 + O(xm)] \qquad (1.14)$$

where $d_A$, $d_B$ and $d_i$ are the dimensions (in units of mass) of A, B and $O_i$ respectively. The higher the dimension of $O_i$ the less singular are the coefficients $C_i(x)$; hence the dominant operators at a short distance are those with the smallest dimensions.

The usefulness of this expansion derives from its universality: the Wilson coefficients are independent of the process under considerations. Process dependence is exhibited in the matrix element of the local operator $O_i$ which is nonsingular at short distances. Another advantage is that in a given theory the expansion usually involves a rather small number of operators. Hence the ensuing calculation is relatively simple.

## 1.7 Chiral Symmetry

Chiral symmetry is an internal symmetry of right and left handed spinors. It has importance in low energy hadronic physics, since its spontaneous breaking generates Goldstone bosons with negative parity, zero spin, unit isospin and zero baryon number called pions. Thus a broken approximate chiral symmetry entails the existence of pions where u and d quarks have small but non-zero masses whereby spontaneous breaking of a symmetry is expressed as the non-vanishing of the vacuum when operated by the charge Q. The transition from the fundamental to the effective level occurs via a phase transition due to spontaneous symmetry breaking generating (pseudo) Goldstone boson. A spontaneously broken symmetry relates processes with different numbers of Goldstone bosons. Since the masses of light quarks are small compared to $\Lambda_{QCD}$, let us set these parameters equal to zero in the first approximation and moreover, make the masses of heavy quarks, $m_c$, $m_b$ and $m_t$ to be infinity.

In this limit QCD Lagrangian $L_{QCD}$ becomes invariant under the following group of (space-time independent) transformations which act on the three flavor indices (u, d, s):

$$q = q_L + q_R \qquad q^{'} = g_R q_R + g_L q_L$$
$$q_R = \frac{1}{2}(1+\gamma_5)q, \tag{1.15}$$

$$q_L = \frac{1}{2}(1-\gamma_5)q$$
$$g_I g_I^{\dagger} = 1, \det g_I = 1; I = L, R \tag{1.16}$$

The above group of transformations (1.15) and (1.16) is $SU(3)_R \times SU(3)_L$ and the resulting symmetry of the QCD Lagrangian is called chiral symmetry of QCD. According to the Noether's theorem, there are $2\times(3^2\text{-}1) = 16$ conserved currents associated with this symmetry.

$$J_I^{\mu a} = \overline{q_I}\gamma_\mu T^a q_I$$

$${}_\mu J_I^{\mu a} = 0; I = L, R \ ; \ a = 1,\ldots\ldots.8 \tag{1.17}$$

The associated conserved charges

$$Q_I^a = {}_{\substack{x^0= \\ const}} J^{oa}{}_I d^3x$$

$$\frac{dQ_I{}^a}{dt} = 0 \tag{1.18}$$

generate the algebra of G= $SU(3)_R \times SU(3)_L$[10]

$$[Q_I^a, Q_I^b] = if^{abc} Q_I^c$$

$$[Q_L^a, Q_R^b] = 0 \tag{1.19}$$

Vector and axial charges can be defined as

$$Q_V^a = Q_R^a + Q_L^a, \ \ Q_A^a = Q_R^a - Q_L^a \tag{1.20}$$

It can be shown that the state of lowest energy is necessarily invariant under the vector charges: $Q_V^a |0\rangle$ =0. For axial charges, however the Wigner-Weyl realization of G, in which $Q_A^a |0\rangle$ =0 is not true, since that would imply that this spectrum contains degenerate parity partners forming multiplets of G. The real word has no parity doublets. For instance, the lightest meson, π(140) and the lowest state with the same spin and flavor, but of opposite parity $a_0$(980) have large mass difference; so is the case with N(940) and N*(1520). Hence it is believed that the alternative possibility called Nambu-Goldstone mode of G, in which $Q_a^A |0\rangle \quad 0$ is realized. In this case, the spectrum contains 8 Goldstone bosons, one for each broken generator, and they form degenerate multiplets of SU(3)   G. The eight lightest hadrons pions, kaons and η have desired quantum numbers of the Goldstone bosons, but they are not massless as required by Goldstone's theorem[11]. Using commutation relations of the vector charge with scalar currents and axial charges with pseudoscalar currents and using the fact that $Q_a^A |0\rangle \quad 0$ it can be shown that

$$\langle 0|u\bar{u}|0\rangle = \langle 0|d\bar{d}|0\rangle = \langle 0|s\bar{s}|0\rangle \quad 0 \qquad (1.21)$$

In the theory of superconductivity, a small electron-electron attraction leads to the appearance of a condensate of electron pairs in the ground state of a metal. In QCD, quark and antiquark have strong attractive interaction, and, if these quarks are massless, the energy cost of creating an extra quark-antiquark pair is small. Thus we expect that the vacuum of QCD will contain a condensate of quark-antiquark pairs. These fermion  pairs must have zero total momentum and angular momentum. They must contain net chiral charge, pairing left-handed quarks with the antiparticles of right-handed quarks. The vacuum state with a quark pair condensate is characterized by a nonzero vacuum expectation value for the scalar operator

$$\langle 0|\bar{q}q|0\rangle = \langle 0|\bar{q}_R q_L + \bar{q}_L q_R|0\rangle \quad 0 \qquad (1.22)$$

and hence is noninvariant with $g_L$ $g_R$. The expectation value signals that the vacuum mixes the two quark helicities. This allows the u and d quarks to acquire effective masses as they move through the vacuum.

Chiral symmetry breaking (CSB) is a nonperturbative phenomenon, which is known to govern the low energy properties of hadrons. The effective chiral Lagrangians have been proposed before the advent of QCD and the phenomenon of CSB and Nambu-Goldstone theorem was established more than 40 years ago.

## 1.8 PCAC

Let $|\pi_a(p)\rangle$ be the state vectors of the Goldstone bosons associated with the spontaneous breakdown of chiral symmetry. We choose the standard normalization $\langle \pi_a(p)|\pi_b(p')\rangle = 2E_p \delta_{ab}(2\pi)^3 \delta^3(\vec{p}-\vec{p}')$. Goldstone's theorem, implies non-vanishing matrix elements of the axial current which connect $|\pi_a(p)\rangle$ with the vacuum:

$$\langle 0|A_a^\mu(x)|\pi_b(p)\rangle = ip^\mu F_0 \delta_{ab} e^{-ip.x} \qquad (1.23)$$

The constant $F_0$ is called the pion decay constant (taken here in the chiral limit, i.e., for vanishing quark mass). Its physical value $f_\pi = (92.4 \pm 0.3)$ MeV is determined from the decay $\pi^+ \rightarrow \mu^+\nu_\mu + \mu^+\nu_\mu\gamma$. The difference between $F_0$ and $f_\pi$ is a correction linear in the quark mass $m_q$. Non-zero quark masses $m_{u,d}$ shift the mass of the Goldstone boson from zero to the observed value of the physical pion mass, $m_\pi$. The relationship between $m_\pi$ and the u and d quark masses is derived as follows. We start by observing that the divergence of the axial current is

$$ {}_{\mu}A_a^{\mu} = i\overline{\psi}\{m, \frac{\tau_a}{2}\}\gamma_5\psi \ , \tag{1.24}$$

where m is the quark mass matrix and {,} denotes the anti-commutator. This is the microscopic basis for PCAC, the Partially Conserved Axial Current (exactly conserved in the limit $m \to 0$) which plays a key role in the weak interactions of hadrons and the low-energy dynamics involving pions [10]. Consider for example the a = 1 component of the axial current:

$$ {}_{\mu}A_1^{\mu} = (m_u + m_d)\overline{\psi} i\gamma_5 \frac{\tau_1}{2}\psi $$

$$ \langle \pi, p | \ {}_{\mu}A_1^{\mu}(0) | 0 \rangle = +m_\pi^2 f_\pi \tag{1.25}$$

and combine this with $[Q_a^A, P_b] = -\delta_{ab}\overline{\psi}\psi$ where $P_a(x) = \overline{\psi}(x)\gamma_5\tau_a\psi(x)$ the pseudoscalar quantity, to obtain

$$ \langle 0 | [Q_1^A, \ A_1^{\mu}] | 0 \rangle = \frac{i}{2}(m_u + m_d)\langle \overline{u}u + \overline{d}d \rangle \tag{1.26}$$

Now insert a complete set of (pseudoscalar) states $|\pi, p\rangle\langle\pi, p|$ in the comutator on the left. Assume, in the spirit of PCAC, that this spectrum of states is saturated by the pion. Then use

Eq.(1.23) to evaluate $\langle 0 | Q_1^A | \pi \rangle \, and \, \langle 0 | \ A^{\mu} | \pi \rangle$ at time t = 0, with $E_p = m_\pi$ at $p$ =0. Since

$$ \frac{d^3p}{2E_p}\langle 0 | Q_1^A | \pi, p\rangle\langle\pi, p| \ {}_{\mu}A_1^{\mu} | 0 \rangle = +i \ \frac{d^3p}{2E_p} E_p f_\pi \delta^3(\vec{p}) m_\pi^2 f_\pi = \frac{i}{2} m_\pi^2 f_\pi{}^2 \ , \tag{1.27}$$

we  arrive at the Gell-Mann, Oakes, Renner (GOR) relation [12]:

$$m_\pi{}^2 f_\pi^2 = -(m_u + m_d)\langle \bar{q}q \rangle + O(m_{u,d}^2)$$

We have set $\langle \bar{q}q \rangle = \langle \bar{u}u \rangle = \langle \bar{d}d \rangle$ making use of isospin symmetry which is valid to a good approximation. Neglecting terms of order $m^2{}_{u,d}$ (identifying $F_0 = f_\pi$ = 92.4 MeV to this order) and inserting $m_u + m_d$ ； 14 MeV [13] at a renormalization scale of order 1GeV, one obtains $\langle \bar{q}q \rangle$ $(0.23 \pm 0.03\text{GeV})^3$ ； 1.6 $\text{fm}^{-3}$.This condensate (or correspondingly, the pion decay constant $f_\pi$) is a measure of spontaneous chiral symmetry breaking. The non-zero pion mass, on the other hand, reflects the explicit symmetry breaking by the small quark masses, with $m_\pi^2$ ： $m_q$ . It is important to note that $m_q$ and $\langle \bar{q}q \rangle$ are both scale dependent quantities. Only their product $m_q\langle \bar{q}q \rangle$ is scale independent, i.e., invariant under the renormalization group.

## 1.9 OZI Rule and its Violation

The phenomenologically-inspired OZI rule [14] states that “disconnected quark diagrams are suppressed relative to connected ones”, and it has served as an excellent guiding principle in the development of strong interaction theory and exception to the OZI rule, which are rare usually signify that some significant  new physics is involved.

The OZI rule was originally invented to explain why dominant decay mode of vector meson $\phi$ [ $\phi$ = $\bar{s}s$ ] is kaon decay, (i.e. $\phi \quad K^+K^-$ ), whereas the dominant decay mode of the vector $\omega$ meson [ $\omega = \frac{1}{\sqrt{2}}(\bar{u}u + \bar{d}d)$ ] is pion decay(i.e., $\omega$ $3\pi$ ), even though the phase space for pion decay mode of the more massive $\phi$ meson is greater than that for the $\omega$ meson. Fig.1.1 (a, b, c) show how the OZI rule explains this experimental fact when the $\omega$ and $\phi$ mesons are represented in terms of their constituents quarks. Thus Fig1.1a allows the $\omega$ $3\pi$ decay process via "connected" quark diagrams whereas Fig.1.1b shows why the decay $\phi$ $3\pi$ , involving as it does "disconnected" quark diagrams, cannot occur; on the other hand, using a quark diagram of the type shown in Fig.1.1c, $\phi$ $K^+K^-$ can take place. It should be pointed out that there is a small width for $\phi$ $3\pi$ decay because, in accordance with the QCD there are gluon lines between the s and u and d quarks Fig.1.1b and such diagram gives rise to reduced pion decay (and the deviation from an absolute OZI rule). As an another example, the preferential decay of the heavy quarkonia $\Psi$ and into c-quark and b-quark containing mesons respectively can be explained with the same type of "OZI rule" argument. Here, J/$\Psi$ = ( $c\bar{c}$ ) is the analog of $\phi(s\bar{s})$ and can decay into $D^0(c\bar{u}) + \bar{D}^0(\bar{c}u)$ provided its mass is sufficient (which is true for the second excited state of J/$\Psi$ and all higher ones), however, J/$\Psi$ can not decay into mesons from which a c quark is absent. Since the ground state of charmonium J/$\Psi$ is not sufficiently massive to allow $D^0 + \bar{D}^0$ decay, its decay width( arising from gluon-induced diagrams.) is of the order of tens of KeV rather than MeV's so that the observed metastability of J/$\Psi$ strongly supports the OZI rule. The OZI rule is rigorous in the large $N_c$ limit. This follows from the fact that an OZI-forbidden process involves at least two closed loops and hence is suppressed (completely suppressed in the large $N_c$ limit) compared to an OZI allowed process which receives contribution from one closed loop, and since we have no way of estimating the degree of accuracy of the OZI rule for finite $N_c = 3$, we must be prepared for violations of the OZI rule.

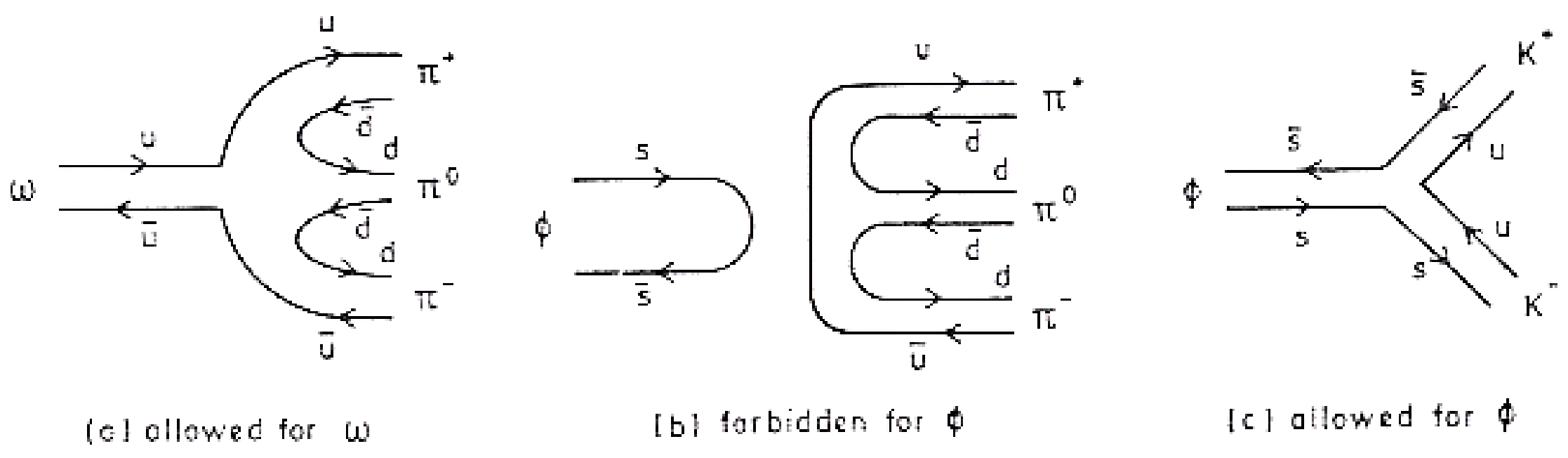

**Figure 1.1:** OZI connected and disconnected quark diagrams for ω and φ.

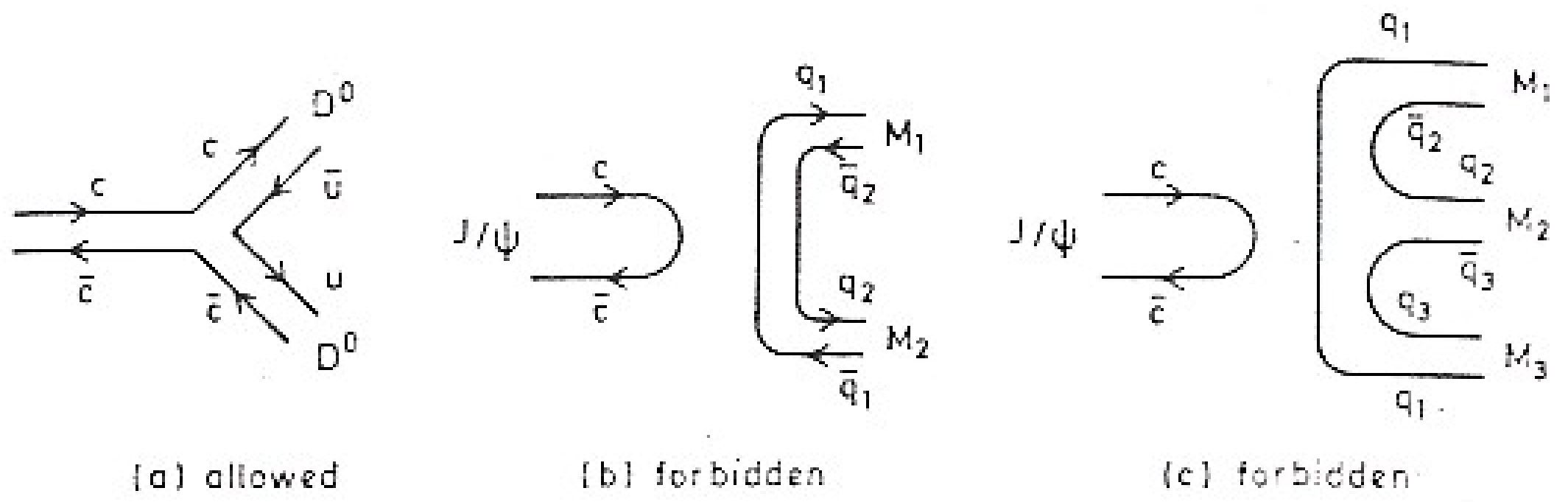

**Figure 1.2:** OZI connected and disconnected quark diagrams for J/ψ and ψ”.

## 1.10 Effective Field Theory

Effective field theory (EFT) is a technique for describing the low energy limit of a theory. It is an effective description because it uses the degrees of freedom and interaction which are relevant at low energy. The basic idea of an effective theory is to introduce the active light particles as collective degrees of freedom, while the heavy particles are frozen and treated as (almost) static sources. The dynamics is described by an effective Lagrangian which incorporates all relevant symmetries of the underlying fundamental theory.

Effective field theories (EFT'S) have long proven to be a powerful tool in particle physics. EFT approach has the promise to establish a relationship of QCD, the theory of strong interaction, to various successful phenomenological models, and has a systematic expansion in a small parameter. Using these interactions one treats the low energy dynamics in a complete field theoretic description. With such treatment one encounters loop diagrams, in which the integration over the momenta includes both low and high energy components. The heavier modes do not appear explicitly, their contribution is somehow included through some parameters in the effective theory. The role of the small parameter is played by the ratio of the typical momentum scale Q in the problem to the scale associated with the physics left out of the effective theory. In the case of nuclear interaction up to momenta of the order of 300 MeV, one can build on effective theory containing nucleons and pions (and delta isobars). However, in those nuclear processes where the typical momentum scale is small compared to the pions mass, one is allowed to use an effective theory without explicit pions, only contact force remains.

For long distances the effective field theory is fully correct since it treats baryons and pions as point particles, but this convention does not provide an accurate representation of physics at distances less than the separation scale. The use of effective field theory technique is an ever growing approach in various fields of theoretical physics. For example, we do not need quantum gravity to understand the hydrogen atom nor does chemistry depend upon the structure of the electromagnetic interaction of quarks.

EFT's are approximate by their very nature. Once the relevant degrees of freedom for the problem at hand have been established, the corresponding EFT is usually treated perturbatively. It does not make much sense to search for an exact solution of the Fermi theory of weak interactions. In the same spirit, convergence of the perturbative expansion in the mathematical sense is not an issue. The asymptotic nature of the expansion becomes apparent once the accuracy is reached where effects of the underlying "fundamental" theory cannot be neglected any longer. The range of applicability of the perturbative expansion depends on the separation of energy scales that define the EFT.

Historically, effective Lagrangians were formulated so as to reproduce the results of current algebra and PCAC at tree level. Basically, the effective Lagrangians were used as convenient alterations to comutator algebra. In 1979, Weinberg [15] extended the scope of the effective Lagrangian formulation by postulating that the use of effective Lagrangians can go beyond current algebra. This assertion was based on the observation that for soft pion processes, chiral Lagrangians offer a powerful parameterization of the S-matrix based on chiral counting arguments and general principles such as symmetries, analyticity, unitarity etc. Weinberg's program has been systematized and extended by Gasser and Leutwyler [16]. The use of effective Lagrangians beyond tree level as a way to understand the hadronic S-matrix in the soft-pion limit, however, side-steps the basic issues of confinement and broken chiral symmetry.

## 1.11 Anomaly

Anomaly arises in quantum field theory when the symmetries of classical field theory are broken by quantum fluctuations inherent in a quantum field theory. Here a symmetry of classical action is not a true symmetry of full quantum theory. Classical Lagrangian in abelian QED with massless fermion or non-abelian QCD with massless fermion posses the property of scale invariance because the gauge field are massless and coupling constants are dimensionless. However, quantum mechanical renormalization introduces a finite renormalization scale for both unbroken QED and QCD and breaks the scale invariance in the process. The “quantum fluctuations” resulting from loop corrections in the renormalization process breaks down the classical chirality invariance and leads to the so called chiral gauge anomalies, where as the axial anomaly follows from the conflicts between gauge invariance and chiral invariance in the process of regulating the theory of quantum level.

Due to this anomaly, Noether current is no longer divergenceless but receives a contribution arising from quantum corrections, and hence is not valid at quantum level after consideration of quantum structures in the corresponding perturbation series. When anomaly arises, the Ward identities relating matrix element, no longer hold, but rather are replaced by a set of anomalous ward identities which take into account the correct current divergences. The QCD anomaly equation for a single flavor can be written as[17]

$$^{\mu}(\bar{q}\gamma_{\mu}\gamma_{5}q) = 2im(\bar{q}\gamma_{5}q) - \frac{\alpha_{s}}{4\pi} G^{a}_{\mu\nu}\tilde{G}^{a\mu\nu} \qquad (1.28)$$

Where $G^a_{\mu\nu}$ is gluon field tensor and $\tilde{G}^{a\mu\nu}$ is its dual. Three- flavor QCD anomaly can be utilized to explain "$U_A(1)$ problem" in QCD, namely the large mass of the η' meson and also the proton spin problem.

It is clear from Eq.(1.28) that the $U_A(1)$ chiral symmetry is explicitly broken by the QCD anomaly. In reality, there is competition between the spontaneous and explicit chiral symmetry breaking because of the anomaly. Since $U_A(1)$ symmetry is broken not spontaneously but explicitly by the anomaly, ή cannot be regarded as a nearly massless Nambu Goldstone boson like the other pseudoscalar mesons. In fact ή mass is as large as the nucleonic mass, i.e., $m_{ή}$ = 958MeV. This is called $U_A(1)$ problem. It can be shown that without the QCD anomaly, the mass of the "non strange" pseudoscalar, $\eta_{ns}$ can only be slightly larger than the mass of the pion and $m_{\eta_{ns}} \leq \sqrt{3}m_\pi$. This inequality becomes a part of the $U_A(1)$ problem in QCD[1]. The resolution to the problem came when it was realized that the anomaly term has been neglected. Denoting $\eta_s$ as the strange pseudoscalar, we can use the relation:

$$m_{\eta'}{}^2 + m_\eta{}^2 = m_{\eta_s}{}^2 + m_{\eta_{ns}}{}^2 \qquad (1.29)$$

Using the current algebra manipulation and SU(3) symmetry in decay constants, it can be shown that

$$m_{\eta'}{}^2 \simeq 2m_k{}^2 - m_\eta{}^2 + A^2 \qquad (1.30)$$

where $A^2$ is the anomaly contribution and A can be expressed in terms of matrix elements of the axial anomaly between vacuum and, $\eta_{ns}$ and $\eta_s$ states. Numerically $A^2 \geq 0.37 GeV^2$. Thus in strong interaction process in which the coupling of the quark current to gluon fields is involved, the three flavor QCD anomaly has a significant role in correcting the deficiencies in current algebra calculations. Also, anomaly makes it possible for spin carried by the gluons to mix the spin by quarks, thus modifying the structure of quark sea.

## 1.12 Proton Spin Problem

According to the nonrelativistic constituent quark model, the whole of the proton spin arises from the quarks. In relativistic quark model, the sum of the z-component of quark spins account for ¾ of the proton spin while rest of the proton spin arises from the quark orbital angular momentum. The 1987 EMC experiment[18] indicated that the first moment of the proton spin structure function $\Gamma_1^P = 0.126 \pm 0.018$ leading to the stunning implication that very little(15%) of the proton spin is carried by the quarks, contrary to the naïve quark model picture. The EMC data implied a substantial sea quark polarization in the region $x < 0.1$, a range not probed by earlier SLAC experiments[19]. In the naïve parton model, the data also implied a large and negative strange sea polarization which is contrary to the basic assumption of the Ellis-Jafe sum rule[20], namely $\Delta s = 0$.

Anomalous gluon effect originating from the axial anomaly provides a plausible and simple solution to the proton spin puzzle. A polarized gluon is preferred to split into a quark-antiquark pair with helicites antiparallel to the gluon spin. Thus a positive gluon spin component $\Delta G$ can give rise to negative sea quark polarization. The lattice calculation indicates that sea polarization is almost independent of light quark flavors. This empirical SU(3) flavor symmetry implies that it is indeed the axial anomaly, which is independent of light quark masses, that accounts for the bulk of helicity contribution of sea quarks. Hence anomaly makes it possible for spin carried by the gluons to mix with the spin carried by quarks, thus modifying the structure of quark sea and explicable for the smallness of the apparent quark contribution to the proton spin.

In chapter II, using a statistical model, in which a nucleon is taken as an ensemble of quark-gluon Fock states, we have calculated the quark contributions to the spin of the nucleon, the ratio of the magnetic moments of nucleons, their weak decay constant, and the ratio of SU(3) reduced matrix elements for the axial current. This has been done neglecting the contribution of s-quark and other heavy quarks, and covering only ~ 86% of the total Fock states. Two modifications of this model has also been worked out with a view to reduce the contributions of the sea components with higher multiplicities .

In chapter III, using the framework of the conventional QCD sum rule, we have studied the isospin splitting in the diagonal pion-nucleon coupling constant by including the quark mass dependent terms, $\pi^0$-$\eta$ mixing and electromagnetic corrections to meson-quark vertices. Some of the implications of the isospin splitting have also been discussed.

In chapter IV, gluonic contributions to the self-energy of a nucleon has been investigated in an effective theory. The couplings of the topological charge density to nucleons give rise to OZI violating η-nucleon and η’-nucleon interactions. The one-loop self-energy of a nucleon arising due to these interactions has been calculated using a heavy baryon chiral perturbation theory. The divergences have been regularized using form factors. The nontrivial structure of the QCD vacuum has also been taken into account.

In chapter V, we calculate the first derivative of the topological susceptibility at zero momentum, $\chi'(0)$ using QCD sum rules. $\chi'(0)$ is useful, among others, in the discussion of the proton spin problem. The mass of η’ and the singlet pseudoscalar decay constant in the chiral limit have also been found as a bonus.

Finally in the last chapter VI, we give summary and concluding remarks.

# CHAPTER-II

## SPIN STUDIES OF NUCLEONS IN A STATISTICAL MODEL

### 2.1 Introduction

The composition of nucleons, in terms of fundamental quarks and gluons degrees of freedom have been modeled variously to account for its observed properties. It is important to calculate as many nucleonic parameters as possible in these models to check their merits and their domains of validity. The naive valence picture of nucleon structure may be regarded as a first order approximation to the real system[1]. Models with one constituent gluon [2] and with one quark-antiquark $q\bar{q}$ pair [3-5], in addition to the three valence quarks, are capable of giving better account of nucleonic properties. In another class of models, it is assumed that nucleons consist of valence quarks surrounded by a "sea" which, in general, contains gluons and virtual quark-antiquark pairs, and is characterized by its total quantum number consistent with the quantum number of nucleons[6,7].

In the chiral quark model of Manohar and Georgi [8], QCD quarks propagate in the nontrivial QCD vaccum having $\bar{q}q$ condensates and this leads to the generation of extra mass to the quarks. As a consequence of this spontaneous chiral symmetry breaking, massless pseudoscalar bound $\bar{q}q$ Goldstone bosons are generated, and this leads to the nontrivial sea structure of the nucleon. In the instanton model [9], the quark-antiquark sea in a nucleon results from a scattering of a valance quark off a nonperturbative vacuum fluctuation of the gluon field, instanton. In the instanton induced interaction described by ’t Hooft effective lagrangian, the flavor of the produced quark-antiquark is different from the flavor of the initial valance quarks, and there is a specific correlation between the sea quark helicity and the valance quark helicity. In the chiral-quark soliton model [10], the large $N_c$ model of QCD becomes an effective theory of mesons with the baryons appearing as solitons. Quarks are described by single particle wave functions which are solutions of the Dirac equation in the field of the background pions. In the statistical approach, the nucleon is treated as a collection of massless quarks, antiquarks and gluons in thermal equilibrium within a finite size volume[11]. The momentum distributions for quarks and antiquarks follow a Fermi-Dirac distribution function characterized by a common temperature and a chemical potential which depends on the flavor and helicity of the quarks.

Recently, a new statistical model has been proposed in which a nucleon is taken as an ensemble of quark-gluon Fock states [12,13]. In this model, using the principle of balance that every Fock state should be balanced with all of the nearby Fock states [13], or using the principle of detailed balance that any two nearby Fock states should be balanced with each other [12], the probability of finding every Fock state of the proton accounting upto ≈ 98% of the total Fock state has been obtained. It has been shown that the model gives an excellent description of the light flavor sea asymmetry (i.e, $\bar{u} \neq \bar{d}$ ) without any parameter.

From the above brief review, it is clear that most of the analytical calculations in the literature on the properties of nucleons related to its spin are done in (i) "minimal" quark model which contains at most a gluon or a quark-antiquark pair, in addition to the three valence quarks, or (ii) models where sea acts only as a background specified only by its quantum numbers with no active role in determining the nucleonic properties. In this article, using statistical ideas, we construct such Fock states of a nucleon which have definite color and spin quantum numbers, and definite symmetry property. The resulting total flavor-spin-color wave function of a spin-up nucleon consists of Fock states with three valence quarks and a sea containing up to five constituents (quark-antiquarks and gluons). We have used this model wave function to calculate the light quark spin content of nucleons, the ratio of their magnetic moments, the semileptonic decay constant of neutron, and the ratio of SU(3) reduced matrix elements for the axial current.

The mode of correlation among the constituents of a Fock state cannot be decided merely by a statistical consideration, and this requires, possibly, some dynamical input. To check the stability of results obtained, under variation in correlation, we introduce two modifications of our primary model and repeat calculations of nucleonic properties in these models as well.

### 2.2. Sea and its Structure

In Ref.[12,13], treating the proton as an ensemble of quark-gluon Fock states, the proton state has been expanded in a complete set of such states as

$$|p\rangle = \sum_{ijk} C_{ijk} |uud, i, j, k\rangle \quad (2.1)$$

where i is the number of $\bar{u}u$ pairs, j is the number of $\bar{d}d$ pairs, and k is the number of gluons. The probability to find a proton in the Fock state $|uud,i,j,k\rangle$ is

$\rho_{ijk} = |C_{ijk}|^2$,

where $\rho_{ijk}$ satisfies the normalization condition,

$$\sum_{ijk} \rho_{ijk} = 1 \quad (2.2)$$

Then, using the detailed balance principle or balance principle, and with sub processes $q \Leftrightarrow q\, g$, $g \Leftrightarrow \bar{q}q$ and $g \Leftrightarrow gg$ considered, all $\rho_{ijk}$ have been calculated explicitly. Interestingly, the model predicts an asymmetry in the sea flavor of $\bar{u}$ and $\bar{d}$ as $\bar{d} - \bar{u} \sim 0.124$ in surprising agreement with the experimental data 0.118±0.012. These quarks and gluons have to be understood as "intrinsic" partons of the proton as opposed to the "extrinsic" partons generated from the QCD hard bremsstrahlung and gluon splitting as a part of the lepton nucleon scattering interaction [14]. The $\bar{q}q$ pairs and gluons, which are multiconnected non-perturbatively to the valence quarks, will collectively be referred to as the sea. Since the proton should be colorless and a $q^3$ state can be in color state $1_c$, $8_c$ and $10_c$, the sea should also be in the corresponding color state to form a color singlet proton.

Furthermore, if the sea is in an S-wave state relative to the $q^3$ core, conservation of angular momentum restricts that the spin of the sea can only be 0,1or 2 to give a spin-1/2 proton. The case of the sea with one $\bar{q}q$ pair, where the sea or at least one of the quarks is needed to be in a relative P-wave to meet the positive parity requirement of the proton, will be treated separately. We take the probabilities of finding various quark-gluon Fock states in a proton from Ref.[13], and assume that the quarks and the gluons can be treated nonrelativistically for our problem, and also that, in general, these are in S-wave motion. The effect of the relativistic motion of the constituents will be discussed later. The case of a neutron will be treated in an analogous way using isospin symmetry.

Nonrelativistic treatments of quarks in nucleon models are well known [1,4-6]. There are phenomenological evidences that gluons also behave as massive particles with mass ≥0.5GeV[15]. There is a firm evidence from lattice calculation also that gluons behave as massive particles at low momenta (≤4GeV)[16]. It has been shown in Ref [5] that the sum of the relativistic quark spin and orbital angular momentum (derived from QCD Lagrangian ) is equal to the sum of the non-relativistic quark spin and orbital angular momentum,

$$\vec{S}_q + \vec{L}_q = \vec{S}^{NR}_q + \vec{L}^{NR}_q \qquad (2.3)$$

Furthermore, it has been shown that on truncating the Fock space to contain only $|q^3\rangle$ and $|q^3\bar{q}q\rangle$ component, the quark orbital angular momentum contribution comes out to be negligible or small [5]. This contribution should decrease on inclusion of Fock states with more "intrinsic" partons, since then each parton will have a lesser linear momentum share, and hence, smaller orbital angular momentum too.

Following Ref.[6], we write the possible combination of $q^3$ and sea wave function, which can give a spin ½ flavor octet, color singlet state as

$\Phi_1^{(1/2)}H_0G_1$ , $\Phi_8^{(1/2)} H_0 G_8$ , $\Phi_{10}^{(1/2)}H_0G_{\overline{10}}$ , $\Phi_1^{(1/2)}H_1G_1$, $\Phi_8^{(1/2)}H_1G_8$, $\Phi_{10}^{(1/2)}H_1G_{\overline{10}}$ and $\Phi_8^{(3/2)}H_1G_8$ , $\Phi_8^{(3/2)} H_2G_8$ . (2.4)

In the above $\Phi^{(1/2,3/2)}{}_{1,8,10}$ is the $q^3$ wave function in obvious notation, while $H_{0,1,2}$ and $G_{1,8,\overline{10}}$ denote spin and color sea wave functions respectively which satisfy

$$\langle H_i \mid H_j\rangle = \delta_{ij} , \quad \langle G_k \mid G_l\rangle = \delta_{kl} .$$

The total flavor-spin-color wave function of a spin up proton which consists of three valence quarks and sea components can be written as :

$$|\Phi_{1/2}{}^{\uparrow}\rangle = (1/N) [\Phi_1^{(1/2\uparrow)}H_0G_1 + a_8\ \Phi_8^{(1/2\uparrow)}H_0G_8 + a_{10}\ \Phi_{10}^{(1/2\uparrow)}H_0G_{\overline{10}} + b_1( \Phi_1^{(1/2)}\otimes H_1)^{\uparrow}G_1 + b_8(\Phi_8^{(1/2)}\otimes H_1)^{\uparrow}G_8 + b_{10} (\Phi_{10}^{(1/2)}\otimes H_1)^{\uparrow}G_{\overline{10}} + c_8 (\Phi_8^{(3/2)}\otimes H_1)^{\uparrow}G_8 + d_8(\Phi_8^{(3/2)}\otimes H_2)^{\uparrow}G_8] \quad (2.5)$$

where $N^2 = 1 + a_8^2 + a_{10}^2 + b_1^2 + b_8^2 + b_{10}^2 + c_8^2 + d_8^2$ , and $(\Phi_1^{(1/2)}\otimes H_1)^{\uparrow}$, etc. have to be written properly with appropriate CG coefficients and by taking into account the symmetry property of the component wave function .

Here, we suggest a possible way to construct the sea wave function using the statistical model of Zhang et al [13]. However, unlike Ref. [6], we will also take into account the "active" sea contribution of the sea in which the relevant operators act on the sea quarks as well. Furthermore, we will use an approximation in which quarks in the $q^3$ core will not be antisymmetrized with the identical quarks appearing in the sea. Use of different labels for valance and sea quarks has been justified with the assumption that the valance and the sea quarks have very different momentum distributions, with the valance quarks being “hard” and the sea quarks “soft”, and that the overlap region between the two momentum distributions is negligible[17]. Consequently, this classification can work where one is concerned with matrix elements having zero momentum transfer and only require that the overlap region between valance and sea quark momentum distribution be negligibly small. Nevertheless, we will use this separation for the problem of quark contribution to the nucleon spin as well.

We assume that the statistical decomposition of the proton state in various quark-gluon Fock states, as obtained by Zhang et al.[13] and which is expected to work at a $Q^2$ ~$1GeV^2$ for the quark system, can be extended down to the proton’s rest frame. Since the quarks and gluons in the Fock states are “intrinsic”, there should be no problem in this extension as far as color quantum numbers are concerned. However, it has been shown by Ma and Zhang [18] that the Melosh rotation [19] generated by the internal transverse momentum spoils the usual identification of the $\gamma^+\gamma_5$ quark current matrix element with the total rest frame spin projection $s_z$, thus resulting in a reduction of $g_A$. It has also been observed [20,21] that the physical value of the anomalous magnetic moment is reduced from its non-relativistic value due to the Melosh rotation. We will estimate the changes in weak decay constant and the ratio of the magnetic moments of the nucleons due to the Melosh rotation towards the end.

Next, we decompose each one of the Fock states $|uud,i,j,k\rangle$ in terms of the set of states appearing in Eq.(2.5) following a statistical approach .

(i) Consider the decomposition of a state $|uud,0,0,2\rangle$ or $|gg\rangle$ sea (two gluons in the sea ).

Spin : uud : $1/2 \otimes 1/2 \otimes 1/2 = 2(1/2) \oplus 3/2$,

gg : $1 \otimes 1 = 0_s \oplus 1_a \oplus 2_s$,

Color : uud : $3 \otimes 3 \otimes 3 = 1 \oplus 8 \oplus 8 \oplus 10$,

gg : $8 \otimes 8 = 1_s \oplus 8_s \oplus 8_a \oplus 10_a \oplus \overline{10}_a \oplus 27_s$.

The subscripts s and *a* denote symmetry and asymmetry respectively under the exchange of two identical bosons (gluons above). Call $\rho_{j1\,j2}$ as the probability that the $q^3$ core and gg sea are in angular momentum states $j_1$ and $j_2$ respectively, and they finally add to give total angular momentum 1/2. Let us compare such probabilities.

$$\rho_{1/2\;0s} / \rho_{1/2\;1a} = \frac{(4/8).(1/9).1}{(4/8).(3/9).(2/6)} = 1,$$

$$\rho_{1/2\;0s} / \rho_{3/2\;2s} = \frac{(4/8).(1/9).1}{(4/8).(5/9).(2/20)} = 2,$$

$$\rho_{3/2\;1a} / \rho_{3/2\;2s} = \frac{(4/8).(3/9).(2/12)}{(4/8).(5/9).(2/20)} = 1,$$

$$\rho_{1/2\;1a} / \rho_{3/2\;1a} = \frac{(4/8).(3/9).(2/6)}{(4/8).(3/9).(2/12)} = 2.$$

The first factor in the numerator or denominator in the r.h.s is the relative probability for the core quarks to have spin $j_1$, the second factor is the same for the two gluons to have spin $j_2$, and finally the third one is the same for $j_1$ and $j_2$ to have resultant 1/2. In future, we will omit the factor which is common in the numerator and the denominator.

Similarly we can compare the probabilities for the $q^3$ core and gg to be in different color substates which finally give a color singlet proton. In obvious notations:

$$\rho_{1\,1s} / \rho_{8\,8s} = \frac{(1/27).(1/64).1}{(16/27).(8/64).(1/64)} = 1/2 = \rho_{1\,1s}/\rho_{8\,8a},$$

$$\rho_{1\,1s} / \rho_{10\,\overline{10}} = \frac{(1/27).(1/64).1}{(10/27).(10/64).(1/100)} = 1 .$$

The product of probabilities in spin and color spaces can be written in terms of one common parameter c as

$\rho_{1/2\ 0s}\,[\rho_{1\,1s}, \rho_{8\,8s}] = 2c\ (1,2)$,

$\rho_{1/2\ 1a}\,[\rho_{8\,8a}, \rho_{10\,\overline{10}}] = 2c\ (2,1)$,

$\rho_{3/2\ 1a}\,[\rho_{8\,8a}] = 2c$ , $\rho_{3/2\ 2s}\,[\rho_{8\,8s}] = 2c$.

There is no contribution to $H_0G_{\overline{10}}$ and $H_1G_1$ sea from two gluon states because $H_0$ and $G_1$ are symmetric whereas $H_1$ and $G_{\overline{10}}$ are antisymmetric under exchange of the two gluons making these product wave functions antisymmetric and hence unacceptable for a bosonic system. The sum of all these probabilities is taken from Ref.[13] and this determines the unknown parameter c :

$\rho_{uud\ gg} = 0.081887$, $c = 0.005118$,

giving us above products of probabilities.

It is clear that the numbers on the r.h.s in above equations containing products of probabilities give the probabilities for finding the Fock state with two gluons in the sea in various substates with specified spin and color quantum numbers. Thus, for instance $\rho_{1/2\ 0s}\,\rho_{11s}$ = 0.01024 means that the probability for finding the three quark core in spin ½ and color singlet state along with the two-gluon sea to be in a scalar and color singlet state is 0.01024.

Similar decomposition will hold good for $|\bar{q}q\,\bar{q}q\rangle$ sea also. By proceeding on a similar line, we get

$\rho_{1/2\ 0s}\,[\rho_{1\,1s}, \rho_{8\,8s}];\ \rho_{1/2\ 1a}\,[\rho_{8\,8a}, \rho_{10\,\overline{10}}];\ \rho_{3/2\ 1a}\,[\rho_{8\,8a}]\ ;\ \rho_{3/2\ 2s}\,[\rho_{8\,8s}]$

$$= 0.000904(1,2;2,1;2;2) \text{ for } |\bar{u}u\bar{u}u\rangle,$$

$$= 0.014571(1,2;2,1;2;2) \text{ for } |\bar{d}d\bar{d}d\rangle.$$

(ii) For decomposition of $|g\,\bar{q}q\rangle$ and $|\bar{u}u\,\bar{d}d\rangle$ sea, symmetry consideration is not needed. Here we have assumed that $\bar{q}q$ carries the quantum numbers of a gluon due to the sub processes $g \Leftrightarrow \bar{q}q$. This gives the relative probability density in color space as $\rho_{1\,1}/\rho_{8\,8}=1/4$. The ratio $\rho_{1\,1}/\rho_{10\,\overline{10}}$ and the relative densities in spin space remain the same as in (i). Proceeding as in the previous case, the products of densities in spin and color spaces come out as

$\rho_{1/2\ 0}\,[\rho_{1\,1}, \rho_{8\,8}, \rho_{10\,\overline{10}}];\ \rho_{1/2\,1}\,[\rho_{11}, \rho_{8\,8}, \rho_{10\,\overline{10}}];\ \rho_{3/2\,1}\,[\rho_{8\,8}];\ \rho_{3/2\,2}\,[\rho_{8\,8}]$

$$= 0.00344\ (1,4,1;1,4,1;2;2) \quad \text{for } |g,\bar{u}u\rangle,$$

$$= 0.00517\ (1,4,1;1,4,1;2;2) \quad \text{for } |g,\bar{d}d\rangle,$$

$$= 0.00366(1,4,1;1,4,1;2;2) \quad \text{for } |\bar{u}u,\bar{d}d\rangle.$$

(iii ) $|gg\,\bar{q}q\rangle$, $|\bar{q}q\,\bar{q}q\,g\rangle$ sea : First we take the product of two spin 1 states and two color octet states as in (i). These are further multiplied with spin 1 and color octet state respectively. The new results needed are

Spin : $1 \otimes 2 = 1 \oplus 2 \oplus 3$,

Color: $10 \otimes 8 = 8 \oplus 10 \oplus 27 \oplus 35$,

$27\otimes 8=8\oplus 10\oplus \overline{10}\oplus 2(27)\oplus 35\oplus 35\oplus 64$ .

Using the subscript s and *a* for symmetry and asymmetry under the exchange of first two bosons, the relative probability densities in spin space are:

$$\rho_{1/2\ 0a}/\ \rho_{1/2\ 1a}=\frac{(1/27).1}{(3/27).(2/6)}=1,\quad \rho_{1/2\ 0a}/\ \rho_{1/2\ 1s}=\frac{(1/27).1}{(6/27).(4/12)}=\frac{1}{2},$$

$$\rho_{1/2\ 1a}/\ \rho_{3/2\ 1a}=\frac{(3/27).(1/3)}{(3/27).(1/6)}=2=\rho_{1/2\ 1s}/\ \rho_{3/2\ 1s},$$

$$\rho_{3/2 1a}/\ \rho_{3/2\ 2a}=\frac{(3/27).(2/12)}{(5/27).(2/20)}=1,\quad \rho_{3/2\ 1s}/\ \rho_{3/2\ 2s}=\frac{(6/27).(4/24)}{(5/27).(2/20)}=2.$$

The ratio of the probability densities in color space are:

$$\rho_{1\ 1s}/\ \rho_{8\ 8s}=\frac{(1/27).(1/512).1}{(16/27).(32/512).(1/64)}=1/8,$$

$$\rho_{1\ 1s}/\ \rho_{10\ \overline{10}\ s}=\frac{(1/27).(1/512).1}{(10/27).(20/512).(1/100)}=\frac{1}{2}=\rho_{1\ 1a}/\ \rho_{10\ \overline{10}\ a},$$

$$\rho_{1\ 1a}/\ \rho_{8\ 8a}=\frac{(1/27).(1/512).1}{(16/27).(32/512).(1/64)}=1/8.$$

The combined probabilities in spin and color space can be written as

$\rho_{1/2\ 0a}[\rho_{1\ 1a},\rho_{8\ 8a},\ \rho_{10\ \overline{10}\ a}]$; $\rho_{1/2\ 1a}[\ \rho_{1\ 1a},\ \rho_{8\ 8a},\ \rho_{10\ \overline{10}\ a}]$; $\rho_{1/2\ 1s}[\rho_{1\ 1s},\ \rho_{8\ 8s},\ \rho_{10\ \overline{10}\ s}]$; $\rho_{3/2\ 1a}[\ \rho_{8\ 8a}]$;

$\rho_{3/2\ 1s}[\rho_{8\ 8s}]$; $\rho_{3/2\ 2a}[\ \rho_{8\ 8a}]$ = 0.00051(1,8,2;1,8,2;2,16,4;4;8;4) for $|gg,\bar{u}u\rangle$ ,

= 0.00076(1,8,2;1,8,2;2,16,4;4;8;4) for $|gg,\bar{d}d\rangle$,

= 0.00007 (1,8,2;1,8,2;2,16,4;4;8;4) for $|\bar{u}u\ \bar{u}u\ ,g\rangle$,

= 0.00025(1,8,2;1,8,2;2,16,4;4;8;4) for $|\bar{d}d\ \bar{d}d\ ,g\rangle$.

(iv) $|\, \bar{u}u\, \bar{d}d\, g \,\rangle$ sea : Here, there is no symmetry requirement. Ratios of probability densities are

$\rho_{1/2\ 0} / \rho_{1/2\ 1} = 1/3, \quad \rho_{1/2\ 0}/\rho_{3/2\ 2} = 1, \quad \rho_{1/2\ 1} / \rho_{3/2\ 1} = 2, \quad \rho_{3/2\ 1} / \rho_{3/2\ 2} = 3/2.$

in spin space, and $\rho_{1\ 1} / \rho_{8\ 8} = 1/8, \quad \rho_{1\ 1} / \rho_{10\ \overline{10}} = 1/2$

in color space. Their products can be written as

$\rho_{1/2\ 0}[\rho_{11}, \rho_{8\ 8}, \rho_{10\ \overline{10}}]; \rho_{1/2\ 1}[\rho_{11}, \rho_{8\ 8}, \rho_{10\ \overline{10}}] ; \rho_{3/2\ 1} [\rho_{8\ 8}]; \rho_{3/2\ 2} [\rho_{8\ 8}]$

$= 0.00048(1, 8, 2;3, 24, 6;12; 8)$ .

(v) $|\, ggg \rangle$ sea :

The wave function for this sea should be completely symmetric under the exchange of any two gluons . Among the product spin function, the total spin S= 0 is completely antisymmetric and one S=1 is completely symmetric. Among the product color functions, there is one color singlet state and one color octet state which are completely antisymmetric; and there is one color singlet state and one color octet state which are completely symmetric. This gives

$\rho_{1/2\ 0}/\rho_{1/2\ 1} = 1, \quad \rho_{1/2\ 1}/\rho_{3/2\ 1} = 2, \quad \rho_{1\ 1a,s}/\rho_{8\ 8} = 1/2$ .

This gives us the product of probabilities in spin and color spaces as

$\rho_{1/2\ 0a}[\rho_{1\ 1a}, \rho_{8\ 8a}]; \rho_{1/2\ 1s}[\rho_{11s}, \rho_{8\ 8s},]; \rho_{3/2\ 1s}[\rho_{8\ 8s}]$

$= 0.00534(1, 2; 1, 2;1)$.

A confined gluon in the sea may be divided into TE (transverse electric ) modes with $J^{pc} = 1^{+-}$ and the TM (transverse magnetic) modes with $J^{pc} = 1^{--}$ . The Fock states with a single gluon in the sea may be considered to be consisting of a TE gluon [22]. Clearly, a gluon in the sea will contribute only to the $H_1G_8$ component of the sea. From this decomposition we get the following numbers for the coefficients in the expansion in Eq.(2.5) of the proton state:

$a_8^2$=0.5043, $a_{10}^2$= 0.0892, $b_1^2$=0.1037, $b_8^2$=1.8133, $b_{10}^2$ = 0.2220,

$c_8^2$ = 0.9067, $d_8^2$ = 0.2630 and $N^2$ = 4.9024.

However, the treatment of a $\bar{q}q$ pair in the sea requires special attention, since as stated earlier, to keep the parity of the system positive, one or a group of the five particles is required to be in a P-wave state. This requires detailed knowledge of spatial wave function. To get the contribution of this particular Fock space, we have borrowed the result from Ref.[5] and scaled it to give the same probability which we are using, as given in Ref.[13]. Thus, we have also introduced non-zero orbital angular momentum states, albeit for only one type of Fock states among the several Fock states considered, in our nucleon wave function. Unlike our treatment, the total wave function in[5] has been properly antisymmetrized. All the above states taken together constitute ≈ 86% of the total Fock space. The cases with three $q\bar{q}$ pairs, four gluons and two $q\bar{q}$ pairs with two gluons, and other higher Fock states have not been considered due to smaller probabilities associated with these Fock states in the statistical model, and due to more involved analysis in their decomposition. We believe, we can get sufficient insight in the problem under consideration even at the cost of directly excluding higher Fock states. We assume that the rest of the quark-gluon sea spanning ~14% of the Fock space of the nucleon also decomposes in color and spin subspaces in approximately the same proportion as the one which we have worked out explicitly above. The number of strange quark-antiquark pairs in the statistical model is 0.05 in the nucleon as compared to the average number of particles which is 5.57[13]; hence we neglect the contributions of the s-quark and other higher mass quarks in calculations of nucleonic properties.

For calculating physical quantities related to the spin of a nucleon, it is useful to introduce two parameters, α and β as [6]

$$\alpha = (1/N^2).(4/9).(2a+2b+3d+\sqrt{2}\,e),$$

$$\beta = (1/N^2).(1/9).(2a-4b-6c-6d+4\sqrt{2}\,e),$$

where,

$$a = (1/2).(1-b_1^2/3),\ b=(1/4).(a_8^2-b_8^2/3),\ c=(1/2).(a_{10}^2-b_{10}^2/3)$$

$$d=(1/18).(5c_8^2-3d_8^2),\ e=(\sqrt{2}/3).b_8c_8\ .$$

The importance of these parameters lies in the fact that they are connected with the numbers of spin-up (n(q↑)) and spin-down (n(q↓)) quarks in the spin-up proton. If $\Delta q = n(q\uparrow)-n(q\downarrow)+n(\bar{q}\uparrow)-n(\bar{q}\downarrow)$, q = u, d ,then $\Delta u = 3\alpha$ and $\Delta d = -3\beta$. Contributions to the parameters α and β from the sea excluding the single $\bar{q}q$ components have been denoted by $\alpha_1$ and $\beta_1$ respectively, whereas those from the single $\bar{q}q$ components have been denoted by $\alpha_2$ and $\beta_2$: $\alpha = \alpha_1+\alpha_2$, $\beta = \beta_1+\beta_2$. Numerical values of all these parameters
have been listed in Table 2.1 (Model C). These can be used to calculate various physical quantities as done in Ref.[6], where the sea plays a role of "passive" background and the relevant operators act only on the three-quark core. When the operator $\sum_i e_i^2\sigma_z^i$ acts on the sea minus the single $\bar{q}q$ component, i.e. when the sea plays the "active" role, the result has been denoted by $\Delta I_1^p$ and $\Delta I_1^n$ for the proton and neutron respectively .There is no such contribution to the magnetic moments due to the “active” sea, since the $\bar{q}q$ pairs carry the quantum numbers of the parent gluons. The total contribution to the nucleon spin from the spins of the quarks, denoted by $I_1^p$ and $I_1^n$, have been displayed in Table 2.2 (Model C) and compared with the revised EMC result [23]. We should note that EMC value is for $Q^2 \approx 10$ GeV$^2$, whereas our result for $I_1^p$ and $I_1^n$ should be considered to work at $Q^2 \sim 1\text{GeV}^2$ where the Fock state decomposition of the nucleon state [13] used in this work applies. To estimate $(g_A/g_V)$, we use Bjorken sum rule

$$(\bar{g}_1^p - \bar{g}_1^n) = \int_0^1 dx[g_1^p(x) - g_1^n(x)] = \frac{1}{6}\left|\frac{g_A}{g_V}\right|[1 - \frac{\alpha_s(Q^2)}{\pi}]$$
$$= 0.191 \quad 0.002$$

written upto $O(\alpha_s^3/\pi^3)$ [24]. We have considered three values of $\alpha_s$ from the recent literature. Authors of Ref.[25] have used $\alpha_s(1\text{GeV}^2) \approx 0.5$ for the same purpose as ours. Particle Data Group[26] average value is $\alpha_s(m_c) = 0.357$, which we modify as $\alpha_s(1\text{GeV}^2) = 0.375$ for our use. Authors of Ref.[27] use $\alpha_s(0) = 0.35$ (to fit the bound states in QCD). The values of $(g_A/g_V)$ obtained for each one of these values have been displayed in Table 2.2 (Model C). The F/D value has been obtained from $\alpha$ and $\beta$ as per the prescription given in Ref.[6].

**TABLE 2.1** : α and β as defined in Ref.[6]: $\alpha_1$ and $\beta_1$ are the contributions from the sea excluding the single $\bar{q}q$ components; $\alpha_2$ and $\beta_2$ are the contribution from the single $\bar{q}q$ components of the sea. $\Delta I_1^p$ and $\Delta I_1^n$ are the contribution to $I_1^p$ and $I_1^n$ respectively when the operator $\sum_i e_i^2 \sigma_z^i$ acts on the sea excluding the single $\bar{q}q$ component. Model C is our first statistical model described in the text. In model P, $\bar{q}q$ pairs have been taken as colorless pseudoscalars, whereas model D is the one in which suppressed higher multiplicity states appear.

| Model Type | $\alpha_1$ | $\alpha_2$ | $\alpha = \alpha_1 + \alpha_2$ | $\beta_1$ | $\beta_2$ | $\beta = \beta_1 + \beta_2$ | $\Delta I_1^p$ | $\Delta I_1^n$ |
|---|---|---|---|---|---|---|---|---|
| Model C | 0.1821 | 0.0417 | 0.2237 | 0.0549 | 0.0186 | 0.0736 | 0.0308 | 0.0406 |
| Model P | 0.2136 | 0.0417 | 0.2552 | 0.0660 | 0.0186 | 0.0846 | 0.0000 | 0.0000 |
| Model D | 0.2223 | 0.0417 | 0.2639 | 0.0521 | 0.0186 | 0.0707 | 0.0151 | 0.0179 |

**TABLE 2.2 :** Comparison of our calculated results of various physical parameters with the experimental numbers. Quantities in [a] are without Melosh rotation and those in [b] are with Melosh rotation for parameters from Ref.[18] $g_A/g_V$ in [a] are obtained using Bjorken sum rule.

| Model Type | $I_1^p$ | $I_1^n$ | $\mu_p/\mu_n$ [a] | $\mu_p/\mu_n$ [b] | $g_A/g_V$ [a] $\alpha_s = 0.35$ | $g_A/g_V$ [a] $\alpha_s = 0.37$ | $g_A/g_V$ [a] $\alpha_s = 0.5$ | $g_A/g_V$ [b] | F/D [a] | F/D [b] |
|---|---|---|---|---|---|---|---|---|---|---|
| Model C | 0.168 | 0.029 | -1.405 | -1.535 | 1.019 | 1.045 | 1.243 | 1.318 | 0.603 | 0.690 |
| Model P | 0.156 | -0.014 | -1.402 | -1.532 | 1.249 | 1.282 | 1.525 | 1.505 | 0.601 | 0.688 |
| Model D | 0.179 | 0.015 | -1.477 | -1.598 | 1.210 | 1.241 | 1.476 | 1.502 | 0.651 | 0.732 |
| Expt. Value [Ref.] | 0.136 [23] | -0.030 [23] | -1.460 [26] | | 1.267 [26] | | | | 0.575 [29] | |

The effect of Melosh rotation on physical quantities related to the spin structure of the nucleon has been discussed in recent literature [18-21].Basically, Melosh rotation effect comes from the relativistic effect of the quark intrinsic transversal motion inside the nucleon. As a result of this, $\Delta q$, measured in polarized deep inelastic scattering and defined as the quark spin in the light-cone formalism, can not be identified with $\Delta q_{QM}$, the spin carried by each quark flavor in the proton rest frame or the quark spin in the quark model . The quark helicity

$$\Delta q = \langle M_q \rangle \, \Delta q_{QM},$$

where $\langle M_q \rangle$ is the averaged value of the (dimensionless) Melosh rotation factor for the quark q, and is less than 1.

In Ref.[20], authors have considered a relativistic three-quark model formulated on the light-cone and concluded that Melosh rotation results in a ≈ 25 % reduction of the non-relativistic predictions for the anomalous magnetic moment, the axial vector coupling, and the quark helicity content of the proton leading to a better agreement with the observation . The model of the nucleon by Ma et al. [21], which include the three quark component and a baryon–meson state with a pseudoscalar meson, is nearer to our case because of its sea. We use their result to estimate the effect due to Melosh rotation on quantities related to the nucleon spin. In effect, it results in a replacement of $\alpha \to \alpha/\langle M_u \rangle$ and $\beta \to \beta/\langle M_d \rangle$, where numerically $\langle M_u \rangle = 0.624$ and $\langle M_d \rangle = 0.912$. This makes the parameters $\alpha$ and $\beta$ closer to the ones used in Ref.[6] on phenomenological ground. For the nucleon spin problem, $\alpha$ and $\beta$, and not their above scaled values, will be used, since the quark helicity $\Delta q$ observed in polarized DIS is actually the quark spin defined in the light-cone formalism for which $\alpha$ and $\beta$ are appropriate quantities.

In order to check the stability of our results against some plausible changes in some physical parameters, we consider two modifications of the above model. It appears reasonable to assume that in determining low energy hadronic observables, the long range and confining forces leading to specific correlations among the constituents, in addition to the statistical consideration, will have a role to play. Based upon this point of view, we introduce the following two models:

## 2.3 Sea with Pseudoscalars

In the statistical formulation of Ref.[12,13], a quark-antiquark pair is created from a gluon splitting: g$\Leftrightarrow \bar{q}q$. This pair, naturally, carries the quantum numbers of the parent gluon. However, this is not an energetically favorable situation even within the hadronic boundary [28] ; the pair on exchange of a soft gluon with the rest of the system, and also possibly on a spin flip, will evolve to a colorless pseudoscalar form, called internal Goldstone boson [28-30]. We will assume that all the $\bar{q}q$ pairs are in one or the other pseudoscalar form practically for whole of their lifetimes giving no contribution to the spin or the color charge of the proton. In case of $|gg\,\bar{q}q\rangle$ state, in order to compensate the odd parity of the $\bar{q}q$ pair, one of the gluons will be assumed to be in TE mode while the other in TM mode. With these assumptions, we can decompose the Fock states, considered earlier, in spin and color spaces as:

$|q\bar{q}\rangle \sim H_1G_1,$

$|u\bar{u}d\bar{d}\rangle$, $|q\bar{q}\;q\bar{q}\rangle \sim H_0G_1,$

$|q\bar{q}, g\rangle$, $|u\bar{u}d\bar{d}, g\rangle$, $|q\bar{q}\;q\bar{q}, g\rangle \sim |g\rangle$,

$|q\bar{q}, gg\rangle \sim |gg\rangle'$.

As a consequence of the last result, we will have

$\rho_{1/2\;0}\,[\rho_{1\,1}, \rho_{8\,8}, \rho_{10\,\overline{10}}]$; $\rho_{1/2\,1}\,[\rho_{11}, \rho_{8\,8}, \rho_{10\,\overline{10}}]$; $\rho_{3/2\,1}\,[\rho_{8\,8}]$; $\rho_{3/2\,2}\,[\rho_{8\,8}]$

$$= 0.00474\;(1,4,1;1,4,1;2;2) \quad \text{for } |q\bar{q}, gg\rangle.$$

The Fock states $|g\rangle$, $|gg\rangle$ and $|ggg\rangle$ can be decomposed as in the previous case. Thus, we get the following contribution to the expansion coefficients in Eq.(2.5) of the proton state:

$a_8^2 = 0.22143$, $a_{10}^2 = 0.02161$, $b_1^2 = 0.04247$, $b_8^2 = 1.25408$, $b_{10}^2 = 0.06825$, $c_8^2 = 0.62704$, $d_8^2 = 0.0898$.

This sea will not "actively" contribute to the spins or the magnetic moments of the nucleons. With this sea, the results of the spin distribution of nucleons come closer to the data as is evident from results of Model P in Table 2.2. There is hardly any change in the values of the ratios $\mu_p/\mu_n$ and F/D from the previous case. Matching the values of $g_A/g_V$ with the experimental numbers favors the smaller values of $\alpha_s$.

## 2.4 Sea with Suppressed Higher Multiplicity States

We propose a second modification of the model in which the contribution to the states with higher multiplicities is suppressed. Within the hadronic boundary, pseudoscalar exchange has been found to dominate over vector exchange and even gluon exchanges [5,28-30]. Although we are not using any dynamical model, we tend to believe that the states with larger number of gluons (having corresponding smaller probabilities) approximate the ones with saturated gluons for which color neutrality is achieved over a certain scale, which is called 'saturation scale'[31,32]. In Landshoff-Nachtmann model, quark–quark and hadron–hadron scatterings are assumed to arise due to exchanges of two non-perturbative gluons having vacuum quantum numbers[33]. It is believed that pomeron and odderon exchanges are associated with the exchanges of a family of glueballs which are colorless but of different spins [33]. It is reasonable to assume that when a set of 'intrinsic' gluons exist in a nucleon, they would prefer to be in a similar state.

Even within the hadronic boundary, Goldstone boson exchange (GBE) model successfully describes diverse phenomenon [28-30]. In color space, singlets are unique due to confinement, but even there the color octet exchange models, and not any higher color states exchange model, have been successfully used [34]. Larger is color multiciplity of a group of particles (here the sea), larger will be the probability of its interaction with the rest of the particles (the core) and smaller will be its probability of survival. Authors of Ref.[6] have, on phenomenological ground, proposed a set of parameters in which states with higher multiplicities occur with lower probabilities.

In view of these phenomenological evidences, it appears reasonable to propose that higher multiplicity states are suppressed. We parameterize this suppression in a simple way by assuming that probability of a system to be in a spin and color state is inversely proportional to the multiplicity (both in spin and color spaces) of the state. This probability factor is additional to the previously incorporated factors in the probabilities. With this new input, we decompose Fock states as follows.

(i) $|gg\rangle$, $|q\bar{q}\, q\bar{q}\rangle$ sea: Equating the sum of the products of probabilities to the probabilities for finding the above Fock states as done in the previous cases, as done in the previous case, we get

$\rho_{1/2\,0s}[\rho_{11s}, \rho_{8\,8s}, ]; \rho_{1/2\,1a}[\rho_{8\,8a}, \rho^{10\,\overline{10}}{}_{a}]; \rho_{3/2\,1a}[\rho_{8\,8a}]; \rho_{3/2\,2s}[\rho_{8\,8s}]$

$$= 0.03903(2,1/16;1/48,1/150;1/192;1/320) \text{ for } |gg\rangle,$$

$$= 0.00345(2,1/16;1/48,1/150;1/192;1/320) \text{ for } |u\bar{u}\, u\bar{u}\rangle,$$

$= 0.00694(2,1/16;1/48,1/150;1/192;1/320)$ for $|\bar{d}d\bar{d}d\rangle$.

(ii) $|g\bar{q}q\rangle$, $|\bar{u}u\,\bar{d}d\rangle$ sea :

$\rho_{1/2\,0}\,[\rho_{11}, \rho_{8\,8}, \rho_{10\,\overline{10}}]$; $\rho_{1/2\,1}\,[\rho_{11},\rho_{8\,8},\rho_{10\,\overline{10}}]$; $\rho_{3/2\,1}\,[\rho_{8\,8}]$; $\rho_{3/2\,2}\,[\rho_{8\,8}]$

$= 0.01912(2, 1/8, 1/50; 2/3, 1/24, 1/150; 1/96; 1/160)$ for $|g\bar{u}u\rangle$,

$= 0.02876 (2, 1/8, 1/50; 2/3, 1/24, 1/150; 1/96; 1/160)$ for $|g\bar{d}d\rangle$,

$= 0.01090 (2, 1/8, 1/50; 2/3, 1/24, 1/150; 1/96; 1/160)$ for $|\bar{u}u\,\bar{d}d\rangle$.

(iii) $|gg\bar{q}q\rangle$, $|\bar{q}q\,\bar{q}q\,g\rangle$ sea :

$\rho_{1/2\,0a}[\rho_{11a}, \rho_{8\,8a}, \rho_{10\,\overline{10}\,a}]$; $\rho_{1/2\,1s}[\rho_{11s}, \rho_{8\,8s}, \rho_{10\,\overline{10}\,s}]$; $\rho_{3/2\,1a}\,[\rho_{8\,8a}]$; $\rho_{3/2\,1s}\,[\rho_{8\,8s}]$; $\rho_{3/2\,2a}$

$[\rho_{8\,8a}]$ $= 0.00328 (1, 1/8, 1/50; 1, 1/8,1/50;1/32;1/32;1/160)$ for $|\bar{u}u\,\bar{u}u\,g\rangle$,

$= 0.00655 (1, 1/8, 1/50; 1, 1/8,1/50;1/32;1/32;1/160)$ for $|\bar{d}d\bar{d}d\,g\rangle$,

$= 0.01952(1, 1/8, 1/50; 1, 1/8,1/50;1/32;1/32;1/160)$ for $|gg\bar{d}d\rangle$,

$= 0.01307(1, 1/8, 1/50; 1, 1/8,1/50;1/32;1/32;1/160)$ for $|gg\bar{u}u\rangle$.

(iv) $|g\bar{u}u\,\bar{d}d\rangle$ sea :

$\rho_{1/2\,0}[\rho_{1\,1}, \rho_{8\,8}, \rho_{10\,\overline{10}}]$ ; $\rho_{1/2\,1}[\rho_{1\,1}, \rho_{8\,8}, \rho_{10\,\overline{10}}]$; $\rho_{3/2\,2}\,[\rho_{8\,8}]$

$= 0.02620 (1/2,1/16,1/100; 1/2,1/16,1/100;1/164;1/160)$ .

(v) $|ggg\rangle$ sea:

$\rho_{1/2\,0a}\,[\rho_{1\,1a}, \rho_{8\,8a}]$; $\rho_{1/2\,1s}[\rho_{1\,1s}, \rho_{8\,8s}]$; $\rho_{3/2\,1s}[\rho_{8\,8s}]$

$= 0.02327(1,1/32; 1/3,1/96;1/384)$ .

We would like to point out that there is nothing special about the use of the inverse of the multiplicity for suppression of higher multiplicity states. One could have fine tuned the power of the multiplicity to fit the data in a better way. It is only a possible way to suppress the contributions of states with higher multiplicities within the nucleon sea, which might be originally due to some dynamics. In the above calculation we have also included the (active) contribution of sea quarks. Numerical results for this case have been displayed in Model D in Tables 2.1 and 2.2 .

## 2.5 Summary and Conclusion

The statistical approach advocated in Ref.[12,13] was successful in describing the large asymmetry between $\bar{u}$ and $\bar{d}$ quark distributions of the proton. We have extended that approach by decomposing various quark-gluon Fock states into states in which the three quark core and the rest of the stuff (called sea ) have definite spin and color quantum numbers, using the assumption of equal probability for each substate of such a state of the nucleon. We have further used the approximation in which a quark in the core is not antisymmetrized with an identical quark in the sea, and have treated quarks and gluons as nonrelativistic particles moving in S-wave (except for a single $\bar{q}q$ sea) motion. Also, we have not taken into account any contribution of the s-quark and other heavy quarks, and we have covered only ≈ 86% of the total Fock state. With these approximations we have calculated the quarks contribution to the spin of the nucleons, the ratio of the magnetic moments of the nucleons , their weak decay constant , and the ratio of SU(3) reduced matrix elements for the axial current. All of these quantities give integrated result of Bjorken variable. We have also considered two modifications of the above statistical approach with a view to reduce the contributions of the sea components with higher multiplicities, and have done the above calculations for those two cases as well.

The effect of Melosh rotation is to increase the values of the physical quantities related to the nucleon spin, which are measured in the rest frame of the nucleon while keeping the quark contribution to the nucleon spin, measured in the light–cone frame, unchanged. If we treat the Melosh rotation as free parameter, we can reproduce the experimental value of $g_A/g_V$ along with $\mu_p/\mu_n$ =1.415, and F/D =0.610 with the Melosh rotation parameters, $(\langle M_u\rangle,\langle M_d\rangle)$ = (0.699, 0.719), (0.797, 0.827) and (0.825, 0.692) for cases (C), (P) and (D) respectively, while keeping the values of $I_1^p$ and $I_1^n$ as listed in Table 2.2 .

Our results of calculation holds good for a typical hadronic energy scale~1 GeV$^2$ [13]. Experimental results for $I_1^p$ and $I_1^n$ apply for $Q^2 \approx 10$ GeV$^2$, and their values will increase when evolved to a lower energy scale. Hence, our calculated results for $I_1^p$ and $I_1^n$ may well be consistent with the data. Our result for the ratio of magnetic moments of nucleons is within few percent of the data. Weak decay constant has been calculated using Bjorken sum rule, written up to $O(\alpha_s^3/\pi^3)$. There is some controversy in the value of $\alpha_s$ at the low energy~1GeV we are working at, and we have chosen three typical values taken from recent literature. The significance of the Melosh rotation connecting the spin states in the light-front dynamics and the conventional instant form dynamics has been widely recognized. We have tried to construct a spin wave function of a nucleon with a non trivial sea in the nucleon rest frame from a statistical model of a nucleon. Such a wave function, along with a Melosh rotation, is capable of giving a reasonable result for several physical quantities related to the nucleon spin.

# CHAPTER-III

## ISOSPIN BREAKING IN DIAGONAL PION-NUCLEON COUPLING CONSTANT: QCD SUM RULE APPROACH

### 3.1 Introduction

Determination of meson-nucleon couplings is of particular interest in particle physics as well as in nuclear physics. In particle physics, estimate of these parameters is useful to test the low energy behaviour of the QCD. In nuclear physics, nucleon-nucleon interactions are traditionally viewed as arising from meson exchanges. Pion exchange is linked to spontaneous and explicit chiral symmetry breaking (CSB) of low-energy QCD. According to Goldstone theorem, pions are Goldstone bosons having point like derivative couplings to the nucleons. For intermediate and short distances, heavy mesons have to be included in the modeling.

On the other hand, at energies much below the scales set by the pion mass, it is sufficient to consider four-nucleon interaction only. Starting from nucleon and pion degrees of freedom, effective field theory has been used for a separation of these scales [1]. Accounting higher order terms in the chiral expansion, a form of two-nucleon potential for the neutron-proton system has been developed in so-called modified Weinberg scheme and shown to be close in accuracy to the so-called modern potentials (in some partial waves) [2]. Isospin symmetry is a good symmetry of low-energy hadronic physics and charge symmetry is even better. In low-energy observables isospin violation is typically much smaller. The study of charge symmetry breaking, which is a special case of isospin breaking, in pion-nucleon coupling is an important step for investigation of charge symmetry breaking effects in nucleon-nucleon interactions.

The effect of isospin violating meson-nucleon couplings has recently seen a strong revival of interest in the investigation of charge symmetry breaking phenomenon. On a microscopical level, isospin symmetry is broken by electromagnetic interaction as well as the mass difference of up and down quarks: $m_u - m_d$. We shall examine the difference between the diagonal pion-nucleon coupling constants, $g_{pp\pi^0} - g_{nn\pi^0}$ using the QCD sum rule method.

## 3.2 A Model Calculation in QCDSR: Proton Mass

QCD sum rule method was originally suggested by Shifman, Vainstein and Zakharov, and has been applied to determine masses and leptonic widths of light mesons (ρ, π, κ*). For these determination, virtuality region is taken of the order of $Q^2$~1GeV$^2$ and $\alpha_s$~0.3-0.4, so that perturbative terms are small i.e $\alpha_s/\pi$~0.1 and hence only leading logarithmic corrections~$[\alpha_s(Q^2)\ln Q^2/\Lambda^2]$ are taken into account. To illustrate the characteristic features of the method and to use it for our main calculation, we shall show a calculation of the mass of the proton using QCD sum rules.

For this purpose we consider the polarization operator as

$$\Pi(p,k) = i\int d^4x\, e^{ipx} \langle 0 | T\{\eta(x), \bar{\eta}(0)\} | 0\rangle \qquad (3.1)$$

where $\eta$ (x) is the quark current with proton quantum numbers and $p^2$ is chosen to be space-like: $p^2 < 0$, $|P^2| \sim 1GeV^2$. The current $\eta$ is colorless product of three quark fields $\eta(x) = \varepsilon_{abc}$ $q^a q^b q^c$, q=u,d, the exact form of the current will be specified below.

Unlike mesons, in baryons there exist several currents with quantum numbers of a given baryon. The choice between them should be done from physical reasons in order to provide: (1) renormcovariance, (2) existence of nonrelativistic limit, (3) the above formulated requirement (for proton) for the functions $f_1$ and $f_2$ (given below in Eq. (3.2) ) to be of the same order, (4) convergence of operator expansion series within accounted terms. Specifically for proton all these requirements are satisfied by the current ,

$\eta = (u^a C\gamma_\mu u^b)\gamma_\mu \gamma_5 d^c \varepsilon_{abc}$ .

The general structure of $\prod$(p) is

$$\prod(p) = \not{p} f_1(p^2) + f_2(p^2) \qquad (3.2)$$

For each of the function $f_i(p^2)$, i=1,2 the following operator expansion can be written:

$$f_i(p^2) = \sum_n C_n^i(p^2) \langle 0|O_n^{(i)}(0)|0\rangle \qquad (3.3)$$

where $\langle 0|O_n^i|0\rangle$ are vacuum expectation values of different operators (vacuum condensates) and $C_n^i(p^2)$ are functions calculable in QCD.

$$f_1(p^2) = C_0 p^4 \ln\frac{\Lambda_u^2}{-p^2} + \sum_k C_2^k \langle 0|\bar{q}\Gamma^{(k)}q.\bar{q}\Gamma^{(k)}q|0\rangle \frac{1}{p^2} +$$
$$C_4 \langle 0|\frac{\alpha_s}{\pi} G_{\mu\nu}^a G^{a\mu\nu}|0\rangle \ln\frac{\Lambda_u^2}{-p^2} + C_6 \langle 0|\frac{\alpha_s}{\pi}\bar{q}q.\bar{q}g_s\sigma^{\mu\nu}G_{\mu\nu}^a\frac{\lambda^a}{2}q|0\rangle \frac{1}{p^4} \qquad (3.4)$$
$$+ polynomials$$

$$f_2(p^2) = C_1 p^2 \ln\frac{\Lambda_u{}^2}{-p^2}\ 0|q\bar{q}|0\ + C_3\ 0|\bar{q}q\frac{\alpha_s}{\pi}G^{a\mu\nu}G^a_{\mu\nu}|0\ \frac{1}{p^2}$$

$$C_5\ {}_k\ 0|\bar{q}q.\bar{q}\Gamma^{(k)}q.\bar{q}\Gamma^{(k)}q|0\ \frac{\alpha_s}{p^4} + polynomials \qquad (3.5)$$

where $C_i$'s are constants, $\Lambda_u$ is the ultraviolet cut-off. The current u-and d-quark masses entering the Lagrangian of QCD are very small, of the order of several MeV, so they can be neglected with very good accuracy, i.e for the time being, we neglect the quark masses, then $L_{QCD}$ is chiral-invariant. If this chiral symmetry would not be spontaneously broken, then $f_2$ $(p^2)$ would remain identically zero. As explained in Chapter 1, the chiral symmetry is spontaneously broken. The first evidence of this is the existence of large baryon masses: $M_B >> \Lambda_{QCD}$. Another signal is the fact that chiral symmetry violating quark condensate $0|\bar{q}q|0$ is non-zero, and is approximately equal to –(240MeV)$^3$.

$$0|q\bar{q}|0\ = -\frac{1}{2}\frac{f_\pi^2 m_\pi^2}{m_u + m_d} = -(240MeV)^3, \qquad (3.6)$$

Since $0|\bar{q}q|0$ is the lowest dimensional chirality violating operator, the operator expansion for $f_2(p^2)$ starts from the term proportional to $0|\bar{q}q|0$ .

For any colorless operator $O_1$ and $O_2$ at large $N_c$

$$0|O_1O_2|0\ =\ 0|O_1|0\ \ 0|O_2|0\ (1+O(\frac{1}{N_c})) \qquad (3.7)$$

i.e in the limit $N_c$ factorization becomes exact. By virtue of factorization and taking into account the relation

$$0|q^a_\alpha(0)\bar{q}^b_\beta(0)|0\ = -\frac{1}{12}\delta^{ab}\delta_{\alpha\beta}\ 0|q\bar{q}|0 \qquad (3.8)$$

(a, b = 1,2,3 are color, α, β are Lorentz indices) all four-quark vacuum expectation values (v.e.v) reduce to the quark condensate square $0|\bar{q}q|0\ ^2$. In order to improve and control the accuracy in the calculation of mass, other v.e.v's will also be taken into account: Gluonic condensate $0|\frac{\alpha}{4\pi}G^a_{\mu\nu}G^{a\mu\nu}|0$, mixed condensate $0|\bar{q}\sigma_{\mu\nu}(\lambda^n/2)G^n_{\mu\nu}q|0$ and higher dimensional v.e.v's $0|\bar{q}q|0\ \ 0|\bar{q}\sigma_{\mu\nu}(\lambda^n/2)G^n_{\mu\nu}q|0$ $,\alpha_s\ \ 0|\bar{q}q|0\ ^3,\ 0|\bar{q}q|0\ \ 0|\frac{\alpha_s}{\pi}G^a_{\mu\nu}G^{a\mu\nu}|0$. The gluonic condensate gives a contribution into chirality preserving structure $f_1(p^2)$.

∏(p) may be expressed using the dispersion relations

$$f_i(s) = -\frac{1}{\pi}\ {}_0\ \frac{\operatorname{Im} f_i(p^2)}{p^2 - s}dp^2 + polynomial \qquad (3.9)$$

In order to extract physical quantities of interest, it is useful, at this stage, to apply Borel transformation on both side of this equation. The Borel (Laplace) transformation is defined as

$$Bf_i(q^2) = \lim_{\substack{-q^2,n \\ -q^2/n=M^2\,fixed}} \frac{(-q^2)^{n+1}}{n!}\frac{d}{dq^{2n}} f_i(q^2)\,. \qquad (3.10)$$

This gives

$$B_{M^2} f_i(s) = \frac{1}{\pi}\ {}_0\ \exp(-\frac{p^2}{M^2})\operatorname{Im} f_i(p^2)dp^2,$$

where $f_i(s)$ is given by dispersion relation (3.9). In particular, we have

$$B_{M^2}\frac{1}{s^n}=\frac{1}{(M^2)^{n-1}(n-1)!} \qquad (3.11)$$

The Borel transform permits to attain three goals at once:

(1) to nullify subtraction term,

(2) to suppress the contribution of the higher excited states compared to the desired lowest state (proton)

(3) to suppress the contributions of high order terms in the operator expansion (owing to factor 1/(n-1)! in (3.11)) .

The lowest state (proton) contribution to imaginary part of $\prod$(p) has the form

$$\mathrm{Im}\,\Pi(p)=\pi\ \langle 0|\eta|p\rangle\langle p|\bar{\eta}|0\rangle\ \delta(p^2-m^2)=\pi\lambda_N{}^2(\not p+m)\delta(p^2-m^2), \qquad (3.12)$$

*where*

$$\langle 0|\eta|p\rangle=\lambda_N u(p)$$

Here $\lambda_N$ is the proton transition constant into quark current and u(p) is the proton spinor.

It is clear from above that proton contribution will dominate in some region of the Borel parameter $M^2$ only when QCD calculated functions $f_1$ and $f_2$ are of same order, and the spontaneous violation of chiral invariance characterized by the value of quark condensate has to explain the numerical value of the proton mass.

Contributions of higher mass states will also be taken into account in order to improve and control the accuracy in the dispersion representation as written above, by replacing Imf($p^2$) by contribution of the simplest quark loops starting from some "continuum threshold" W. Taking into account all points stated above, the sum rule for the calculation of the proton mass is given as

$$M^6 E_2\left(\frac{W^2}{M^2}\right)L^{-4/9} + \frac{4}{3}a^2 L^{4/9} + \frac{1}{4}bM^2 E_0\left(\frac{W^2}{M^2}\right).L^{-4/9} - \frac{1}{3}a^2 \frac{m_o^2}{M^2}$$ , (3.13)

$$= \tilde{\lambda}_n^2 \exp\left(-\frac{m^2}{M^2}\right)$$

$$2aM^4 E_1\left(\frac{W^2}{M^2}\right)L^{-4/9} + \frac{272}{81}\frac{a^3}{M^2}L^{4/9} + \frac{1}{4}bM^2 E_0\left(\frac{W^2}{M^2}\right).L^{-4/9} - \frac{1}{3}a^2 \frac{m_o^2}{M^2}$$ (3.14)

$$= \tilde{\lambda}_n^2 \exp\left(-\frac{m^2}{M^2}\right)m$$

Here

$a_q = -(2\pi)^2 \langle 0|\bar{q}q|0\rangle$ ,

b = $(2\pi)^2 \; 0|\frac{\alpha_s}{\pi}G^a_{\mu\nu}G^{a\mu\nu}|0 \quad 0.5GeV^2$ and

$-g \; 0|\bar{q}\sigma_{\alpha\beta}\frac{\lambda^n}{2}G_{\alpha\beta}{}^n q|0 \quad m_0^2 \; 0|q\bar{q}|0$

$m_0^2 \quad 0.8GeV^2$

The factors

$E_0(x) = 1-e^{-x}$ , $E_1(x)= 1-(1+x)e^{-x}$ and $E_2(x)= 1-(1+x+x^2/2)e^{-x}$ (3.15)

take into account the continuum contribution. We also have

$\tilde{\lambda}_N = 32\pi^4\lambda_N^2$,

the factors $L = \frac{\ln(M/\lambda)}{\ln(\mu/\lambda)}$ take into account the anomalous dimensions of the operators (Λ is the QCD parameter, μ is the renormalization point, numerical values hereafter corresponds to μ=1.0GeV).

Thus proton mass m and the constant $\lambda_N$ may be found from the sum rules by the best fit, which is to be made within a restricted interval of $M^2$. The small contributions of continuum (say less than 50%) and small percentage of higher power correction (say,<10%) restrict $M^2$ from above and below respectively. Outside this interval the accuracy of the theory is noncontrollable. Estimation of contribution of higher mass states and higher power corrections makes the result obtained by this method reliable. The best fit in the permissible interval $0.7<M^2<1.2GeV^2$ at the chosen values of v.e.v's and W=1.5GeV gives m. The value of m can be determined by dividing Eq.(3.14) by Eq.(3.13) and plotting the l.h.s with respect to $M^2$. At the chosen values of v.e.v's and W=1.5GeV, one gets

m =1.0±0.1GeV.

Similarly $\tilde{\lambda}_n^2$ can be determined from, say, Eq.(3.13) by taking exponential on l.h.s and plotting it with respect to $M^2$ and finding the best fit in the permissible interval:

$\tilde{\lambda}_n^2 = 2.1 \pm 0.2 GeV^6$.

## 3.3 Isospin Splitting in Pion-Nucleon Coupling Constant

Many strong interaction-processes involve meson-baryon coupling constants as the main ingredient. The determination of the fundamental quantities such as masses, decay constants and form factors of hadrons requires information about the physics at large distances. In other words, for a reliable determination of these parameters we need some non-perturbative approach. Among all non-perturbative approaches, QCD sum rule is one of the most powerful method designed to estimate low energy characteristics of hadrons and studying their properties. As explained earlier, this method is based on the short distance OPE of vacuum-vacuum correlation function in terms of condensates. Within this frame work we can go into the region of intermediate momentum transfers and take due care about non-perturbative effects. This kind of QCD sum rule approach significantly reduces model dependence of the results obtained.

NN scattering length in $^1S_0$ state plays a special role because it is extremely sensitive to small differences in the strength of the force. The values of pp, nn, np scattering length i.e $a_{pp}$, $a_{nn}$, $a_{pn}$ respectively after the coulomb forces are removed, indicate the pn force is, in average, slightly more attractive than nn and pp forces, whereas nn interaction is more than pp interaction. Experimental values for the different NN scattering length in $^1S_0$ state shows that charge symmetry is slightly broken in nature. This kind of breaking is mostly attributed to the proton-neutron mass difference, which appears in the two pion exchange graphs involving one nucleon and one delta. This mass difference affects the kinetic energy of the nucleon and besides, it also influences all meson-exchanges diagrams, mainly through the propagation of nucleon intermediate states with the correct mass in 2π-exchanges diagrams.

Knowledge of these couplings, along with the isospin breaking in them, from QCD may be used for construction of nucleon-nucleon (NN) potential [3]. Introduction of charge symmetry breaking in NN potential models by hand may not be unique. The NN scattering data used in the fitting processes are not precise enough to pick out a specific mesonic channel. Isospin violation of the nucleon-nucleon interaction has been studied in an effective field theory approach, where leading order charge independence breaking is explained in terms of one-pion exchange together with a four-nucleon contact term [4]. Therefore, it is useful to constrain the isospin violation in the pion-nucleon couplings directly from QCD based non-perturbative methods such as QCD sum rule.

At the fundamental level, isospin violation takes place due to charge difference and mass difference of up- and down-quarks. At the phenomenological level, the effect of these differences may get augmented due to strong interaction, and in practice, this may appear in the form of isospin splitting of other phenomenological parameters such as quark condensates. QCD sum rules have been used in past to study pion-nucleon couplings and also their isospin breaking[5-11]. Three different methods have been used to investigate pion-nucleon coupling constant in the framework of the conventional QCD sum rule. In the three-point function method, one studies the vacuum-to-vacuum matrix element of the correlation function of the interpolating fields of the two nucleons and a meson[6]. However, it has been argued that the method suffers from contamination of higher resonance states [12].

In the two-point function external field method, one studies the correlation function of the interpolating fields of the two nucleons in the presence of an external pion field [7]. However, the induced condensates appearing in this method are not as reliably known, as the other more commonly used condensates. In the following we shall follow the third, the two-point function method [5, 8-10] in which one studies vacuum-to-pion matrix element of the correlation function of the interpolating fields of two nucleons:

$$\Pi(p,k)=i\int d^4x\, e^{ipx} \langle 0 | T\{\eta(x), \bar{\eta}(0)\} | \pi^0(k)\rangle \qquad (3.16)$$

Here, $\eta$ is the interpolating field of a nucleon and $|\pi^0(k)\rangle$ is the neutral pion state with momentum k. Isospin is suppressed for simplicity. For η, Ioffe's interpolating field [13] will be used; for proton, it is written as

$$\eta(x)=\varepsilon_{abc}\,(u_a^T(x)C\gamma_\mu u_b(x))\gamma_5\gamma^\mu d_c(x) \qquad (3.17)$$

The correlation function is calculated, on the one hand, by Wilson's operator product expansion(OPE), and on the other hand it is evaluated using hadronic physical states. The two descriptions are matched in the deep Euclidean region via the dispersion relation and the physical quantity of interest is extracted.

The expression (3.16) is known to have four Dirac structures [11]. Among these, the coefficient of the double pole of $i\gamma_5\hat{p}$ structure on the mass shell vanishes, and the sum rule obtained at the Dirac structure $i\gamma_5$ substantially underestimates the ratio F/D compared to its value known in SU(3) symmetry limit [5]. The Dirac structure $i\gamma_5\hat{k}$ has been found not to be reliable for calculating the πNN coupling as it contains large contribution from the continuum [8]. The sum rules for the meson-baryon coupling constant at the structures $i\gamma_5\hat{k}$ and $\gamma_5\sigma_{\mu\nu}p^\mu k^\nu$ have been studied extensively in [5,8-10]. Kim et al. [8,9] have claimed to find nice features in the sum rule at the $\gamma_5\sigma_{\mu\nu}p^\mu k^\nu$ Dirac structure for calculation of πNN coupling constant. It was observed that for this sum rule the coupling constant comes out to be independent of the choice of the effective Lagrangian, i.e, independent of pseudoscalar and axial vector schemes [10], and is stable against the variation of the continuum parameter due to cancellation of contributions from higher resonances of different parities [8]. We use this sum rule to calculate isospin splitting in the diagonal pion-nucleon coupling constant $g_{\pi NN}$. In the existing result for the correlation function (3.16), we also include quark mass dependent terms and do renormalization group improvement. In addition, we also take into account the effect of $\pi^0$–$\eta$ mixing and electromagnetic correction to the $\pi^0$– quark couplings.

In order to reduce the dependence of the splitting in the coupling on the isospin splitting in the quark condensate, which is rather poorly known, we take the ratio of the sum rule for the coupling $g_{\pi NN}$ to the corresponding chiral-odd sum rule for the nucleon mass, and then consider the difference and the average of this ratio for proton and neutron. This resulting sum rule is fitted to a straight line form, which directly gives the difference and the average of the couplings:

$$\delta g = g_{\pi^0 pp} - g_{\pi^0 nn}, \qquad g_{\pi NN} = (g_{\pi^0 pp} + g_{\pi^0 nn})/2 \,. \qquad (3.18)$$

## 3.4 Sum Rules for Pion-Nucleon Couplings

As stated above, in order to construct sum rules for the coupling $g_{\pi NN}$ at the structure $\gamma_5\sigma_{\mu\nu}p^\mu k^\nu$, in addition to the results already derived in Ref.[5], we calculate contributions coming from the quark mass dependent terms of Figs.3.1(a) and 3.1(b). We enumerate below the Fourier transforms and the Borel transforms of the coefficients of $\gamma_5\sigma_{\mu\nu}p^\mu k^\nu$, of these contributions for the proton:

$$\text{Fig 3.1(a)} \xrightarrow{F.T.} -(1/2\pi^2) m_d f_\pi \gamma_5\sigma_{\mu\nu} p^\mu k^\nu \ln(-p^2)$$

$$\xrightarrow{B.T.} (M^2/2\pi^2) m_d f_\pi \gamma_5\sigma_{\mu\nu} p^\mu k^\nu \qquad (3.19a)$$

$$\text{Fig 3.1(b)} \xrightarrow{F.T.} -(1/9f_\pi)(m_u/p^4)\langle \bar{u}u\rangle\langle \bar{d}d\rangle \gamma_5\sigma_{\mu\nu} p^\mu k^\nu$$

$$\xrightarrow{B.T.} (1/9f_\pi)(m_u/M^2)\langle \bar{u}u\rangle\langle \bar{d}d\rangle \gamma_5\sigma_{\mu\nu} p^\mu k^\nu \qquad (3.19b)$$

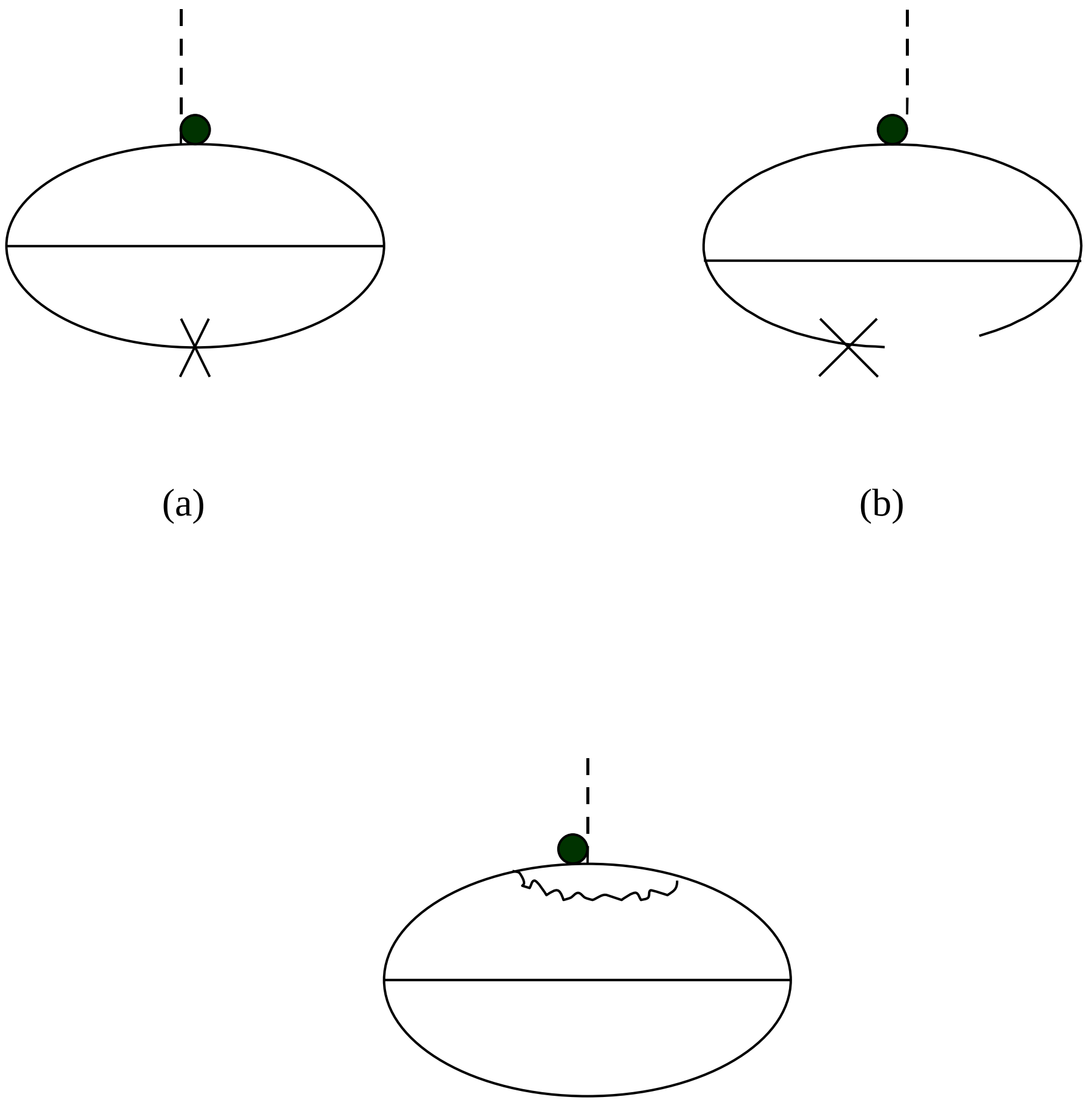

(a) (b)

(c)

**Figure 3.1**: The additional diagrams considered in this work. The solid lines denote the quark propagator, dashed line is the pion propagator, and the blob denotes their interaction vertex. Cross denotes quark mass insertion.

We have checked that the coefficient of the operator $m_q (\langle \bar{u}u \rangle, \langle \bar{d}d \rangle) \langle (\alpha_s/\pi)G^2 \rangle$ is zero. So far we have assumed that $\pi^0$ mass eigen state is a pure isovector state. However, it is well known that the mass eigenstates $\pi^0$ and $\eta$ are not pure octet states [14], rather they are mixtures of flavor octet eigenstates $\pi_3$ and $\pi_8$. Denoting $\pi - \eta$ mixing angle by $\theta$, the mass eigenstates may be written as :

$$|\pi^0\rangle = |\pi_3\rangle + \theta|\pi_8\rangle, \quad |\eta\rangle = |\pi_8\rangle - \theta|\pi_3\rangle.$$

Since $\theta$ is small $\cong$ 0.01, this amounts to the replacement for the couplings:

$$g_{\pi^0 pp} = g_{\pi 3pp} + \theta\, g_{\pi 8pp}, \quad g_{\pi^0 nn} = g_{\pi 3nn} - \theta\, g_{\pi 8nn}. \qquad (3.20)$$

Here, we ignore any possible mixings of $\pi^0$ and $\eta$ with $\eta'$. We use the sum rules for the couplings of pure octet states, $g_{\pi 3NN}$ and $g_{\pi 8NN}$ [5] in the above Eq. (3.20) to get the couplings of the physical state $\pi^0$ with nucleons.

It has been pointed out in [7] that the vertex corrections to $\pi^0$uu and $\pi^0$dd couplings, due to photon exchanges, can give rise to non negligible isospin breaking in $g_{\pi NN}$. Specifically, it has been found that in the minimum subtraction scheme the following electromagnetic corrections arise to the pion-quark couplings:

$$g_{\pi^0 uu} \to g_{\pi^0 uu}\{1+\frac{\alpha}{4\pi}(\frac{52}{9} - \frac{4}{3}\gamma_E)\}, \quad g_{\pi^0 dd} \to g_{\pi^0 dd}\{1+\frac{\alpha}{4\pi}(\frac{13}{9} - \frac{1}{3}\gamma_E)\} \qquad (3.21)$$

The most important correction arising due to these vertex corrections is shown in Fig.3.1(c) for perturbative contribution. This correction will be different for proton and neutron because $\pi^0$ couples to different quark lines for the two cases. Similar correction will also arise in other terms. In effect, the coefficient of each term in the OPE is multiplied with a factor which depends on the charge of the quark to which $\pi^0$ couples in the nucleon.

Combining the sum rule for the meson-nucleon couplings as obtained in Ref.[5], but with the specification of the flavor of the quark condensate, at the Dirac structure $\gamma_5\sigma_{\mu\nu} p^\mu k^\nu$ with the above three types of corrections, we get the following sum rules, after Borel transformation and renormalization group improvement, for the diagonal pion-nucleon couplings:

$$g_{\pi^0 pp}\lambda_p^2(1+D_{\pi p}M^2)$$

$$= - e^{(m_p^2/M^2)} \frac{\bar{d}d}{f_\pi} [\{ \frac{M^4 E_0(S_o^p/M^2)}{12\pi^2} + \frac{1}{216} \langle (\alpha_s/\pi)G^2 \rangle + \frac{\bar{u}u}{9} m_u\}(1 - \frac{\theta f_\pi}{\sqrt{3} f_\eta})$$

$$\{1+\frac{\alpha}{4\pi}(\frac{13}{9} - \frac{1}{3}\gamma_E)\} + f_\pi^2 (\frac{4M^2}{3} - \frac{m_0^2 L^{-14/27}}{6} + \frac{26\delta^2}{27} - \frac{M^4 L^{-8/9} m_d E_0(S_o^p/M^2)}{2\pi^2\langle \bar{d}d\rangle} \quad 1+\frac{10\delta^2}{9M^2} \quad )($$

$$1+\frac{\theta f_\eta}{\sqrt{3} f_\pi}) \{1+\frac{\alpha}{4\pi}(\frac{52}{9} - \frac{4}{3}\gamma_E)\}], \qquad (3.22a)$$

$$g_{\pi^0 nn}\lambda_n^2(1+D_{\pi n}M^2)$$

$$= - e^{(m_n^2/M^2)} \frac{\bar{u}u}{f_\pi} [\{ \frac{M^4 E_0(S_o^n/M^2)}{12\pi^2} + \frac{1}{216} \langle (\alpha_s/\pi)G^2 \rangle + \frac{\bar{d}d}{9} m_d \}(1 + \frac{\theta f_\pi}{\sqrt{3} f_\eta})$$

$$\{1+\frac{\alpha}{4\pi}(\frac{52}{9} - \frac{4}{3}\gamma_E)\} + f_\pi^2 (\frac{4M^2}{3} - \frac{m_0^2}{6} L^{-14/27} + \frac{26\delta^2}{27} - \frac{M^4 L^{-8/9} m_u E_0(S_o^n/M^2)}{2\pi^2\langle \bar{u}u\rangle}) \times \; 1+\frac{10\delta^2}{9M^2}$$

$$\left(1-\frac{\theta f_\eta}{\sqrt{3} f_\pi}\right)\{1+\frac{\alpha}{4\pi}\left(\frac{13}{9}-\frac{1}{3}\gamma_E\right)\}] \qquad (3.22b)$$

Here, $L = \ln(M^2/\Lambda_{QCD}^2)/\ln(\mu^2/\Lambda_{QCD}^2)$, μ is the renormalization scale, and $\Lambda_{QCD}$ is the QCD scale parameter. The anomalous dimensions of various operators have been taken into account through the appropriate powers of L [15]. It is understood that within this truncated series, v.e.v's of the quark and the gluon operators have to be taken at the scale μ. $D_{\pi N}$ are unknown constants arising from resonance states $\langle \bar{q} g\sigma.Gq\rangle = m_0^2 \langle \bar{q}q\rangle$, $\gamma_E$ is the Euler's constant. $S_o^{p,n}$ are the continuum thresholds which take care of contributions of excited states in a standard way, and $E_o(x)=1-e^{-x}$. It is clear from the sum rules (3.22a) and (3.22b) that the isospin splitting in the diagonal coupling constant, δg, has a direct dependence on the isospin splitting of the light quark condensate $\langle \bar{q}q \rangle$ and on the same of the coupling of the nucleon interpolating field to the nucleon state, $\lambda_N$. Both these splittings are rather poorly known. However, if we divide these sum rules by the chiral-odd mass sum rules of the respective nucleons, then the $\lambda_N$ dependence will get cancelled and the dependence of δg on the isospin splitting of the quark condensate will get minimized. We use the sum rule for the calculation of neutron-proton mass difference derived by Yang et al.[15]:

$$\lambda_p^2 = -(\frac{1}{m_p})\exp(\frac{m_p^2}{M^2}) < \bar{d}d > [\frac{1}{4\pi^2} M^4 E_1(\frac{S_{oN}^p}{M^2})(1+\frac{4\chi}{9}) - \frac{1}{18} < (\frac{\alpha_s}{\pi})G^2 > +$$

$$\frac{544}{81}\pi\alpha_s < \bar{u}u >^2 \frac{L^{-1/9}}{M^2} - (\frac{M^6}{16\pi^2})(\frac{m_d}{<\bar{d}d>})L^{-8/9} E_2(\frac{S_{oN}^p}{M^2}) + (\frac{M^2}{32\pi^2})(\frac{m_d}{<\bar{d}d>})$$

$$< (\frac{\alpha_s}{\pi})G^2 > L^{-8/9} E_2(\frac{S_{oN}^p}{M^2}) - \frac{4}{3} m_d \frac{< \bar{u}u >^2}{< \bar{d}d >} - 2m_u < \bar{u}u > +(\frac{m_{em}^2 M^2}{24\pi^2})E_0(\frac{S_{oN}^p}{M^2})]$$

(3.23a)

$$\lambda_n^2 = -(\frac{1}{m_n})\exp(\frac{m_n^2}{M^2}) < \bar{u}u > [\frac{1}{4\pi^2} M^4 E_1(\frac{S_{oN}^n}{M^2})(1+\frac{\chi}{9}) - \frac{1}{18} < (\frac{\alpha_s}{\pi})G^2 > +$$

$$\frac{544}{81}\pi\alpha_s < \bar{d}d >^2 \frac{L^{-1/9}}{M^2} - (\frac{M^6}{16\pi^2})(\frac{m_u}{<\bar{u}u>})L^{-8/9} E_2(\frac{S_{oN}^n}{M^2}) + (\frac{M^2}{32\pi^2})(\frac{m_u}{<\bar{u}u>})$$

$$< (\frac{\alpha_s}{\pi})G^2 > L^{-8/9} E_2(\frac{S_{oN}^n}{M^2}) - \frac{4}{3} m_u \frac{< \bar{d}d >^2}{< \bar{u}u >} - 2m_d < \bar{d}d >]$$

(3.23b)

The terms with χ in Eqs.(3.23a) and (3.23b) take into account perturbative electromagnetic contribution, and $m_{em}^2 a_q \quad (2\pi)^2 \langle e\bar{q}\sigma.Fq\rangle$, and $\langle\bar{u}\sigma.Fu\rangle = \frac{2}{3}\langle\bar{q}\sigma.Fq\rangle$ with $\langle\bar{d}\sigma.Fd\rangle = -\frac{1}{3}\langle\bar{q}\sigma.Fq\rangle$, has been introduced ($F_{\mu\upsilon}$ is the electromagnetic-field strength tensor). In order to attain fit, $\chi$ = 0.0036 and $m_{em}^2$ = 0.048GeV$^2$ has been chosen [15].

Here, $S_{oN}^{p,n}$ are the continuum thresholds for the mass sum rules, and these may, in general be different from $S_o^{p,n}$. $E_1(x) = 1-(1+x)e^{-x}$ and $E_2(x) = 1-(1+x+x^2/2)e^{-x}$. Eliminating $\lambda_p^2$ of Eq.(3.22a) with the above $\lambda_p^2$ of Eq.(3.23a) and $\lambda_n^2$ of Eq.(3.22b) with the above $\lambda_n^2$ of Eq.(3.23b), we get the sum rules for $g_{\pi^0 pp}$ and $g_{\pi^0 nn}$. Finally, on taking the difference and the average of these two sum rules we get sum rules for $\delta g$ and $g_{\pi NN}$, which we express as:

$$\delta g\,(1+ D^a_{\pi N} M^2) = F_a(M^2), \quad g_{\pi^0 NN}\,(1+D^s_{\pi N} M^2) = F_s(M^2), \qquad (3.24)$$

where $D^a_{\pi N}$ and $D^s_{\pi N}$ are constants. We shall study the sum rule for $g_{\pi^0 NN}$ also, in parallel with that for $\delta g$, and compare the result for $g_{\pi^0 NN}$ with that derived earlier [5] in a similar approach. Thus a straight line fit of $F_{a,s}(M^2)$ will directly give $\delta g$ and $g_{\pi^0 NN}$ in the form of intercepts.

## 3.5 Analysis of Result and Discussion

Let us define $a_q = -(2\pi)^2 \langle \bar{q}q \rangle$, $b = \langle g_s^2 G^2 \rangle$, $\gamma = \dfrac{\bar{d}d}{\bar{u}u} - 1$ and set $\langle \bar{q}q \rangle = \dfrac{1}{2}[\,\bar{d}d + \bar{u}u\,]$.

Normally, for the calculation of $g_{\pi NN}$, light quark mass dependent terms are not included. However, we find that the perturbative quark mass dependent term is numerically more important than the power corrections in quark mass independent terms. To get an idea of the errors involved in values of $\delta g$ and $g_{\pi NN}$, we vary the values of condensates and the continuum threshold consistent with their values used in literature: the value of $a_u$ has been varied from 0.45 $GeV^3$[17] to 0.55 $GeV^3$ [7,15,18], while that of b has been varied between 0.47 $GeV^4$[5,15] to 1.0 $GeV^4$[16,17]; $\delta^2$ has been varied from 0.2 $GeV^2$[19] to 0.35 $GeV^2$ [20]. Most of our analysis is based on $\gamma$ = -0.01 which is the upper limit from a range given in Ref.[6] based on various sources: 0.002<- $\gamma$ <0.010. For the sake of comparison, we have also given results obtained for $\gamma$ =-0.00657[14] .The variation of the continuum threshold $S_0$ from 2.07$GeV^2$ to 2.57$GeV^2$ [5], for a given set of condensates, changes $g_{\pi NN}$ by a maximum of 3% and changes δg at most by 7%. The range of the Borel mass squared is 0.8 $GeV^2 \leq M^2 \leq 1.0$ $GeV^2$. This range is chosen so as to ensure that the contribution of excited states remains less than 50% and that of the operator of the highest dimension considered remains less than 10% of the total. Smaller range of Borel mass parameter, such as the above, is normally used whenever QCD sum rules are applied for calculating isospin splittings of nucleonic parameters [15]. Moreover, this range is within the ones used in Refs. [5,15].

We have looked for the values of $g_{\pi NN}$ and δg in the parameter space spanned by $a_u$, b, $\delta^2$ and $S_0$ within the ranges stated above. The highest and the lowest values of δg and $g_{\pi NN}$ along with the values of the parameters for which they arise are displayed in Table 3.1. In all, we get $\delta g = -(4.92 \pm 1.90) \times 10^{-2}$ and $-(5.09 \pm 1.87) \times 10^{-2}$, $g_{\pi NN} = 11.76 \pm 2.43$ and $11.13 \pm 2.45$ for $\Lambda_{QCD} = 0.1$ GeV[15] and 0.15 GeV respectively. For a given set of values of $a_u$, b, $\delta^2$ and $S_0$, the maximum variation occurring in δg due to change in $\Lambda_{QCD}$ from 0.1 GeV to 0.15 GeV is 6.3% while that for $g_{\pi NN}$ is 7.5%. The values of $\delta g/g_{\pi NN}$ obtained are $-(4.17 \pm 1.42) \times 10^{-3}$ and $-(4.55 \pm 1.42) \times 10^{-3}$ for $\Lambda_{QCD} = 0.1$ GeV and 0.15 GeV respectively. The lowest numerical range of δg, for $\gamma = -0.00657$ in the same parameter space, gets pushed down to $-1.13 \times 10^{-2}$. In Table 3.2, we have displayed a set of values of parameters ($a_u$, b, $\delta^2$ and $S_0$) which give rise to central values of δg for the two values of $\Lambda_{QCD}$ and γ.

Contributions to δg for its central value coming from various symmetry breaking parameters are displayed in Table 3.3. We observe that the contributions coming from the non vanishing values of each of γ, α, θ, $\Delta m_q$, and $\Delta m_N$ individually add up almost linearly to give the final value of δg when all of these parameter are non zero. It is well known that $m_q < \bar{q}q >$ is renormalization group invariant quantity [15]. From the contributions of $\Delta m_q$ and γ to δg ,

it is evident that $\delta g$ will remain stable for a variation in $\Delta m_q$ and the corresponding variation in $\gamma$ in accordance with the renormalization group equation. The largest contribution to $\delta g$ ($\delta g/g_{\pi NN} = -2.0 \times 10^{-2}$ ) comes from $\Delta m_N \neq 0$ alone. Naively, one may expect its contribution to be ~ $\Delta m_N/m_N \sim 10^{-3}$ . However, r.h.s of Eqs (3.22a,b), when divided by r.h.s of mass sum rules of Eqs.(3.23a,b) contain electromagnetic contribution from phenomenological parameters $\chi$ and $m_{em}$ . Moreover, $D_{\pi N}$ in eqs.(3.22a,b), which decide the slope of the straight line, arises due to the transitions $N \rightarrow N^*$, and depends on the nucleon mass nonlinearly due to its dependence on $\lambda_N$ and $g_{\pi NN^*}$. The resulting fractional change in $D_{\pi N}$, due to change in $m_N$, is larger, and is in opposite direction ($D_{\pi p} = 7.35\times10^{-2}$, $D_{\pi n} = 6.86\times10^{-2}$ ) compared to that in $g_{\pi NN}$ ($g_{\pi^0 pp}$ =11.602, $g_{\pi^0 nn}$=11.810 ) in the region of interest $M^2 \sim m_N^2$. Finally, it should be kept in mind that the separation of contributions to $\delta g$ , as shown in Table 3.3, is not very clear cut. As is evident from Eqs.( 3.23a,b), $\Delta m_N$ itself arises due to $\gamma$ and $\Delta m_q$, in addition to its dependence on purely phenomenological parameters $\chi$ and $m_{em}$. The smallest contribution to $\delta g$ ($\delta g/g_{\pi NN} \sim \pm 10^{-4}$ ) comes from $\Delta S_0 \neq 0$ and $\Delta S_{0N} \neq 0$. The continuum for the proton may come from a combination of $p\pi^0$ and $n\pi^+$, while that for the neutron may come from a combination of $n\pi^0$ and $p\pi^-$. This is well supported by the fact that the first $\frac{1}{2}^+$ state [N(1440)] decays (60-70)% of the time to $N\pi$. Hence, in an average sense we expect $S_0^p = S_0^n$ for the sum rules (3.22a) and (3.22b), and $S_{0N}^p = S_{0N}^n$ for the sum rules (3.23a) and (3.23b). To get an idea of the effect of the difference of the above continuum thresholds for proton and neutron on $\delta g$, in view of the above argument, we consider this difference to be typically in the range of 0.1% [15].

The resulting value of δg, for the choice $(S_0^p - S_0^n)/\overline{S_0} = (S_{0N}^p - S_{0N}^n)/\overline{S}_{0N} = \pm 0.1\%$, has been displayed in Table 3.3. In view of the very small contribution of $\Delta S_0$ and $\Delta S_{0N}$ to δg, we set them zero in our further analysis.

To sum up, taking into account uncertainties in the quark condensate, the gluon condensate, the twist-4 parameter $\delta^2$, the continuum threshold $S_0$ and the QCD scale parameter, $\Lambda_{QCD}$, we obtain for γ= -0.01 the following estimate of δg and $g_{\pi NN}$:

$$\delta g = -(4.99 \pm 1.97)\times 10^{-2},$$

$$g_{\pi NN} = 11.44 \pm 2.76,$$

$$\delta g / g_{\pi NN} = -(4.36 \pm 1.62)\times 10^{-3}. \qquad (3.25)$$

**TABLE 3.1:** The maximum and minimum values obtained for δg and $g_{\pi NN}$ in the parameter space spanned by $a_u$ = (0.45－0.55) GeV$^3$, b= (0.47－1.0) GeV$^4$, $\delta^2$ = (0.2－0.35) GeV$^2$, $S_0^{p,n}$ = (2.07－2.57) GeV$^2$, $\Lambda_{QCD}$=(0.1－0.15) GeV and $M^2$ = (0.8－1.0)GeV$^2$. The fixed parameters are $S_{0N}^{p,n}$ (the continuum threshold in the mass sum rule)=2.25[15], $m_u$=0.0051, $m_d$=0.0089, $m_0^2$=0.8, μ=0.5, $m_p$=0.93827,$m_n$=0.93957, $f_\pi$=0.093 (all in GeV units), $f_\eta/f_\pi$ =1.1[21].

| $a_u$ GeV$^3$ | b GeV$^4$ | $\delta^2$ GeV$^2$ | $S_0$ GeV$^2$ | $\Lambda_{QCD}$=0.1GeV | | | | $\Lambda_{QCD}$ =0.15 GeV | | | |
|---|---|---|---|---|---|---|---|---|---|---|---|
| | | | | δg×10$^2$ | | $g_{\pi NN}$ | | δg×10$^2$ | | $g_{\pi NN}$ | |
| | | | | γ = －0.01 | γ = －0.00657 | γ = －0.01 | γ = －0.00657 | γ = －0.01 | γ = －0.00657 | γ = －0.01 | γ = －0.00657 |
| 0.55 | 0.47 | 0.35 | 2.57 | -3.02 | -3.35 | 11.03 | 11.02 | -3.23 | -1.98 | 10.32 | 10.30 |
| 0.45 | 1.00 | 0.20 | 2.07 | -6.82 | -4.66 | 12.20 | 12.19 | -6.96 | -4.49 | 11.66 | 11.64 |
| 0.55 | 0.47 | 0.20 | 2.57 | -3.80 | -1.36 | 9.33 | 9.31 | -3.88 | -1.13 | 8.68 | 8.66 |
| 0.45 | 1.00 | 0.35 | 2.07 | -5.79 | -3.41 | 14.19 | 14.18 | -6.06 | -3.35 | 13.58 | 13.56 |

**TABLE 3.2 :** Values of parameters (in GeV units) used for determining central values of δg for different values of $\Lambda_{QCD}$ and γ.

| Central values | $\Lambda_{QCD}$=0.1GeV | | $\Lambda_{QCD}$ =0.15 GeV | |
|---|---|---|---|---|
| | $\gamma = -0.01$ | $\gamma = -0.00657$ | $\gamma = -0.01$ | $\gamma = -0.00657$ |
| | $\delta g = -4.92\times10^{-2}$ $g_{\pi NN}=11.80$ | $\delta g = -3.01\times10^{-2}$ $g_{\pi NN}=12.81$ | $\delta g = -5.09\times10^{-2}$ $g_{\pi NN}=11.15$ | $\delta g = -2.81\times10^{-2}$ $g_{\pi NN}=12.25$ |
| Parameters | | | | |
| $a_u$ | 0.543 | 0.461 | 0.534 | 0.470 |
| b | 0.914 | 0.900 | 0.912 | 0.850 |
| $\delta^2$ | 0.310 | 0.300 | 0.310 | 0.310 |
| $S_0$ | 2.520 | 2.520 | 2.560 | 2.160 |

**TABLE 3.3:** Contribution to $\delta g$ from various symmetry breaking parameters (SBP's) taken to be non zero, one at a time, and also when all SBP's are non-zero. The values of $a_u$, b, $\delta^2$, and $S_0$ have been taken from Table 3.2, so as to give central values of $\delta g$ and $g_{\pi NN}$ as obtained there. $\Delta m_q$= 0.0 means, $m_u = m_d$ = 0.007, $\Delta m_N$ = 0.0 means $m_p = m_n$= 0.93892 (average nucleon mass) along with the coefficients of $\chi$ in the Eqs. (3.23a) and (3.23b) being 5/18 and $m_{em}^2=0$. In the row with $\Delta S_0 \neq 0$ and $\Delta S_{0N} \neq 0$, the two results are for the two signs of $\Delta S_0$ and $\Delta S_{0N}$ respectively.

| Parameters | $\Lambda_{QCD}$ =0.1 GeV | | $\Lambda_{QCD}$=0.15 GeV | |
|---|---|---|---|---|
| | $\delta g\times10^2$ | $g_{\pi NN}$ | $\delta g\times10^2$ | $g_{\pi NN}$ |
| $\alpha = \theta = \Delta m_q = \Delta S_0 = \Delta S_{0N} = \Delta m_N = 0$, $\gamma = -0.01$ | -8.48 | 11.87 | -9.36 | 11.22 |
| $\alpha = \theta = \Delta m_q = \Delta S_0 = \Delta S_{0N} = \Delta m_N = 0$, $\gamma = -0.00657$ | -4.47 | 12.89 | -5.18 | 12.32 |
| $\alpha = 1/137, \gamma = \theta = \Delta S_0 = \Delta S_{0N} = \Delta m_q = \Delta m_N = 0$ | 1.90 | 11.82 | 1.86 | 11.17 |
| $\theta = 0.01, \gamma = \alpha = \Delta S_0 = \Delta S_{0N} = \Delta m_q = \Delta m_N = 0$ | 11.42 | 11.80 | 11.11 | 11.15 |
| $\Delta m_q \neq 0, \gamma = \alpha = \Delta S_0 = \Delta S_{0N} = \theta = \Delta m_N = 0$ | 11.10 | 11.80 | 10.35 | 11.15 |
| $\Delta m_N \neq 0, \gamma = \alpha = \Delta S_0 = \Delta S_{0N} = \theta = \Delta m_q = 0$ | -20.81 | 11.71 | -19.01 | 11.07 |
| $\alpha = \theta = \Delta m_q = \Delta m_N = \gamma = 0.0$ $(S_0^p - S_0^n)/\overline{S_0} = \pm 0.1\%$ $(S_{0N}^p - S_{0N}^n)/\overline{S}_{0N} = \pm 0.1\%$ | 0.26 -0.25 | 11.83 11.83 | 0.30 -0.28 | 11.17 11.17 |

| | | | | | |
|---|---|---|---|---|---|
| All symmetry breaking parameters are non zero | $\gamma = -0.01$ | -4.92 | 11.80 | -5.09 | 11.15 |
| | $\gamma = -0.00657$ | -3.01 | 12.81 | -2.81 | 12.25 |

Using QCD sum rules, in which pion field has been treated as the external field, authors of Ref.[7] have found $\delta g/g_{\pi NN}$ as –0.008, and in the cloudy bag model [22] it is –0.006. As already stated, bulk of the contribution to $\delta g$ comes from the nucleon mass difference $\Delta m_N$. The quark mass difference $\Delta m_q$, and $\pi$-$\eta$ mixing angle $\theta$ contribute to $\delta g$ in the opposite direction, as obtained in[23] also; but these are almost cancelled by the contribution coming from $\delta m_N$. The sign of our result for $\delta g$ differs from that of the three-point function method [6], the chiral bag model [24] and quark-gluon model [25]. In chiral effective field theory, the underlying effective Lagrangian has been extended to include strong isospin-violating and electromagnetic four fermion contact interactions [4]. In these works there is no direct derivation of $\delta g$ or $g_{\pi NN}$, but isospin violation in N-N scattering has been worked out. These authors find that the leading charge symmetry breaking (CSB) effects are four nucleon contact terms of order $\alpha$ and order $\Delta m_q$, while the contribution due to $\Delta m_N$ is rather small. Since this formulation is based on a two-nucleon problem, a direct comparison with our result is difficult. In contrast to chiral effective field theory, in QCD sum rule approach, results based on QCD dynamics are matched to those obtained from effective field theory in an appropriate

Borel window and the quantity of the interest is extracted. Our result of $g_{\pi NN}$ is consistent with that of Ref.[5], and the results of recent measurements [26 ]: $g_{\pi NN} \sim 13-13.5$.

Finally, we will discuss some of the implications of the isospin breaking in the diagonal pion-nucleon coupling constant. Obviously it will contribute to the long range part of the charge asymmetric nuclear potential $V_{CA}=V_{nn}-V_{pp}$ for the $^1S_0$ state. In order to calculate its effect on the difference of pp and nn scattering lengths, we use the phenomenological Argonne $v_{18}$ potential [27] disregarding the electromagnetic potential part. With this potential, using $g_{\pi^0 nn}$ and $g_{\pi^0 pp}$, obtained for $\delta g$ from $-6.0\times10^{-2}\leq \delta g \leq -3.8\times10^{-2}$ and the corresponding $g_{\pi NN}$ from $8.7\leq g_{\pi NN} \leq 13.5$, in the OPEP part of $v_{18}$, we find using the standard method [28] that

$$0.8\ \text{fm} \leq |a_{nn}| - |a_{pp}| \leq 2.3\ \text{fm} \quad (3.26)$$

as against the experimental result [29]:

$$|a_{nn}| - |a_{pp}| = (1.6\pm 0.6)\ \text{fm}. \quad (3.27)$$

Earlier, we had observed that the nucleon mass difference gives the dominant contribution to $\delta g$. Reversing the problem, one may ask how much of the nucleon mass difference arises due to $\delta g$? Analysis of the effect of pion loops on nucleon mass has been done by several authors in effective theories of meson-nucleon interaction [30]. Hecht et al. have concluded that the $\pi N-$loop reduces the nucleon's mass by $\sim(10-20)\%$. Assuming that half of this is due to $\pi^0-$loop, we find that $\delta g$ will give rise to a mass difference $\delta m_n - \delta m_p \approx -(0.25 - 0.5)$ MeV, which is a shift in opposite direction to the actual mass difference of the nucleons. Obviously in this case, we cannot neglect the effect of other heavier meson exchanges, and what we have got is far from the end of the story.

# CHAPTER-IV

## GLUONIC CONTRIBUTION TO THE SELF-ENERGY OF NUCLEON

### 4.1 Introduction

The axial anomaly is known to be one of the most subtle effects of the quantum field theory. In QCD, the most important consequence of the axial anomaly is the fact that the would-be ninth Goldstone boson, the $\eta'$, is massive even in the chiral limit[1]. This extra mass is induced by non-perturbative gluon dynamics[2] and the axial anomaly. The role of gluonic degrees of freedom and OZI violation in the $\eta'$- nucleon system has been investigated through, among others, the flavour–singlet Goldberger-Treiman relation [3], which, in the chiral limit, reads

$$Mg_A^{(0)} = \sqrt{\frac{3}{2}}F_0(g_{\eta' NN} - g_{QNN}) \qquad (4.1)$$

Here $g_A^{(0)}$ is the flavour-singlet axial-charge measured in polarized deep inelastic scattering, $g_{\eta' NN}$ is the $\eta'$-nucleon coupling constant, $g_{QNN}$ is the one-particle irreducible coupling of the topological charge density Q = $\frac{\alpha_s}{4\pi}G\tilde{G}$ to the nucleon. In Eq.(4.1), M is the nucleon mass and $F_0$ renormalizes[4] the flavor-singlet decay constant. The coupling constant $g_{QNN}$ is, in part, related [3] to the amount of spin carried by polarized gluons in a polarized proton. The large mass of $\eta'$ and the small value of $g_A^{(0)}$ (=0.2–0.35), extracted from deep inelastic scattering [5], point to substantial violations of the OZI rule in the flavour-singlet $J^P = 1^+$ channel [6]. A large positive $g_{QNN} \sim 2.45$ is one possible explanation of small value of $g_A^{(0)}\Big|_{\mathrm{DIS}}$.

It is important to look for other significant consequences which are sensitive to $g_{QNN}$. OZI violation in the $\eta'$-nucleon system is a probe of the role of the gluons in dynamical chiral symmetry breaking in low-energy QCD. It will be interesting to calculate the nucleon self-energy due to this kind of gluonic interaction. The gluonic contribution to the nucleon self-energy will be over and above the contributions associated with meson exchange models. It is known[7] that the pion self energy to the nucleon is negative, and it alone contributes (10%-20%) of the nucleon mass. Our objective in this work is to calculate self-energy due to this kind of gluonic interaction.

The perturbation theory used most extensively to study QCD at low energies is generically referred to as chiral perturbation theory, with inclusion of baryons, the effective theory is called baryon chiral perturbation theory, whose non-relativistic limit with respect to baryon is referred to as heavy chiral perturbation theory.

In the conventional chiral perturbation theory, the masses of the ground state baryon octet can be expanded in quark mass as [8] ($m_P^2 \sim m_q$) :

$$M_B = M_0 + \sum_q a_q m_q + \sum_q b_q m^{3/2}{}_q + \sum_q c_q m^2{}_q + \ldots\ldots \qquad (4.2)$$

Borasoy [9] has shown that $\eta'$ can also be included in baryon chiral perturbation theory in a systematic way. Chiral perturbation theory has been a useful tool in the understanding of low-energy QCD hadron dynamics. Its application to baryons through a new formulation of the low energy chiral effective Lagrangian in which the baryons appear as a heavy static field has been introduced. In this approach, $\eta'$ is included as a dynamical field variable instead of integrating it out from the effective field theory. It has a justification in 1/$N_C$ expansion where the axial anomaly is suppressed by powers of 1/$N_C$ and $\eta'$ appears as a ninth Goldstone boson[2].

## 4.2 The Low-Energy Effective Lagrangian

Independent of the detailed QCD dynamics, one can construct low-energy effective chiral Lagrangians which include the effect of the anomaly and axial U(1) symmetry, and use these Lagrangians to study low-energy processes involving the $\eta$ and $\eta'$ with OZI violation. In the meson sector, the $U_A(1)$ extended low energy effective Lagrangian can be written as[10]

$$L_{meson} = \frac{F_\pi^2}{4} Tr(\ ^{\mu}U\ \ _{\mu}U^{+}) + \frac{F_\pi^2}{4} Tr[\chi_0 (U + U^{+})] + \frac{1}{2} iQTr[\log U - \log U^{+}] + \frac{3}{m^2_{\eta_0} F_0^2} Q^2 , \qquad (4.3)$$

where $U = \exp(i\frac{\phi}{F_\pi} + i\sqrt{\frac{2}{3}}\frac{\eta_0}{F_0})$ and $\phi = \ _{k} \phi_k \lambda_k$ with $\phi_k$ denoting the octet of would-be Goldstone bosons($\pi, K, \eta_8$) arising out of spontaneous breaking of chiral $SU(3)_L$ $SU(3)_R$ symmetry. $\eta_0$ is the singlet boson and Q is the topological charge density; $\chi = diag[m_\pi{}^2, m_\pi{}^2, (2m_k{}^2 - m_\pi{}^2)]$ is the meson mass matrix, the pion decay constant $F_\pi$ =92.4 MeV and $F_0$ renormalizes the flavor-singlet decay constant. The $U_A(1)$ gluonic potential involving Q is constructed to reproduce the axial anomaly in the divergence of the renormalized axial-vector current [11]:

$$\partial^{\mu} J_{\mu 5}^{(0)} = \sum_{k=1}^{f} 2i(m_k \bar{q}_k \gamma_5 q_k) + N_f [\frac{\alpha_s}{4\pi} G^{\mu\nu} \tilde{G}_{\mu\nu}] \qquad (4.4)$$

and to generate the gluonic contribution to the $\eta$ and $\eta'$ masses. Here $J_{\mu 5}^{(0)} = \bar{u}\gamma_\mu \gamma_5 u + \bar{d}\gamma_\mu \gamma_5 d + \bar{s}\gamma_\mu \gamma_5 s$, $N_f = 3$, $G_{\mu\nu}$ is the gluon field strength tensor, $\tilde{G}_{\mu\nu} = \frac{1}{2} \varepsilon^{\mu\nu\alpha\beta} G_{\alpha\beta}$, and $Q(z) = \frac{\alpha_s}{4\pi} G^{\mu\nu}(z) \tilde{G}_{\mu\nu}(z)$.

The low-energy effective Lagrangian $L_{meson}$ is readily extended to include $\eta$-nucleon and $\eta'$-nucleon couplings. The chiral Lagrangian for the meson-baryon coupling upto O(p) in the meson momentum is [4]

$$L_{mb} = Tr\bar{B}(i\gamma_\mu D^\mu - M_0)B + FTr(\bar{B}\gamma_\mu \gamma_5 [a^\mu, B]) + DTr(\bar{B}\gamma_\mu \gamma_5 \{a^\mu, B\}) +$$

$$\frac{i}{3} KTr(\bar{B}\gamma_\mu \gamma_5 B) Tr(U^+ \partial^\mu U) - \frac{G_{QNN}}{2M_0} \partial^\mu Q Tr(\bar{B}\gamma_\mu \gamma_5 B) + \frac{C}{F_0^4} Q^2 Tr(\bar{B}B) \qquad (4.5)$$

Here B denotes the baryon octet and $M_0$ denotes the baryon mass in the chiral limit. $D_\mu = \partial_\mu - i v_\mu$ is the chiral covariant derivative, $v_\mu = -\frac{i}{2}(\xi^+ \partial_\mu \xi + \xi \partial_\mu \xi^+)$ and $a_\mu = -\frac{i}{2}(\xi^+ \ _\mu\xi - \xi \ _\mu\xi^+)$ where $\xi = U^{1/2}$. The SU(3) couplings are F= $0.459 \quad 0.008$ and D= $0.798 \quad 0.008$. The axial-vector current has an expansion $a_u = -\frac{1}{2F_\pi} \ _\mu\phi - \frac{1}{2F_0}\sqrt{\frac{2}{3}} \ _\mu\eta_0 + .......$.

In continuum QCD, dynamical chiral symmetry breaking is normally studied using Dyson-Schwinger equation for quark and gluon Green's functions [12]. In low-energy effective theory given by Eqs.(4.3) and (4.5), a flavor independent self-energy of baryons will arise due to interactions of baryons with the topological charge density Q which is a flavor singlet as well as color singlet object. This gluonic term Q has no kinetic energy term, but it mixes with $\eta_0$ to generate gluonic mass term for the $\eta'$. The determination of masses of the physical $\eta$ and $\eta'$ mesons also requires diagonalization of the $(\eta_{8,}\eta_0)$ mass matrix. Thus, part of the $\eta$ mass is also generated by the gluonic term Q [13].

The relativistic framework including baryons poses problem due to the existence of a new mass scale, namely the baryons mass in the chiral limit $M_0$; a strict chiral counting scheme, i.e., a one-to-one correspondence between the meson loops and the chiral expansion does not exist. In order to overcome this problem one integrates out the heavy degrees of freedom of the baryons, similar to a Foldy-Wouthuysen transformation, so that a chiral counting scheme emerges. Observables can then be expanded simultaneously in the Goldstone boson octet masses and the $\eta'$ mass that does not vanish in the chiral limit. One obtains a one-to-one correspondence between the meson loops and the expansion in their masses and derivatives both for octet and singlet [9].

After integrating out the heavy degrees of freedom of the baryons from the effective theory [14] and assigning a four-velocity v to the baryons, the heavy baryon Lagrangian to the order we are working , reads as

$$L_{mb} = Tr(\overline{B}iv.DB) + 2FTr(\overline{B}S_{\mu}[a^{\mu}, B]) + 2DTr(\overline{B}S_{\mu}\{a^{\mu}, B\}) +$$

$$2\frac{i}{3}KTr(\overline{B}S_{\mu}B)Tr(U^{+} \quad ^{\mu}U) - \frac{G_{QNN}}{M_0}( \quad ^{\mu}Q)Tr(\overline{B}S_{\mu}B) + \frac{C}{F_0^4}Q^2Tr(\overline{B}B) \qquad (4.6)$$

where $S_{\mu} = \frac{i}{2}\gamma_5\sigma_{\mu\nu}v^{\nu}$ is the Pauli-Lubanski spin vector .

In this work, our objective is to calculate the masses of baryon octets arising due to gluonic terms within the framework of heavy baryon chiral perturbation theory (HBChPT) including the $\eta'$ .

In baryon chiral perturbation theory, the transition between short and long distance occurs around a distance scale of ~1Fermi which corresponds to the measured size of a baryon, or a momentum scale of 200 MeV which is said to be the separation scale. For long distances the effective field theory is fully correct since it treats baryons and pions as point particles, but this convention does not provide an accurate representation of the physics at distance less than the separation scale. In general, it is not a problem, since high energy effect has the same structure as the terms in the general local lagrangian, so that any incorrect loop contribution can be compensated be a shift of parameters of the Lagrangian. Structure of loops can be understood in this effective theory by separating the short distance and long distance physics within the loop integral.

In practice, such loop effects can cause problem, when the residual short distance contributions are large even after renormalization. But they can be removed by the adjustment of parameters which must be consequently large.

HBChPT is the effective field theory of the standard model at low energies in the hadronic sector which can be successfully applied within the sector of Goldstone bosons. However, traditional SU(3) heavy baryon chiral perturbation theory does not appear to work well. The leading non-analytic component from loop corrections destroy the good experimental agreement which exists at lower order.

The additional contributions have to be compensated by higher order counter terms. This leads to the problems with convergence of chiral series, and problem can be solved using some kind of cutoff regularization instead of common dimensional regularization scheme. Here dimensionally regularized Feynman diagrams carry implicit and large contributions from short distance physics. In contrast, the cutoff scheme picks out the long distance part of the integral, which behaves, as expected, on physical grounds.

We restrict ourselves to the one-loop diagrams of the $\eta$ and $\eta'$ with the vertices arising due to gluonic interactions with the baryons. For this purpose, we use the following matrix elements [15,16]:

$$\langle 0|Q|\eta\rangle = \frac{1}{\sqrt{3}} m^2{}_{\eta} (f_8 \cos\theta - \sqrt{2} f_0 \sin\theta) \tag{4.7}$$

$$\langle 0|Q|\eta'\rangle = \frac{1}{\sqrt{3}} m^2{}_{\eta'} (f_8 \sin\theta + \sqrt{2} f_0 \cos\theta) \tag{4.8}$$

where

$\langle 0|J_{\mu 5}^{(8)}|\eta(p)\rangle = 2if_8 \cos\theta\, p_\mu$ , $\langle 0|J_{\mu 5}^{(8)}|\eta'(p)\rangle = 2if_8 \sin\theta\, p_\mu$ ,

$\langle 0|J_{\mu 5}^{(0)}|\eta(p)\rangle = -\sqrt{6}if_0 \sin\theta\, p_\mu$ , $\langle 0|J_{\mu 5}^{(0)}|\eta'(p)\rangle = \sqrt{6}f_0 \cos\theta\, p_\mu$ , and

$$J_{\mu 5}^{(8)} = \frac{1}{\sqrt{3}}(\bar{u}\gamma_\mu\gamma_5 u + \bar{d}\gamma_\mu\gamma_5 d - 2\bar{s}\gamma_\mu\gamma_5 s)\ . \qquad (4.9)$$

## 4.3 Regulaization of the Self-Mass

Both the one-loop diagrams given by Figs. 4.1(a) and 4.1(b) are divergent. However, we must remember that we are working in an effective field theory which uses the degrees of freedom and the interactions which are correct only at low energy. It has been shown by Donoghue et al.[8] that any incorrect loop contribution coming from short distance physics can be compensated for by a shift of the parameters of the Lagrangian.

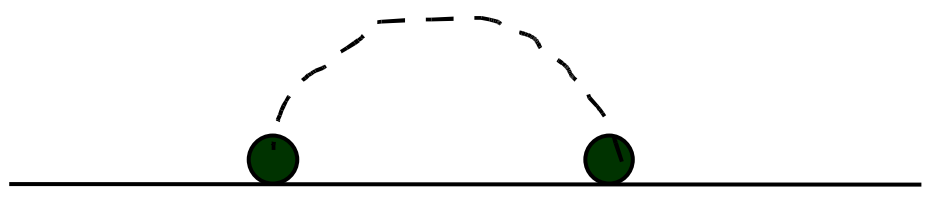

(a)

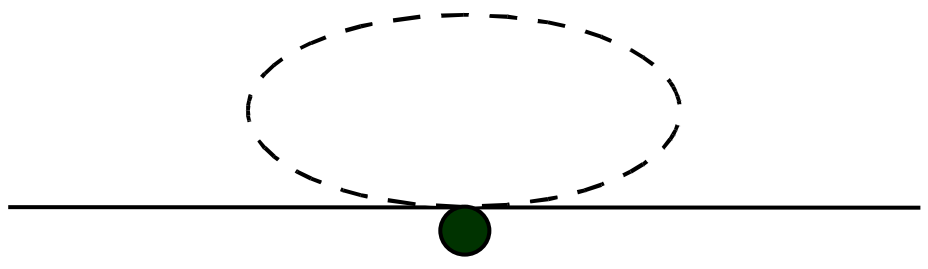

(b)

**Figure 4.1:** (a) Self-energy diagram; (b) Tadpole diagram

Our choice of ultraviolet regulator, which represents a separation scale of long distance physics from the short distance physics, will be dictated by phenomenological considerations. In baryon chiral perturbation theory, which deals with baryons and Goldstone bosons, the separation scale is taken as~1 fm[8] corresponding to the measured size of a baryon. For our problem, we consider an average "gluonic transverse size" of nucleon[17] $\langle\rho^2\rangle$  0.24fm$^2$ corresponding to a dipole parameterization:

$$H_g(x,t) \;\alpha\; (1-\frac{t}{m_g^2})^{-2} \;,\; m_g^2 = 1.1GeV^2 ,\; x \sim 10^{-1} \qquad (4.10)$$

This gives a two-gluon form factor, which we denote by u, of a nucleon [18] and can be used in the self-energy diagram. Another way to look at this problem is that the $U_A(1)$ gluonic potential involving the topological charge density leads to a contact interaction at a "short distance" (~0.2 fm) where glue is excited in the interaction region [4] of the proton-proton collision and then evolves to become an $\eta$ or $\eta'$ in the final state. This will lead to a sharp cutoff at an energy scale ~1GeV. In the tadpole diagram, we may use u, u$^2$, u$^{3/2}$ (geometric mean of the first two), since the phenomenology does not provide any clear-cut rule for this. Similarly, three types of form factors will be used in the tadpole diagram for exponential regularization also. In the monopole case, use of u in the tadpole diagram does not remove the divergence while u$^{3/2}$ remains analytic in a restricted region; hence we use only u$^2$. Specifically, our form factor u(k) for monopole, dipole and exponential regularization has the form:

$u(k) = \Lambda^2/(\Lambda^2 - k^2)$, $\Lambda^4/(\Lambda^2 - k^2)^2$, $\exp(k^2/\Lambda^2)$ . (4.11)

As stated earlier, dimensional regularization scheme is not particularly suitable for effective field theories since it gives large contributions from short distance physics[8]. We have displayed our numerical results for the self-mass of the nucleon coming from both Figs.(4.1a, 4.1b), $\delta m$, in Table 4.1. As discussed above, if we consider the regulator mass $\Lambda$ ； 1GeV for the dipole and the sharp cut-off regularization schemes on phenomenological ground, we observe that $\delta m$ for dipole ($u^{3/2}$-column), exponential ($u^2$-coloumn) and sharp cut-off schemes are approximately same for each mixing angle. Furthermore, $\delta m$ for monopole form factor is related to that for exponential form factor (both for $u^2$-columns) by their regulator scales [7]: $\Lambda_{exp}$ $\sqrt{2}\,\Lambda_{mon}$. Hence, we take

$\Delta m$ ； -0.076 GeV ($\theta = -18.5^o$),

； -0.030 GeV ($\theta = -30.5^o$).

If we take the nontrivial structure of the QCD vacuum into account then in the last term of Eq.(4.6), we can make the replacement $Q^2 \rightarrow \langle Q^2\rangle + Q^2$. $\langle Q^2\rangle$ can be calculated using vacuum saturation hypothesis:

$$\langle Q^2 \rangle = (-1/384)\langle \frac{\alpha_s}{\pi} G^2 \rangle^2 \text{；} -(1/384)(0.012)^2 GeV^8, \qquad (4.12)$$

where for the gluon condensate, we have used the numerical values used by ITEP group[19]. This gives a positive contribution to the nucleon mass:

$\delta m^{(0)}$ ； +0.007 GeV.

Taking this into account, we get the total contribution to the nucleon mass coming from its interaction with the topological charge density $\delta m_{tot}$ ； -(2.5-7.5)% of the nucleon mass. It is known that the one-loop pion contribution to the nucleon mass is $\delta m_{pion}$ ; -(10-20)% of the nucleon mass[7]. Unlike $\delta m_{pion}$ , $\delta m_{tot}$ is flavor independent and is same for all the members of the octet baryon family. This kind of contribution to baryon mass will not arise in models with quark-meson interaction only. It is known that the color-magnetic-field energy in the nucleon is negative[20]. We have not talked about the role of scalar and tensor gluoniums in effective field theories. In particular, scalar gluonium can give rise to Higgs- type mechanism, but this is beyond the scope of the present work.

**TABLE 4.1:** Self- energy of a nucleon, δm, arising due to its interactions with the topological charge density in monopole, dipole, exponential and sharp cut-off schemes as a function of regulator scale Λ. $\eta$ - $\eta'$ mixing angle θ is taken as -18.5°. In dipole and exponential regularizations, in the tadpole diagram the form factor u (which appears at each vertex in the self-energy diagram), $u^{3/2}$ or $u^2$ has been used. Numerical values of Λ and δm are in GeV unit.

| Λ | Monopole | Dipole | | | Exponential | | | Sharp cut-off |
|---|---|---|---|---|---|---|---|---|
| | | u | $u^{3/2}$ | $u^2$ | u | $u^{3/2}$ | $u^2$ | |
| 0.6 | -0.033 | -0.025 | -0.015 | -0.004 | -0.050 | -0.027 | -0.017 | -0.013 |
| 0.8 | -0.078 | -0.055 | -0.036 | -0.010 | -0.112 | -0.062 | -0.040 | -0.036 |
| 1.0 | -0.148 | -0.100 | -0.071 | -0.021 | -0.206 | -0.118 | -0.078 | -0.076 |
| 1.2 | -0.248 | -0.161 | -0.121 | -0.037 | -0.334 | -0.198 | -0.135 | -0.136 |

**TABLE 4.2:** Self- energy of a nucleon, δm, as a function of regulator scale Λ for the same form factors as in Table 4.1, but for θ = -30.5°.

| Λ | Monopole | Dipole | | | Exponential | | | Sharp cut-off |
|---|---|---|---|---|---|---|---|---|
| | | u | $u^{3/2}$ | $u^2$ | u | $u^{3/2}$ | $u^2$ | |
| 0.6 | -0.012 | -0.044 | -0.006 | -0.002 | -0.019 | -0.011 | -0.007 | -0.006 |
| 0.8 | -0.027 | -0.088 | -0.015 | -0.004 | -0.042 | -0.025 | -0.017 | -0.016 |
| 1.0 | -0.049 | -0.148 | -0.027 | -0.009 | -0.073 | -0.045 | -0.032 | -0.030 |
| 1.2 | -0.080 | -0.225 | -0.044 | -0.014 | -0.115 | -0.073 | -0.052 | -0.051 |

**TABLE 4.3:** Self- energy of a nucleon, δm, in dimensional regularization ( $\overline{MS}$ ) scheme, as a function of renormalization point μ .

| μ | θ=-18.5º | θ=-30.5º |
|---|---|---|
| 0.5 | -0.260 | -0.099 |
| 0.7 | -0.163 | -0.062 |
| 1.0 | -0.060 | -0.023 |

# CHAPTER-V

## THE DERIVATIVE OF THE TOPOLOGICAL CHARGE SUSCEPTIBILITY AT ZERO MOMENTUM AND AN ESTIMATE OF η’ MASS IN THE CHIRAL LIMIT

## 5.1 Introduction

The axial vector current in QCD has an anomaly

$$\partial^{\mu}\bar{q}\gamma_{\mu}\gamma_{5}q = 2im_{q}\bar{q}\gamma_{5}q + \frac{\alpha_{s}}{4\pi}G^{a}_{\mu\nu}\tilde{G}^{a\mu\nu} \qquad (5.1)$$

$$\tilde{G}^{a\mu\nu} = \frac{1}{2}\varepsilon^{\mu\nu\rho\sigma}G^{a}_{\rho\sigma}\ . \qquad (5.2)$$

The topological susceptibility $\chi(q^2)$ defined by

$$\chi(q^2) = i\int d^4x e^{iqx}\left\langle 0\middle|T\{Q(x),Q(0)\}\middle|0\right\rangle \qquad (5.3)$$

$$Q(x) = \frac{\alpha_{s}}{8\pi}G^{a}_{\mu\nu}\tilde{G}^{a\mu\nu} \qquad (5.4)$$

is of considerable theoretical interest and is of considerable theoretical interest and has been studied using a variety of theoretical tools like lattice gauge theory, QCD sum rules, chiral perturbation theory etc. In particular the derivative of the susceptibility at $q^2 = 0$

$$\chi'(0) = \left.\frac{d\chi(q^2)}{dq^2}\right|_{q^2=0} \qquad (5.5)$$

enters in the discussion of the proton-spin problem [1-5]. As is well known the first moment of $g_1^p$ can be expressed in terms of the axial charges of the proton:

$$\int_0^1 dx g_1^p(x,Q^2) = \frac{1}{12} C_1^{NS}(\alpha_s(Q^2)(a^3 + \frac{1}{3}a^8) + \frac{1}{9} C_1^s((\alpha_s(Q^2)a^0(Q^2) \quad (5.6)$$

$$\left\langle 0 \middle| A_\mu^{(3)} \middle| p,s \right\rangle = \frac{1}{\sqrt{2}} a^3 s_\mu, \left\langle 0 \middle| A_\mu^{(3)} \middle| p,s \right\rangle = \frac{1}{\sqrt{6}} a^8 s_\mu$$
$$\left\langle p,s \middle| A_\mu^{(0)} \middle| p,s \right\rangle = \frac{1}{\sqrt{2}} a^0(Q^2) s_\mu \quad (5.7)$$

In QCD parton model, the axial charges are represented in terms of moments of parton distribution as

$$a^3 = \Delta u - \Delta d, a^8 = \Delta u + \Delta d - 2\Delta s$$
$$a^0(Q^2) = \Delta u + \Delta d + \Delta s - n_f \frac{\alpha_s}{2\pi} \Delta g(Q^2) \quad (5.8)$$

In naïve parton model $a^0 = a^8$ , the OZI prediction. The 'proton spin' problem is a question of understanding the dynamical origin of the OZI violation $a^0(Q^2) < a^8$. Shore, Veneziano and Narison [1] have shown that

$a^0(Q^2) = \frac{1}{2m_N} 6\sqrt{\chi'(0)}\hat{\Gamma}_{\eta^0 NN} : \eta^0 : OZI$ Goldstone boson, the unphysical state which would become Goldstone boson for spontaneously broken $U_A(1)$ in the absence of anomaly. Ioffe et al.[2] have calculated the part of the proton spin carried by u,d,s quarks in the framework of the QCD sum rules in the external fields. An important contribution comes from the operator, which is the limit of massless u,d,s quarks is equal to $\chi'(0)$.

## 5.2 Calculation of the First Derivative of Topological Susceptibility

In the QCD sum rule approach, one can determine $\chi'(0)$ as follows. Using dispersion relation one can write

$$\frac{\chi'(q^2)}{q^2}-\frac{\chi'(0)}{q^2}=\frac{1}{\pi}\int dsIm\chi(s)[\frac{1}{(s-q^2)^2 s}+\frac{1}{(s-q^2)s^2}]+subtractions \quad (5.9)$$

Defining the Borel transform of a function f(q$^2$) by

$$\hat{B}f(q^2)=\lim_{\substack{-q^2,n\\ -q^2/n=M^2\ fixed}}\frac{(-q^2)^{n+1}}{n!}\frac{d}{dq^{2n}}f(q^2) \quad (5.10)$$

one gets from Eq.(5.9)

$$\chi'(0)=\frac{1}{\pi}\int\frac{ds\,\mathrm{Im}\,\chi(s)}{s^2}(1+\frac{s}{M^2})e^{-s/M^2}-\hat{B}[\frac{\chi'(q^2)}{q^2}] \quad (5.11)$$

According to Eq.(5.3) Im χ(s) receives contribution from all states $|n\rangle$ such that $\langle 0|Q|n\rangle\neq 0$. In particular we have [6]

$$\langle 0|Q|\pi^0\rangle = i\, f_\pi m_\pi{}^2(\frac{m_d-m_u}{m_d+m_u})\frac{1}{2\sqrt{2}} \quad (5.12)$$

The matrix elements, when $|n\rangle$ is $|\eta\rangle$ or $|\eta'\rangle$, can be determined as follows. It is known from both theoretical considerations based on chiral perturbation theory as well as phenomenological analysis that one needs two mixing angles $\theta_8$ and $\theta_0$ to describe the coupling of the octet and singlet axial vector currents to η and η' [7-9] . Introducing the definition

$$\langle 0|J^a_{\mu 5}|P(p)\rangle = if^a_P p_\mu\ ,\quad a=0,8;\ P=\eta,\eta'\ , \quad (5.13)$$

where $J^{8,0}_{\mu 5}$ are the octet and singlet axial currents :

$$J^{8}_{\mu 5} = \frac{1}{\sqrt{6}}(\bar{u}\gamma_{\mu}\gamma_{5}u + \bar{d}\gamma_{\mu}\gamma_{5}d - 2\bar{s}\gamma_{\mu}\gamma_{5}s) \qquad (5.14)$$

$$J^{0}_{\mu 5} = \frac{1}{\sqrt{3}}(\bar{u}\gamma_{\mu}\gamma_{5}u + \bar{d}\gamma_{\mu}\gamma_{5}d + \bar{s}\gamma_{\mu}\gamma_{5}s) \qquad (5.15)$$

The $|P(p)\rangle$ represents either η or η' with momentum $p_{\mu}$. The couplings $f_P^{a}$ can be equivalently represented by two couplings $f_8$, $f_0$ and two mixing angles $\theta_8$ and $\theta_0$ by the matrix identity

$$\begin{pmatrix} f_{\eta}^{8} & f_{\eta}^{0} \\ f_{\eta'}^{8} & f_{\eta'}^{0} \end{pmatrix} = \begin{pmatrix} f_8\cos\theta_8 & -f_0\sin\theta_0 \\ f_8\sin\theta_8 & f_0\cos\theta_0 \end{pmatrix} \qquad (5.16)$$

Phenomenological analysis of the various decays of η and η' to determine $f_p^{a}$ has been carried out by a number of authors [7-9]. In a recent analysis [9] Escribano and Frere find with

$f_8=1.28\, f_{\pi}$ $\quad (f_{\pi}=130.7\text{MeV})$, (5.17)

the other three parameters to be

$\theta_8=(-22.2\pm1.8)^{o}$ , $\theta_0=(-8.7\pm2.1)^{o}$ , $f_0=(-1.18\pm0.04)f_{\pi}$ (5.18)

The divergence of the axial currents are given by

$${}^{\mu}J^{8}_{\mu 5} = \frac{2}{\sqrt{6}}(m_u\bar{u}i\gamma_5 u + m_d\bar{d}i\gamma_5 d - 2m_s\bar{s}i\gamma_5 s) \qquad (5.19)$$

$${}^{\mu}J^{0}_{\mu 5} = \frac{2}{\sqrt{3}}(m_u\bar{u}i\gamma_5 u + m_d\bar{d}i\gamma_5 d + m_s\bar{s}i\gamma_5 s) + \frac{1}{\sqrt{3}}\frac{3\alpha_s}{4\pi}G^{a}_{\mu\nu}\tilde{G}^{a\mu\nu} \qquad (5.20)$$

Since $m_u$, $m_d$ <<$m_s$ one can neglect them [10] to obtain

$$\langle 0|\frac{3\alpha_s}{4\pi}G^{a}_{\mu\nu}\tilde{G}^{a\mu\nu}|\eta\rangle = \sqrt{\frac{3}{2}}m^{2}_{\eta}(f_8\cos\theta_8 - \sqrt{2}f_0\sin\theta_0) \qquad (5.21)$$

$$\langle 0| \frac{3\alpha_s}{4\pi} G^a_{\mu\nu} \tilde{G}^{a\mu\nu} |\eta'\rangle = \sqrt{\frac{3}{2}} m^2_{\eta'} ( f_8 \sin\theta_8 + \sqrt{2} f_0 \cos\theta_0 ) . \tag{5.22}$$

Using Eqns.(5.12), (5.21) and (5.22) we get the representation of $\chi(q^2)$ in terms of physical states as

$$\chi(q^2) = -\frac{m^4_{\pi}}{8(q^2 - m^2_{\pi})} f_\pi^2 (\frac{m_d - m_u}{m_d + m_u})^2 - \frac{m^4_{\eta}}{24(q^2 - m^2_{\eta})} ( f_8 \cos\theta_8 - \sqrt{2} f_0 \sin\theta_0 )^2 - \frac{m^4_{\eta'}}{24(q^2 - m^2_{\eta'})} ( f_8 \sin\theta_8 + \sqrt{2} f_0 \cos\theta_0 )^2 \tag{5.23}$$

+ higher mass states .

On the other hand $\chi(q^2)$ has an operator product expansion [11,12,1,5]

$$\chi(q^2)_{OPE} = -(\frac{\alpha_s}{8\pi})^2 \frac{2}{\pi^2} q^4 \ln(\frac{-q^2}{\mu^2})[1 + \frac{\alpha_s}{\pi}(\frac{83}{4} - \frac{9}{4}\ln(-\frac{q^2}{\mu^2}))]$$

$$-\frac{1}{16}\frac{\alpha_s}{\pi}\left\langle 0|\frac{\alpha_s}{\pi} G^2 |0\right\rangle (1 - \frac{9}{4}\frac{\alpha_s}{\pi}\ln(\frac{q^2}{\mu^2})) + \frac{1}{8q^2}\frac{\alpha_s}{\pi}\left\langle 0|\frac{\alpha_s}{\pi} g_s G^3 |0\right\rangle$$

$$-\frac{15}{128}\frac{\pi\alpha_s}{q^4}\left\langle 0|\frac{\alpha_s}{\pi} G^2 |0\right\rangle^2 + 16(\frac{\alpha_s}{4\pi})^3 \sum_{i=u,d,s} m_i \left\langle \overline{q}_i q_i \right\rangle [\ln(-\frac{q^2}{\mu^2}) + \frac{1}{2}]$$

$$-\frac{1}{2}\int d\rho\, n(\rho)\rho^4 q^4 K_2^2(Q\rho)$$

+ screening correction to the direct instantons. (5.24).

In Eqn.(5.24), the first term arises from the perturbative gluon loop with radiative correction [12], the second, third and the fourth term are from the vacuum expectation values of $G^2$, $G^3$ and $G^4$. The $\left\langle 0|G^4|0\right\rangle$ term has been expressed as $\left\langle 0|G^2|0\right\rangle^2$ using factorization [11]. The fifth term proportional to the quark mass has been computed by us and is indeed quite small compared to other terms numerically. Finally, the last two terms represent the contribution to $\chi(q^2)$ from the direct instantons [11] n(ρ) is the density of instanton of size ρ ,

$K_2$ is the Mc Donald function and $Q^2=-q^2$. In a recent work [13] Forkel has emphasized the importance of screening correction which almost cancels the direct instanton contribution (cf. especially Fig.8 and Secs. V and VI of Ref.[13]. For this reason we shall disregard the instanton terms for the present and return to it later.

From Eq.(5.11), we now obtain

$$\chi'(0) = \frac{f_\pi{}^2}{8}\left(\frac{m_d - m_u}{m_d + m_u}\right)^2 \left(1+\frac{m_\pi^2}{M^2}\right) e^{-\frac{m_\pi^2}{M^2}} + \frac{1}{24}\left(f_8 \cos\theta_8 - \sqrt{2} f_0 \sin\theta_0\right)^2$$
$$\left(1+\frac{m_\eta^2}{M^2}\right) e^{-\frac{m_\eta^2}{M^2}} + \frac{1}{24}\left(f_8 \sin\theta_8 + \sqrt{2} f_0 \cos\theta_0\right)^2 \left(1+\frac{m_{\eta'}^2}{M^2}\right) e^{-\frac{m_{\eta'}^2}{M^2}}$$
$$-\left(\frac{\alpha_s}{4\pi}\right)^2 \frac{1}{\pi^2} M^2 E_0\left(\frac{w^2}{M^2}\right)\left[1+\frac{\alpha_s}{\pi}\frac{74}{4}+\frac{\alpha_s}{\pi}\frac{9}{2}\left(\gamma - \ln\frac{M^2}{\mu^2}\right)\right] \tag{5.25}$$
$$-16\left(\frac{\alpha_s}{4\pi}\right)^3 \frac{1}{M^2} \sum_{i=u,d,s} m_i \left\langle \bar{q}_i q_i \right\rangle - \frac{9}{64}\frac{1}{M^2}\left(\frac{\alpha_s}{\pi}\right)^2 \left\langle \frac{\alpha_s}{\pi} G^2 \right\rangle$$
$$+\frac{1}{16}\frac{1}{M^4}\frac{\alpha_s}{\pi}\left\langle g_s \frac{\alpha_s}{\pi} G^3 \right\rangle - \frac{5}{128}\frac{\pi^2}{M^6}\frac{\alpha_s}{\pi}\left\langle \frac{\alpha_s}{\pi} G^2 \right\rangle^2 .$$

Here $E_0(x)= 1-\exp(-x)$ and takes into account the contribution of higher mass states, which has been summed using duality to the perturbation term in $\chi_{OPE}$ , and W is the effective continuum threshold. We take $W^2 =2.3GeV^2$, and in Fig.5.1 plot the r.h.s of Eq.(5.25) as a function of $M^2$. We take $\alpha_s$=0.5 for μ=1GeV and

$$\left\langle 0 \right| g_s{}^2 G^2 \left| 0 \right\rangle = 0.5\,\mathrm{GeV}^4 \tag{5.26}$$

$\left\langle 0 \right| \bar{s}s \left| 0 \right\rangle =0.8\left\langle 0 \right| \bar{u}u \left| 0 \right\rangle$=-0.8(240MeV)$^3$, $m_s$=150MeV and $m_u/m_d \approx 0.5$ . Writing

$$\left\langle 0 \right| g_s{}^3 G^3 \left| 0 \right\rangle = \frac{\varepsilon}{2}\left\langle 0 \right| g_s{}^2 G^2 \left| 0 \right\rangle , \tag{5.27}$$

we take $\varepsilon=1GeV^2$ . We also have PCAC relation

$$-2(m_u+m_d)\left\langle 0 \right| \bar{u}u \left| 0 \right\rangle = f_\pi{}^2 m_\pi{}^2 \tag{5.28}$$

For $f_0$, $f_8$, $\theta_8$ and $\theta_0$ we use the central values given in Eqs.(5.17) and (5.18).

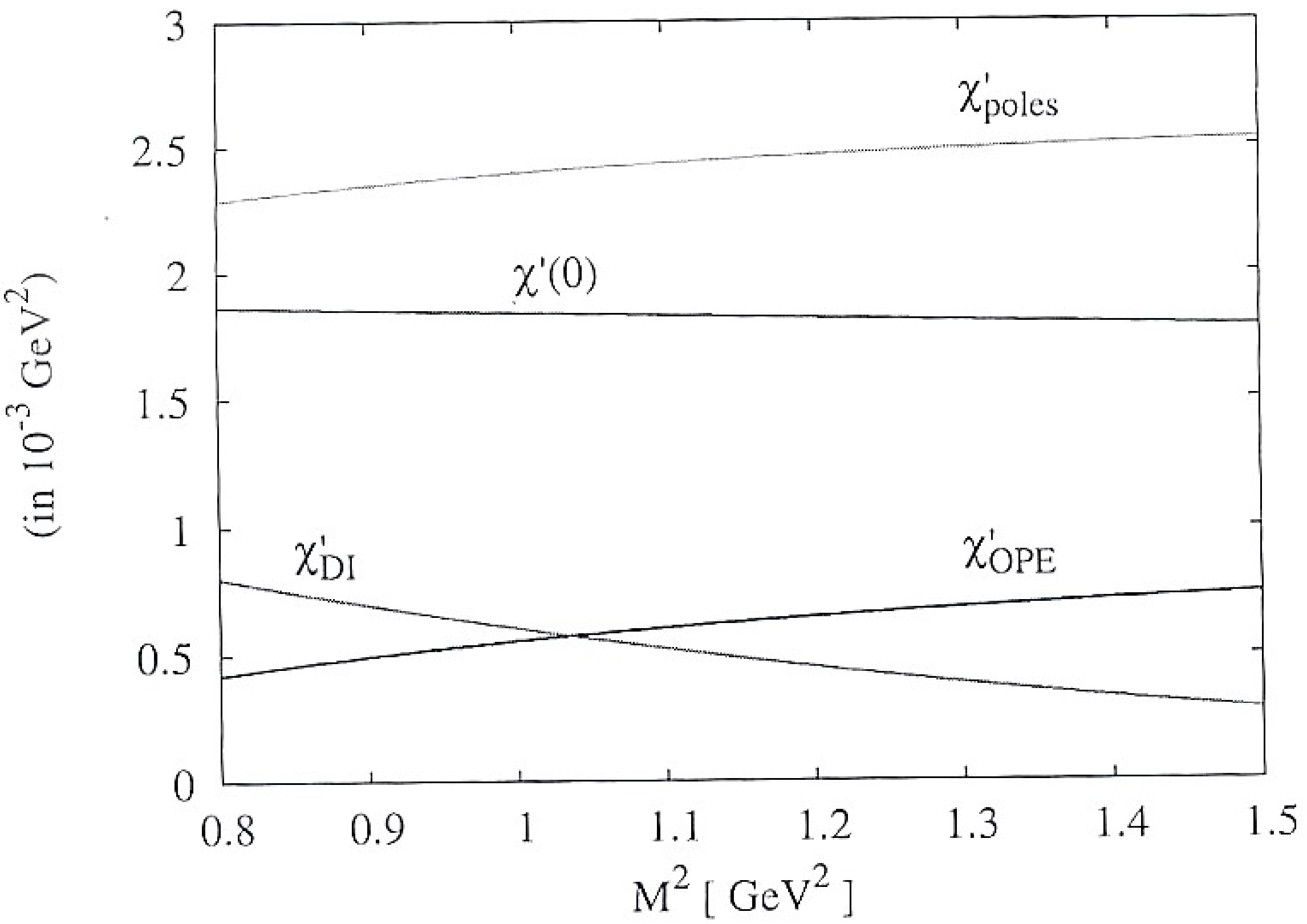

**Figure 5.1:** Various terms contributing to $\chi'(0)$, Eq.(5.25). The value of $\chi'(0)$ is the one obtained without the direct instantons. The latter, see Eq.(5.34), is given by $\chi'_{DI}$, which is larger than $\chi'_{OPE}$ and also has the wrong behaviour suggesting that screening corrections are important.

Let us now examine how the various terms in the r.h.s of Eq.(5.25) add up to remain a constant. The pion term is small and has little variation because of the low mass, η and η' are significantly larger and η is even larger than η'. In Fig.5.1 the upper line gives the combined contribution of π, η, and η' which we denote as $\chi'_{poles}$ and it is seen that it has gentle increase with $M^2$. The OPE terms given by the last three lines in Eq.(5.25), which we denote by $\chi'_{OPE}$, so that

$$\chi'(0) = \chi'_{poles} - \chi'_{OPE}$$

is also plotted in Fig.5.1. It is seen that $\chi'_{OPE}$ is roughly about 25% of $\chi'_{poles}$ also increases with $M^2$, with the result that $\chi'(0)$ is nearly constant w.r.t $M^2$.

We expect this trend of compensating variation in $\chi'_{poles}$ and $\chi'_{OPE}$ to be maintained when variation in $\chi'_{poles}$ due to uncertainties in $\theta_8$, $\theta_0$, $f_8$, $f_0$ [see Eqs. (5.17) and (5.18)] and the variations in $\chi'_{OPE}$ due to uncertainties in the estimates of the vacuum condensates are taken into account. We can then obtain from Fig. 5.1 the value

$$\chi'(0) \approx 1.82\times 10^{-3}\,\text{GeV}^2 \qquad (5.29)$$

We note that the determination, Eq.(5.29) is in agreement with an entirely different calculation by two of us from the study of the correlator of isoscalar axial vector currents

$$\pi_{\mu\nu}{}^{I=0} = \frac{i}{2} \quad d^4 x e^{iq.x} \left\langle 0 \left| \{\bar{u}\gamma_\mu\gamma_5 u(x) + \bar{d}\gamma_\mu\gamma_5 d(x), \bar{u}\gamma_\mu\gamma_5 u(0) + \bar{d}\gamma_\mu\gamma_5 d(0)\} \right| 0 \right\rangle$$

$$\pi_{\mu\nu}^{I=0} = -\pi_1^{I=0}(q)^2 g_{\mu\nu} + \pi_2^{I=0}(q^2) q_\mu q_\nu \qquad (5.30)$$

$\pi_1^{I=0}(q^2=0)$ can be computed from the spectrum of axial vector meson. In Ref.[14] a value

$$\pi_1^{I=0}(q^2=0) = -0.0152\,\text{GeV}^2 \qquad (5.31)$$

was obtained. It is not difficult to see that when $m_u = m_d = 0$

$$\pi_1^{I=0}(q^2=0) = -8\,\chi'(0) \qquad (5.32)$$

which shows consistency between Eqs.(5.29). Let us now return to Eq.(5.24) and consider the effect of incorporating the direct instanton term Eq.(5.25) in the spike approach [5]

$n(\rho) = n_0 \delta(\rho - \rho_c)$ (5.33)

with $n_0 = 0.75\times10^{-3}$ GeV$^4$ and $\rho_c = 1.5$GeV$^{-1}$ The contribution of the instanton to $\hat{B}[\frac{\chi'(q^2)}{q^2}]$ can be found using the asymptotic expansion for $K_2(z)$ and $K'_2(z)$ and we find it to be

$$\chi'_{DI} = \frac{n_0}{4}\sqrt{\pi}\rho_c{}^4 M^2 [M\rho_c + \frac{9}{4}\frac{1}{M\rho_c} + \frac{45}{32}\frac{1}{M^3\rho_c{}^3}] e^{-M^2\rho_c{}^2} \quad (5.34)$$

We have plotted this term separately in Fig.5.1. It is not difficult to see that $\chi'(0)$ will no longer remain constant. This strongly suggests that screening corrections to $[\frac{\chi'(q^2)}{q^2}]$ are important just as they are for $[\frac{\chi(q^2)}{q^2}]$ as found by Forkel [13].

## 5.3 η’-Mass in the Chiral Limit

We now turn to an estimate of η’ mass in the chiral limit. $m_u = m_d = m_s = 0$. In this limit SU(3) flavor symmetry is exact and , we have $m_\pi = m_\eta = 0$ while η’ is a singlet. Let us denote by $\eta_\chi$ = η’($m_s$=0) and $m_\chi = m_{\eta'}$(ms=0) the singlet particle and its mass in the chiral limit. Returning to Eq.(5.24), we first note that the explicitly quark mass dependent term in $\chi_{OPE}$

$$-16(\frac{\alpha_s}{4\pi})^3 \sum_{i=u,d,s} m_i \langle \bar{q}_i q_i \rangle \approx 1.85\times10^{-6}(\text{GeV})^4$$

is numerically much smaller than for example

$$\frac{9}{64}(\frac{\alpha_s}{\pi})^2 \left\langle \frac{\alpha_s}{\pi} G^2 \right\rangle \approx 4.5\times10^{-5}(\text{GeV})^4$$

which is itself much smaller than the perturbative term. In the chiral limit $\langle 0|Q|\pi\rangle = \langle 0|Q|\eta\rangle = 0$

If we assume that the quark mass dependence of $\chi'(0)$ is negligible then $\chi'(0)$ in Eq.(5.25) can also be expressed in terms of $f_{\eta_\chi}$ and $m_\chi$

$$\chi'(0) = \frac{1}{12} f_{\eta_\chi}{}^2 (1+\frac{m^2{}_\chi}{M^2}) e^{-\frac{m^2{}_\chi}{M^2}} \, ) - B[\frac{\chi'_{OPE}(q^2)}{q^2}] \qquad (5.35a)$$

We may then write from Eqs.(5.25) and (5.35a) for $0.8\text{GeV}^2 < M^2 < 1.2\text{GeV}^2$

$$\frac{1}{12} f_{\eta_\chi}{}^2 (1+\frac{m^2{}_\chi}{M^2}) e^{-\frac{m^2{}_\chi}{M^2}} \quad \frac{f_\pi{}^2}{8}(1+\frac{m_\pi{}^2}{M^2}) e^{-\frac{m_\pi{}^2}{M^2}} + \frac{1}{24}(f_8 \cos\theta_8 - \sqrt{2} f_0 \sin\theta_0)^2 (1+\frac{m^2{}_\eta}{M^2}) e^{-\frac{m^2{}_\eta}{M^2}}$$

$$+\frac{1}{24}(f_8 \sin\theta_8 + \sqrt{2} f_0 \cos\theta_0)^2 (1+\frac{m^2{}_{\eta'}}{M^2}) e^{-\frac{m^2{}_{\eta'}}{M^2}} \qquad (5.35b)$$

In Fig.5.2 we have plotted the l.h.s and r.h.s of Eq.(5.35b) in the interval $0.8\ \text{GeV}^2 < M^2 < 1.2\text{GeV}^2$ for $m_\chi$ 723MeV. From this we obtain $f_{\eta_\chi}$ =178MeV which is of the same order as physical decay constants $f_8$ and $f_0$.

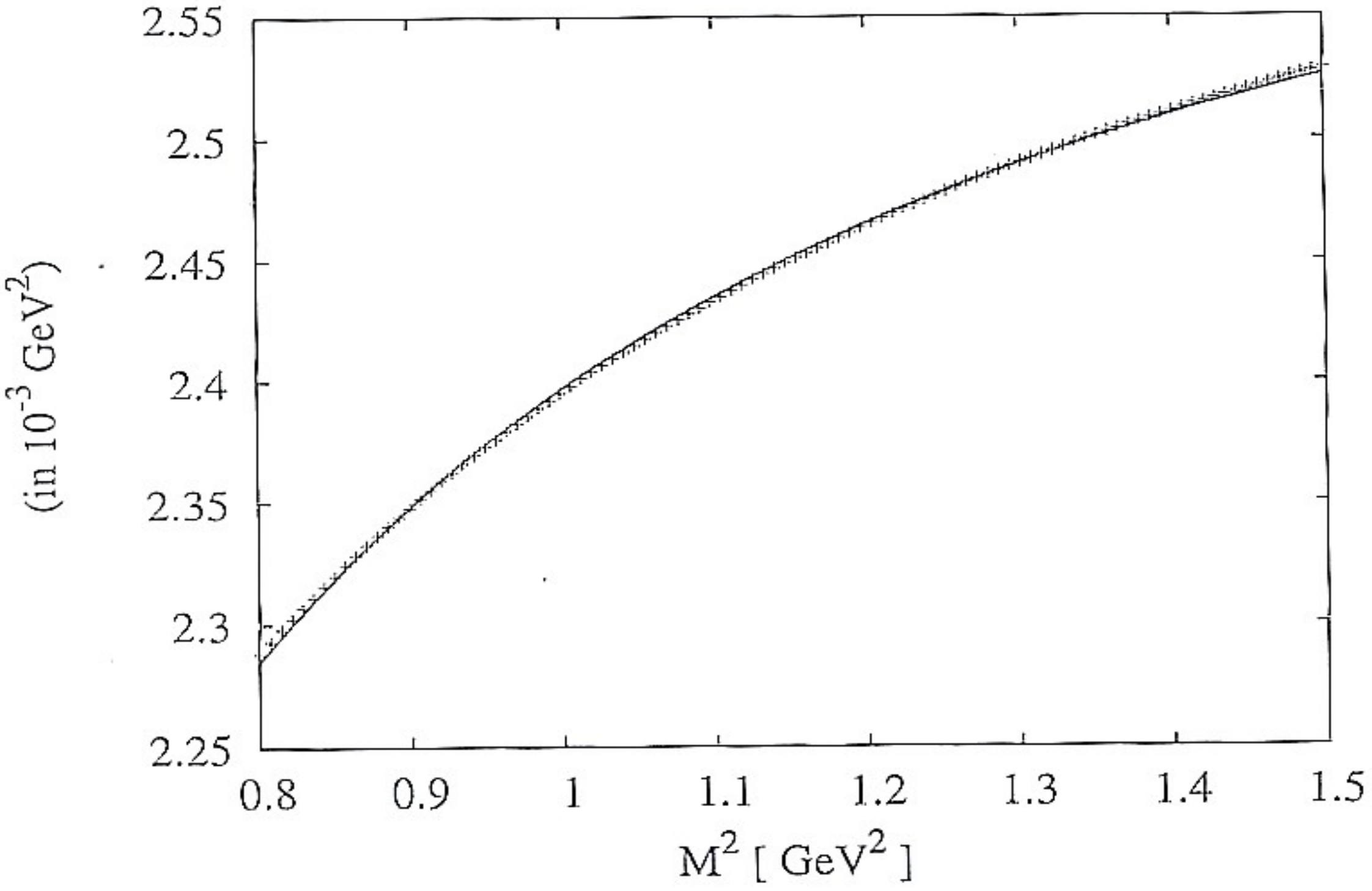

**Figure 5.2**: Estimate of η' mass and coupling in the chiral limit, see Eq.(5.35b). The continuous curve corresponds to $m_\chi$=723 MeV. The continuous line is for l.h.s of Eq.(5.35b) and line with croses is for r.h.s of Eq.(5.35b).

## 5.4 Result and Discussion

We now compare our result for $\chi'(0)$ with some earlier results. In Ref.[1] Narison et al obtained a value for $\chi'(0) \approx 0.7\times10^{-3}(\mathrm{GeV})^2$ substantially different from the value derived here. Since the expression for $\chi_{OPE}$ used by us is identical to theirs, albeit the estimate used for the gluon condensate is slightly different, we need to explain the difference in $\chi'(0)$. The most important difference is in expression of $\chi(q^2)$ in terms of physical intermediate states. We have seen that both η and η' contribute, and in fact η makes a larger contribution than η'. In Ref.[1] only η'(950) state is taken into account. We have also seen that if we were to take the chiral limit then η and η' contribution to $\chi(q^2)$ is representable by $\eta_\chi$ with mass $m_\chi \approx 750$MeV which is substantially different from the physical η' mass. This also explains why Narison et al find stability in the sum rule for rather larger $W^2$=6GeV$^2$ instead of $W^2$=2.3GeV$^2$. We must also add that while our Eq.(5.11) involves only $[\frac{\chi'(q^2)}{q^2}]$, Narision et al use the linear combination of two sum rules (cf. Eq.(6.22 ) of Ref[1]). Comparing with Ref.[5] we note the following: The radiative corrections to the perturbative loop given in Eq.(5.25) viz. $\frac{\alpha_s}{\pi}\frac{74}{4}$, which is large, is ignored in Ref.[5]. We also note that the coefficient of the $\left\langle \frac{\alpha_s}{\pi} G^2 \right\rangle$ arises from radiative corrections, which is also ignored in Ref.[5]. As already remarked, they use physical η' mass even when $m_s$ =0, the chiral limit. Since in the sum rules squares of the masses exp [-(723)$^2$/M$^2$] as against exp[-(958)$^2$/M$^2$] occur, this is a serious error both in Ref.[5] and [1]. Even disregarding all the drawbacks, the sum rule in Ref.[5] for $\tilde{f}_{\eta'}^2$ works rather poorly. It is easy to read off from Fig.1 of Ref.[5] that $\tilde{f}_{\eta'}^2 = 12\chi'(0)$ varies from 0.019 GeV$^2$ at M$^2$=1.5GeV$^2$ to 0.034GeV$^2$ at M$^2$=1.1 GeV$^2$, and grows even faster at lower M$^2$, hardly a constant. This is to be contrasted $\chi'(0)$ as computed here, where it changes barely by 2% within the same range of M$^2$.

In Ref [3], Ioffe and Khodzhamiryan's claim that the OPE for $\chi(q^2)$ does not converge is based on the following. They computed the correlators.

$$q_\mu q_\nu i \quad d^4 x e^{iqx} \left\langle 0 \middle| T\{J^0_{\mu 5}(x), J^q_{\nu 5}(0)\} \middle| 0 \right\rangle \tag{5.36}$$

where $J^q_{\mu 5} = \bar{q}\gamma_\mu \gamma_5 q$ with $m_u$=$m_d$=0 but $m_s$≠0

and $J^0_{\mu 5}$ is flavor singlet current. Introducing the definition

$\left\langle 0 \middle| J^q_{\mu 5}(x) \middle| \eta'(p) \right\rangle$ = i $p_\mu g_{\eta'}{}^q$

they estimated $\dfrac{g_{\eta'}{}^s}{g_{\eta'}{}^u} \approx 2.5$. (5.37)

If SU(3) symmetry was exact this ratio would be unity. Insisting that the ratio in Eq.(5.37) should be close to unity even when $m_s$≠0, they concluded that their result signals a breakdown of OPE[3]. As discussed earlier, $\left\langle 0 \middle| J^8_{\mu 5} \middle| \eta' \right\rangle \neq 0$. In fact using the phenomenological values given in Eqs.(5.17) and (5.18), it is easy to obtain

$$\frac{g_{\eta'}{}^s}{g_{\eta'}{}^u} = \frac{\sqrt{2}[f_0 \cos\theta_0 - \sqrt{2} f_8 \sin\theta_8]}{[f_8 \sin\theta_8 + \sqrt{2} f_0 \cos\theta_0]} \quad 2.24 \tag{5.38}$$

which is enough close to the estimate of Ref.[3]. In Ref.[5] $\theta_8$ was estimated to be -18.8º assuming $\dfrac{f_8}{f_0} = 1.12$ and $\theta_0 = -2.7$ using QCD sum rules. With these values one will still find that the ratio $\dfrac{g^s_{\eta'}}{g^u_{\eta'}}$ =1.96, far different from unity as may be naively expected. As in the case of Narison et al [1], Ioffe and Samsonov [5] and, Forkel [13] also do not take into account the π, η matrix element of the anomaly in their sum rules involving $\chi(q^2)$. We also note that $\chi'(0)$ was estimated in Refs.[2,4] to be $\chi'(0) = (2.3 \quad 0.6) \quad 10^{-3}$ by fitting the QCD sum rule for singlet axial vector matrix element of the proton. We must add, $\chi'(0)$ coincides with the longitudinal part of the SU(3) singlet axial vector current correlator only in the limit zero strange quark mass.

In conclusion we find a value of $\chi'(0) \approx 1.82 \times 10^{-3} GeV^2$ without incorporating direct instantons. Screening corrections to the latter appears to be significant. We also obtained an estimate $m_\chi$=723MeV and $f_{\eta\chi}$=178MeV .

# CHAPTER-VI

## SUMMARY AND CONCLUSION

Nucleon is a microscopic system with complicated structure. The dominant role, in determining its constitution and properties, is played by QCD. At low energies QCD is a non perturbative and intractable. Hence models plays a major role in the studies of nucleonic properties. Nucleonic parameters are calculated in various models appropriate for particular parameter and for particular energy range.

QCD allows existence of rich sea of virtual quarks, antiquarks and gluons accompanied by the valence quarks. Naturally this sea plays an important role in determining the properties of nucleons. Since a fully dynamical calculation with several constituents is a difficult task, the statistical ideas may be important in understanding some of the properties of nucleons. We have used a statistical model in which a nucleon is taken as an ensemble of quark-gluon Fock states and where a probability to find a Fock state is decided by the principle of balance or principle of detailed balance. These quarks and gluons have to be understood as 'intrinsic' partons of the nucleon.

The total flavor-spin-color wave function of a spin-up nucleon has been decomposed in a three-quark core and a sea with definite spin and color quantum numbers and the respective expansion coefficients have been determined. The sea is taken to be flavorless but with angular momentum and color quantum numbers which, when combined with the corresponding quantum numbers of three-quark core makes nucleon spin 1/2 and colorless system respectively. We have used the simplifying approximations in which a quark in the core is not antisymmetrized with an identical quark in the sea, and have treated quarks and gluons as nonrelativistic particles moving in S-wave (except for a single $\bar{q}q$ sea which has been treated separately) motion. Some justification to this approximation comes from the fact that the sum of relativistic quark spin and orbital angular momentum is equal to the sum of nonrelativistic quark spin and orbital angular momentum, and the fact that the quark orbital angular momentum contribution has been shown by some authors to be small. We have also not taken into account any contribution of the s-quark and other heavy quarks, and accounted for only ; 86% of the total Fock states. The number of strange quark-antiquark pairs in the statistical model is 0.05 in the nucleon as compared to the average number of particles which is 5.57. We assume that the rest of the quark-gluon sea spanning~14% of the Fock states of the nucleon also decomposes in color- and spin-subspaces in approximately the same proportion as what we have done explicitly. Furthermore, for the decomposition of a particular Fock state into substates with definite spin and color quantum numbers, it was assumed, in the spirit of statistical approach, that each of these substates occur with equal probability. With these substates, we have calculated the quarks' contribution to the spin of the nucleon, the ratio of the magnetic moments of the nucleons, their weak decay constant and the ratio of SU(3) reduced matrix elements of the axial current. All of these quantities refer to the integrated result of the Bjorken variable. The Melosh rotation effects, which come from the relativistic effect of the quarks' intrinsic transversal motion inside the nucleon, has also been taken into account. The stability of our result against some plausible changes in some physical parameters have been checked by considering two modifications of our above model. In one model $\bar{q}q$ pairs in the sea have been assumed to appear in colorless pseudoscalar form, and in the other states with higher multiplicities have been assumed to be suppressed, their probability being inversely proportional to their multiplicities.

Pion-nucleon coupling plays an important role in investigation of low energy properties of nucleons. Pion being the lightest meson, provides the longest range of the nuclear force. In effective Lagrangian, the pions and nucleons provide the lightest degrees of freedom in the respective categories and the study of their interactions can provide a basis for the study of interaction among heavier hadrons. Hence determination of $g_{\pi NN}$ and its isospin splitting is of interest for particle physics as well as nuclear physics. Even within the hadronic boundary, Goldstone boson exchange model successfully describes diverse phenomenon. The pions generate the non trivial sea to nucleons. This has been used to study the flavor symmetry breaking and the spin structure of the nucleons.

The study of charge symmetry breaking in $g_{\pi NN}$ is an important step for the investigation of charge symmetry breaking effects in NN interaction. We have used QCD sum rule, a nonperturbative method, to study the charge splitting in $g_{\pi NN}$. Within this approach, we have made a systematic study of various factors contributing to $\delta g$: the nucleon mass difference, the quark mass difference, the isospin splitting in $\langle \bar{q}q \rangle$, mixing between $\pi^0$ and $\eta$, and the charge difference between u and d quarks. The nucleon mass difference has been found to make the largest contribution while electromagnetic interaction of quarks makes the lowest contribution.

$g_{\pi NN}$ and $\delta g$ (the splitting in the diagonal pion-nucleon coupling constant) are quantities which can not be measured directly from experiments, but these are phenomenologically important quantities that appear in numerous problems related to nucleons in particle physics and nuclear physics. The successful application of this approach for $g_{\pi NN}$ and $\delta g$ will encourage us to apply it for other hadronic couplings.

In the final part of the thesis, we have studied the effect of the gluonic topological charge density, Q, on nucleon's mass and spin. The coupling of Q to a nucleon gives rise to OZI-violating $\eta$-nucleon and $\eta'$-nucleon interactions. We have studied one-loop self-energy of a nucleon arising due to these interactions using heavy baryon chiral perturbation theory. The small value of $g_A^0\big|_{DIS}$, the flavor-singlet axial charge measured in deep inelastic scattering, points to substantial violations of the OZI rule in the flavor- singlet $J^P=1^+$ channel. One possible explanation for this from the flavor-singlet Goldberger-Treiman relation is the existance of large positive value of the one-particle irreducible coupling of the topological charge density to the nucleon $g_{QNN}$ ~2.45. We have calculated the nucleon self-energy due to this kind of gluonic interaction, which will be over and above the contributions associated with meson exchange models. Conventional type of form factors have been used to regularize the divergences appearing in one loop calculation of the self-energy. The nontrivial structure of the QCD vacuum also contributes to the self-energy of the nucleon through non vanishing value of $\langle Q^2 \rangle$. Taking all this into account, we estimate the total contribution to the nucleon mass from its interaction with the topological charge density to be around -(2.5-7.5)% of the nucleon mass, as compared to -(10-20)% of the nucleon mass coming from one-loop pion diagrams.

The 'proton spin' problem arises due to lack of our understanding the dynamical origin of the OZI violating inequality wherein the measurement shows that the singlet axial charge of proton is less than the corresponding octet charge. It can be shown that the singlet axial charge of a nucleon is related to the first derivative of the susceptibility at zero momentum transfer, $\chi'(0)$. It is in this context that, our evaluation of $\chi'(0)$, is relevant for the study of nucleonic properties.

Our statistical model calculation of nucleonic properties can be improved by incorporating more Fock states and by involving strange quark-antiquark pairs in the sea of the nucleon. We can extend our calculation of charge symmetry breaking of pion-nucleon coupling to the full isospin symmetry breaking by including the couplings of charged pions to the nucleons as well. In the nucleon self-energy calculation due to interaction of topological charge density with the nucleon, the divergences can be renormalized by introducing counterterms in the effective Lagrangian. Finally, one can try to find the contribution of quarks to the nucleon spin by using the value of $\chi'(0)$ that we have obtained.

We conclude that nucleon is a many body complex system whose low-energy behaviour is determined mainly by strong interaction. Non-perturbative approach to QCD, such as QCD sum rule and the QCD based effective theory, and the models such as a statistical model, have a complementary role in exposing different aspects of nucleonic properties.

We conclude that nucleon is a many body complex system whose low-energy behaviour is determined mainly by strong interaction. Non-perturbative approach to QCD, such as QCD sum rule and the QCD based effective theory, and the models such as a statistical model, have a complementary role in exposing different aspects of nucleonic properties.